\documentclass[fleqn,usenatbib]{mnras}

\usepackage[T1]{fontenc}

\DeclareRobustCommand{\VAN}[3]{#2}
\let\VANthebibliography\thebibliography
\def\thebibliography{\DeclareRobustCommand{\VAN}[3]{##3}\VANthebibliography}

\usepackage{graphicx}	
\usepackage{amsmath}	
\usepackage{amssymb}	
\usepackage{bm}
\usepackage{multirow}

\newcommand{\aref}[1]{\hyperref[#1]{Appendix~\ref{#1}}}
\newcommand{\tensor}[1]{\bm{\mathsf{#1}}}
\newcommand{\tf}{\tilde{f}}
\newcommand{\That}{\hat{\mathbf{t}}}
\newcommand{\Nhat}{\hat{\mathbf{n}}}
\newcommand{\Bhat}{\hat{\mathbf{b}}}
\newcommand{\criptic}{\textsc{criptic}}
\newcommand{\Criptic}{\textsc{Criptic}}

\usepackage{scalerel,tikz}
\usetikzlibrary{svg.path}
\definecolor{orcidlogocol}{HTML}{A6CE39}
\tikzset{orcidlogo/.pic={
 \fill[orcidlogocol] svg{M256,128c0,70.7-57.3,128-128,128C57.3,256,0,198.7,0,128C0,57.3,57.3,0,128,0C198.7,0,256,57.3,256,128z};
 \fill[white] svg{M86.3,186.2H70.9V79.1h15.4v48.4V186.2z}
 svg{M108.9,79.1h41.6c39.6,0,57,28.3,57,53.6c0,27.5-21.5,53.6-56.8,53.6h-41.8V79.1z M124.3,172.4h24.5c34.9,0,42.9-26.5,42.9-39.7c0-21.5-13.7-39.7-43.7-39.7h-23.7V172.4z}
 svg{M88.7,56.8c0,5.5-4.5,10.1-10.1,10.1c-5.6,0-10.1-4.6-10.1-10.1c0-5.6,4.5-10.1,10.1-10.1C84.2,46.7,88.7,51.3,88.7,56.8z};
}}
\newcommand\orcidicon[1]{\href{https://orcid.org/#1}{\mbox{\scalerel*{
\begin{tikzpicture}[yscale=-1,transform shape]
\pic{orcidlogo};
\end{tikzpicture}
}{|}}}}

\title[CRIPTIC]{Cosmic Ray Interstellar Propagation Tool using It\^o Calculus (\criptic): software for simultaneous calculation of cosmic ray transport and observational signatures}

\author[Krumholz, Crocker, \& Sampson]{
Mark R. Krumholz$^{\orcidicon{0000-0003-3893-854X}1,2}$\thanks{E-mail: mark.krumholz@anu.edu.au (MRK)}, Roland M. Crocker$^{\orcidicon{0000-0002-2036-2426}1}$, Matt L. Sampson$^{\orcidicon{0000-0001-5748-5393}1}$
\\
$^{1}$Research School of Astronomy and Astrophysics, Australian National University, Canberra ACT 2601 Australia\\
$^{2}$ARC Centre of Excellence for Astronomy in Three Dimensions (ASTRO-3D), Canberra ACT 2601 Australia
}

\date{Accepted XXX. Received YYY; in original form ZZZ}

\pubyear{2022}

\begin{document}
\label{firstpage}
\pagerange{\pageref{firstpage}--\pageref{lastpage}}
\maketitle

\begin{abstract}
We present \criptic, the Cosmic Ray Interstellar Propagation Tool using It\^o Calculus, a new open-source software package to simulate the propagation of cosmic rays through the interstellar medium and to calculate the resulting observable non-thermal emission. \Criptic~solves the Fokker-Planck equation describing transport of cosmic rays on scales larger than that on which their pitch angles become approximately isotropic, 
and couples this
to a rich and accurate treatment of the microphysical processes by which cosmic rays in the energy range $\sim$MeV to $\sim$PeV
lose energy and produce emission. \Criptic~is deliberately agnostic as to both the cosmic ray transport model and the state of the background plasma through which cosmic rays travel. It can solve problems where cosmic rays stream, diffuse, or perform arbitrary combinations of both, and the coefficients describing  these transport processes can be arbitrary functions of the background plasma state, the properties of the cosmic rays themselves, and local integrals of the cosmic ray field itself (e.g., the local cosmic ray pressure or pressure gradient). The code is parallelised using a hybrid OpenMP-MPI paradigm, allowing rapid calculations exploiting multiple cores and nodes on modern supercomputers. Here we describe the numerical methods used in the code, our treatment of the microphysical processes, and the set of code tests and validations we have performed.
\end{abstract}




\section{Introduction}

The last few years have seen an explosion of interest in cosmic rays (CRs) from two distinct angles. The star- and galaxy-formation communities have engaged in intense study of CRs as a form of stellar feedback in star and galaxy formation \citep[e.g.,][]{Enslin07a, Socrates08a, Uhlig12a, Salem14a, Girichidis16a, Wiener17a, Chan19a, Hopkins20a, Crocker21a, Crocker21b}, while the high-energy astrophysics community has paid increasing attention to star-forming galaxies and the CRs within them as important sources at both radio and $\gamma$-ray wavelengths \citep[e.g.,][]{Thompson06a, Lacki10a, Lacki10b, Yoast-Hull16a, Peretti19a, Krumholz20a, Roth21a, Werhahn21b, Werhahn21c, Hopkins22a}. However, studies in both of these areas are hampered by our lack of understanding of the fundamental plasma processes by which CRs couple to the background gas in galaxies through which they propagate. While most plasma physics models predict that CRs should be self-confined at relatively low energy and confined by extrinsic turbulence at high energy \citep[e.g.,][and references therein]{Zweibel17a}, the energy at which this transition occurs, and the normalisation and energy-dependence of the rate of transport in each of these two regimes, remains fundamentally uncertain (and likely dependent on the environment), and simple models have proven challenging to reconcile with observational constraints \citep[e.g.,][]{Hopkins21c, Hopkins21a}. Until we gain a better understanding of how CRs couple to the gas in galaxies, it will be difficult to make definitive statements about either the role of CRs in regulating galaxy formation or the contribution of star-forming galaxies to the non-thermal sky.

One promising avenue for progress in this area is making detailed comparisons between the predictions made by different CR transport models and observations of galaxies' non-thermal emission. During their transport through gas, CRs suffer repeated collisions with the molecules, atoms, and nuclei they encounter. Low-energy CRs ionise the gas, altering its chemistry. High-energy CRs collide directly with gas nuclei, producing sprays of unstable secondary particles that decay almost immediately into final-state particles, including $\gamma$-ray photons, neutrinos, and relativistic electrons and positrons. The $\gamma$-rays and neutrinos are (in principle) directly observable, while the electrons and positrons go on to produce their own radiative signatures at radio and $\gamma$-ray wavelengths. These signatures hold open the possibility of distinguishing between CR transport models, because different models predict different behaviours as a function of CR energy and galactic environment, which in turn manifest as changes in galaxies' non-thermal spectra \citep[e.g.,][]{Krumholz20a, Crocker21a, Ambrosone22a}.

Our observational knowledge of these signatures will expand radically in the next few years as new instruments come online. The Cherenkov Telescope Array (CTA; \citealt{CTA19a}) will be able to see $\gamma$-ray sources an order of magnitude fainter, and with an order of magnitude better resolution, than current instruments. The radio sky, and the galaxies that populate it, will be increasingly-well revealed by the Square Kilometre Array (SKA) and its pathfinders. Finally, the upgrade of the South Polar IceCube neutrino telescope and the commissioning of the KM3NeT neutrino telescope will significantly improve our knowledge of the neutrino sky. This overall improvement in instrumentation will grant us the ability to directly probe the CR populations of star-forming galaxies with unprecedented depth and precision. This will, in turn, illuminate our understanding of CR transport.

However, there are a limited range of software tools capable of predicting the diverse signals that will be observable by the next generation of telescopes. Until very recently, CR-hydrodyamics simulations followed only a single CR energy bin; the most recent generation of simulations include a few distinct CR energies \citep[e.g.,][]{Armillotta21a, Armillotta22a, Girichidis22a, Hopkins22a}, but the computational cost of following multiple energies in the context of a fully self-consistent CR-hydrodynamics calculation means that these simulations achieve very limited spectral resolution in the observable signatures they predict, and rely on highly-simplified treatments of the microphysical interaction between CRs and their environment (e.g., treating $pp$ collisional losses as continuous, ignoring Klein-Nishina effects when calculating inverse Compton scattering). Fluid treatments that involve integration over broad bins in CR energy also necessarily have difficulty in treating sharp spectral features, for example sharp changes between streaming and diffusion at particular CR energies. Moreover, all of these methods have thus far proven too expensive to use in carrying out an extensive parameter study of how emission changes as one makes differing assumptions about the microphysics of CR interactions with the background plasma.

Conversely, a range of tools exist to predict CR observables using detailed microphysics, but generally only for highly-simplified, time-independent background plasma states. The most prominent of these is \textsc{galprop} \citep{Strong98a}, but other examples include 
\textsc{picard} \citep{Kissmann14a},
\textsc{dragon} \citep{Evoli17a}, and \textsc{crpropa} \citep{Merten17a}. These codes generally offer a much more detailed treatment of microphysics than is achieved in the CR-hydrodynamics simulations, and thus a correspondingly better prediction of observables, but at the price that they are not very flexible in terms of the assumed model of CR propagation, or in terms of the way that the background gas is described. Thus for example it would not be straightforward to use one of these tools to post-process a full 3D simulation and predict the observable, CR-driven emission from it. Nor for example can these codes easily handle a situation where the transport of CRs switches from primarily streaming to primarily diffusion as a function of CR energy.

This situation motivates us to introduce Cosmic Ray Interstellar Propagation Tool using It\^o Calculus (\criptic), a new software tool for the purpose of calculating CR transport and observable emission. \Criptic~attempts to balance the advantages of the dedicated CR propagation codes -- high-accuracy microphysics, high spectral resolution, accurate treatment of secondary particles, relatively high speed -- with those of the CR-hydrodynamics codes -- complex, multi-phase gas backgrounds, with CR propagation properties that vary depending on the gas state. Of course it also has disadvantages, in that its complexity level and computational cost are higher than for \textsc{galprop} and similar software, and it does not achieve the full consistency between CR propagation and hydrodynamic evolution that comes from solving CRs and hydrodynamics together. Nonetheless, \criptic~is unique in that it offers the ability to make realistic predictions for CR-driven emission at high spectral resolution and high physical accuracy from 3D simulations of a complex, multi-phase galactic gas ecosystem. In this regard the intended use of \criptic~for CRs is analogous to that of tools such as \textsc{RadMC-3D} \citep{Dullemond12a} or \textsc{Powderday} \citep{Narayanan21a} for photons: while these tools are too expensive to use in real time as part of a self-consistent radiation-hydrodynamics calculation, they offer much more realistic predictions of observable emission than would be possible using radiation-hydrodynamics simulations alone.

The remainder of this paper is laid out as follows. In \autoref{sec:model} we describe the basic system of equations that \criptic~solves, and in \autoref{sec:numerics} we describe the numerical method by which we solve them. We present validation tests in \autoref{sec:tests}, and summarise and discuss future prospects in \autoref{sec:discussion}.

\section{Physical model}
\label{sec:model}

\subsection{Formulation of the problem}

\Criptic~is intended to simulate the transport of CRs on scales significantly larger than the CR mean free path to pitch angle scattering. It therefore solves the Fokker-Planck Equation (FPE) for the evolution of the pitch angle-averaged CR distribution function $f(\mathbf{x},p)$ as a function of position $\mathbf{x}$ and 
the (magnitude of the)
CR momentum $p$; future extensions may also include explicit evolution of the pitch angle distribution, but are beyond the scope of the present paper. We solve a separate FPE for each species of CR tracked in a simulation. The equation we solve is \citep{Skilling75a, Zweibel17a}
\begin{eqnarray}
    \frac{\partial f}{\partial t} & = &
    \frac{\partial}{\partial x_i} 
    \left(
    K_{ij} \frac{\partial f}{\partial x_j}
    \right) + 
    \frac{1}{p^2}\frac{\partial}{\partial p}\left(p^2 K_{\rm pp} \frac{\partial f}{\partial p}\right)
    \nonumber \\
    & & {} - \frac{\partial}{\partial x_i}\left[\frac{\partial}{\partial p^3}\left(p^3 u_i\right) f\right] + \frac{\partial}{\partial p^3}\left[\frac{\partial}{\partial x_i} \left(p^3 u_i\right) f \right] 
    \nonumber \\
    & & {} - \frac{1}{p^2} \frac{\partial}{\partial p}\left(p^2 \dot{p}_{\rm cts} f\right) - L f + S,
    \label{eq:fp}
\end{eqnarray}
where we have added extra terms describing losses to the original transport equation derived by \citet{Skilling75a}; note that in this equation, although by assumption $f$ depends only on the magnitude of CR momentum $p$, it still describes the number of CRs per unit volume in momentum space, i.e., the number of CRs with momentum from $p$ to $p+dp$ is $4\pi p^2 f \, dp$ rather than $f\, dp$. Here and throughout \autoref{sec:model}, we adopt the convention that italic indices ($i,j$) go from 1 to 3 and Greek indices ($\alpha,\beta$) go from 1 to 4, and we make use of the Einstein summation convention whereby repeated indicates are understood to be summed over. The quantities appearing in this equation are the spatial diffusion tensor $\tensor{K}$, the momentum diffusion coefficient\footnote{Note that the subscript $\mathrm{pp}$ here indicates diffusion in momentum space, and should not be confused with proton-proton collisions; to avoid confusion, we will generically refer to proton interactions in general as nuclear inelastic collisions, so that $\mathrm{pp}$ is used only to indicate the momentum direction in phase space.} $K_{\rm pp}$, the total (streaming plus advection) velocity of the CRs $\mathbf{u}$, the rate at which the magnitude of the CR momentum decreases due to continuous processes $\dot{p}_{\rm cts}$ (those where the change in momentum per interaction is small), the catastrophic loss term $L$ (describing processes that destroy CRs completely or cause large changes in their momenta), and the source function $S$. In general these quantities are functions of the properties of the background plasma at position $\mathbf{x}$, the CR momentum $p$, and the CR distribution $f$ at that point in space -- for example, the direction of CR streaming may depend on the gradient of $f$ -- in which case the problem is non-linear. The background plasma is threaded by a large-scale magnetic field $\mathbf{B}$, where large-scale here means that $\mathbf{B}$ is averaged on scales much larger than the CR isotropisation scale.

Before proceeding further, we pause to explain our formulation of the continuous and catastrophic loss terms. We approximate continuous loss processes as causing a continuous decrease in the momentum of each CR particle at a rate $\dot{p}_{\rm cts}$. This corresponds to advection in momentum space at a velocity $-\dot{p}_{\rm cts} \hat{r}_p$, where $\hat{r}_p$ is a radial (in momentum space) unit vector. The term we have included in the \autoref{eq:fp} is simply the divergence of $-\dot{p}_{\rm cts} f \hat{r}_p$. By contrast, the term $-Lf$ we introduce for catastrophic losses corresponds to direct removal of CRs from the population, at a rate proportional to the number density of CRs at a given position; in \autoref{eq:fp}, $L$ is the probability per unit time of such a loss. If the loss process represents a discrete jump in momentum rather than complete destruction, there will also be corresponding source term $S \propto -Lf$. 
We describe in \autoref{ssec:microphysics}
the full set of microphysical processes we include and their mathematical 
representation
in terms of continuous loss terms, catastrophic loss terms, and source terms.

\subsection{From Fokker-Plank to stochastic differential equation}

Our basic approach to solving \autoref{eq:fp} is to transform it from a PDE to a stochastic differential equation (SDE). This part of our treatment is similar to the methods proposed by \citet{Kopp12a} and \citet{Merten17a}. To simplify the problem of transforming the FPE into a SDE, we will always work in a local coordinate system defined relative to the magnetic field. We therefore set up a TNB coordinate frame defined by the unit vectors
\begin{eqnarray}
    \That & = & \frac{\mathbf{B}}{|\mathbf{B}|} \\
    \Nhat & = & \frac{\left(\That\cdot\nabla\right)\That}{k} \\
    \Bhat & = & \That \times \Nhat.
\end{eqnarray}
Here $\That$, $\Nhat$, $\Bhat$, and $k$ are the tangent vector, normal vector, binormal vector, and curvature of the local magnetic field. We therefore adopt the convention in \autoref{eq:fp} that indices $i=(1,2,3)$ correspond to the $(\That,\Nhat,\Bhat)$ directions, respectively. In our chosen frame the diffusion tensor $\tensor{K}$ is diagonal, with elements
\begin{equation}
    K_{ij} = \left\{\begin{array}{ll}
    K_{\parallel}, & i=j=1 \\
    K_{\perp}, & i=j=2\mbox{ or }i=j=3 \\
    0, & \mbox{otherwise.}
    \end{array}\right.
\end{equation}
Note that we explicitly allow for the possibility of diffusion perpendicular to field lines, with diffusion coefficient $K_{\perp}$, and in this respect our equation differs from that given by \citet{Skilling75a} or \citet{Zweibel17a}. Our reason for including this term is that the scale on which $\mathbf{B}$ is measured is \textit{not} assumed to be smaller that the turbulent dissipation scale, and thus there may be turbulent fluctuations in the magnetic field on top of the large-scale guide field $\mathbf{B}$; this will often be the case in galactic or cosmological simulations, for example, where the simulation resolution is insufficient to resolve the magnetic dissipation scale. Thus we wish to leave open the possibility that, while CRs do not diffuse perpendicular to magnetic field lines (to leading order), the field lines themselves can wander perpendicular to the large-scale guide field, and this will induce an effective diffusion in the CRs perpendicular to the large-scale guide field \citep{Beattie22a, Sampson22a}.

In order to transform \autoref{eq:fp} into an SDE, we must first recast it in standard Fokker-Planck form
\begin{equation}
    \frac{\partial f}{\partial t} = - \frac{\partial}{\partial q_{\alpha}} \left(A_\alpha f\right) + \frac{1}{2} \frac{\partial^2}{\partial q_\alpha \partial q_\beta}\left(D_{\alpha\beta} f\right),
    \label{eq:fp_general}
\end{equation}
where $\mathbf{A}$ is the drift vector, $\tensor{D}$ is the diffusion tensor (which must be symmetric), and $\mathbf{q}$ is the vector of phase-space variables upon which $f$ depends, which we take to be $\mathbf{q} = (x_1, x_2, x_3, p)$. In order to do so, we make a change of variables from $f(\mathbf{x},p)$ to
\begin{equation}
    \tf(\mathbf{x},p) = 4 \pi p^2 f(\mathbf{x},p),
\end{equation}
i.e., $\tf$ represents the probability density per radial distance in momentum space, while $f$ is the probability density per unit volume in momentum space. The advantage of this change is that, unlike $f$ itself, $\tf$ is invariant under a steady flow of CRs in momentum space.\footnote{Note that this step is omitted in the derivation given by \citet{Kopp12a} and \citet{Merten17a}, and as a result their equations for the momentum distribution are not correct. However, since none of the tests reported in their papers consider CR momentum evolution, the problem does not manifest in the published results.} With this change of variables, and making some further algebraic simplification, \autoref{eq:fp} becomes
\begin{eqnarray}
    \frac{\partial \tf}{\partial t} & = &
    \frac{\partial^2}{\partial x_i\partial x_j} \left(K_{ij} \tf\right) 
    + 
    \frac{\partial^2}{\partial p^2}\left(K_{\rm pp} \tf \right)
    \nonumber \\
    & & {} - \frac{\partial}{\partial x_i} \left[\left(\frac{\partial K_{ij}}{\partial x_j} + u_i + \frac{p}{3} \frac{\partial u_i}{\partial p}\right) \tf\right] 
    \nonumber \\
    & & - \frac{\partial}{\partial p} \left[\left(
    \frac{\partial K_{\rm pp}}{\partial p} + 2\frac{K_{\rm pp}}{p} - \frac{p}{3}\frac{\partial u_i}{\partial x_i} - \dot{p}_{\rm cts}\right) \tf\right]
    \nonumber \\
    & & {} - L \tf + \tilde{S},
    \label{eq:fp_spherical}
\end{eqnarray}
where $\tilde{S} = 4\pi p^2 S$ is the rate per unit volume per unit linear momentum at which new CRs are injected.\footnote{By contrast, $S$ is the injection rate per unit volume in space per unit volume in momentum space.}
For computational purposes it is convenient to explicitly write out the total velocity $\mathbf{u} = \mathbf{v} + w\That$, where $\mathbf{v}$ is the advection velocity of the background gas and $w \That$ is the streaming velocity of the CRs along the field, in the frame comoving with the gas. Doing so, \autoref{eq:fp_spherical} becomes
\begin{eqnarray}
    \frac{\partial \tf}{\partial t} & = &
    \frac{\partial^2}{\partial x_i\partial x_j} \left(K_{ij} \tf\right) 
    + 
    \frac{\partial^2}{\partial p^2}\left(K_{\rm pp} \tf \right)
    \nonumber \\
    & & {} - \frac{\partial}{\partial x_i} \left\{\left[\frac{\partial K_{ij}}{\partial x_j} + v_i + \left(w + \frac{p}{3} \frac{\partial w}{\partial p}\right)\hat{t}_i\right] \tf\right\} 
    \nonumber \\
    & & - \frac{\partial}{\partial p} \left\{\left[
    \frac{\partial K_{\rm pp}}{\partial p} + 2\frac{K_{\rm pp}}{p} - \dot{p}_{\rm cts}
    \right.\right.
    \nonumber \\
    & & \qquad\quad \left.\left. {} -
    \frac{p}{3}\left(
    \frac{\partial v_i}{\partial x_i} + \frac{dw}{dx_i} \hat{t}_i + w \frac{d\hat{t}_i}{dx_i}
    \right)
    \right] \tf\right\}
    \nonumber \\
    & & {} - L \tf + \tilde{S}.
    \label{eq:fp_spherical2}
\end{eqnarray}

Ignoring the catastrophic loss and source terms for the moment, \autoref{eq:fp_spherical2} corresponds to \autoref{eq:fp_general} with drift vector 
\begin{equation}
    A_\alpha = \left\{
    \begin{array}{ll}
    \frac{\partial K_{\parallel}}{\partial x_\alpha} + v_\alpha + w + \frac{p}{3}\frac{\partial w}{\partial p}, & \alpha=1 \\
    \frac{\partial K_{\perp}}{\partial x_\alpha} + v_\alpha,
    & \alpha=2\mbox{ or }3 \\
    \frac{\partial K_{\rm pp}}{\partial p} + \frac{2}{p} K_{\rm pp} - \dot{p}_{\rm cts} 
    \\
    \quad {} - \frac{p}{3} \left(
    \frac{\partial v_i}{\partial x_i} + \frac{\partial w}{\partial x_i}\hat{t}_i +
    w \frac{\partial \hat{t}_i}{\partial x_i}
    \right), & \alpha=4
    \end{array}\right.
    \label{eq:drift}
\end{equation}
and diffusion tensor 
\begin{equation}
    D_{\alpha\beta} = \left\{
    \begin{array}{ll}
        2K_{\parallel}, & \alpha=\beta=1 \\
        2K_{\perp}, & \alpha=\beta=2\mbox{ or }3 \\
        2K_{\rm pp}, & \alpha=\beta=4 \\
        0, & \alpha \neq \beta
    \end{array}
    \right..
    \label{eq:diffusion}
\end{equation}
Note that, in writing out the components of the drift vector and diffusion tensor, we have made use of the fact that $K_{ij}$ is diagonal in our chosen coordinate system.

The It\^{o} SDE corresponding to \autoref{eq:fp_spherical} is \citep[section 6.1]{Gardiner09a}
\begin{equation}
    dq_\alpha(t) = A_\alpha(\mathbf{q},t) \, dt + d_{\alpha\beta}(\mathbf{q},t) \, dW_\beta(t),
    \label{eq:ito_sde}
\end{equation}
where
\begin{equation}
    d_{\gamma\alpha} d_{\gamma\beta} = D_{\alpha\beta}
\end{equation}
and $d\mathbf{W}(t)$ is a four-dimensional Wiener process. Since $\tensor{D}$ is diagonal in our chosen coordinate frame, we trivially have $d_{\alpha\beta} = \sqrt{D_{\alpha\beta}}$, so $\tensor{d}$ is diagonal as well. For notational convenience we define the vector of diffusion coefficients $\bm{\kappa} = (2 K_\parallel, 2 K_\perp, 2 K_\perp, 2 K_{\rm pp})$, i.e., $\bm{\kappa}$ is just the vector of diagonal elements of $\tensor{d}$, so the SDE becomes
\begin{equation}
    dq_\alpha(t) = A_\alpha(\mathbf{q},t)\,dt + \left(\sqrt{\kappa(\mathbf{q},t)} \, dW\right)_\alpha.
    \label{eq:sde}
\end{equation}
We refer to $\bm{\kappa}$ as the diffusion vector from this point forward, keeping in mind that it in fact just the vector of eigenvalues of a rank-2 tensor.

\subsection{Microphysical processes: losses, secondaries, and observables}
\label{ssec:microphysics}

\subsubsection{Formalism}

The continuous and catastrophic loss terms $\dot{p}_{\rm cts}$ and $L\tf$ can represent a range of microphysical interactions between CRs and their environment. In turn, the catastrophic loss terms generally have corresponding source terms $\tilde{S}$ that represent either new particles produced in the interaction, or existing CRs jumping discontinuously from a higher to a lower momentum; we will generically write these source terms in the form
\begin{equation}
    \tilde{S}_s(p) = \sum_{s'} \int L_{s'}(p') \xi_{s,s'}(p') \left(\frac{d\phi}{dp}\right)_{s,s'}(p') \tf_{s'}(p') \, dp',
    \label{eq:source_generic}
\end{equation}
where $\tilde{S}_s(p)$ is the source function for members of species $s$ with momentum $p$, the sum runs over all species $s'$, $L_{s'}(p')$ is the catastrophic loss rate for members of species $s'$ with momentum $p'$, $\xi_{s,s'}(p')$ is the mean multiplicity for production of species $s$ by the loss of members of species $s'$ with momentum $p'$,\footnote{That is, $\xi_{s,s'}(p') = 2$ means that, on average, a loss of one particle of species $s'$ with momentum $p'$ leads to the production of two particles of species $s$.} $d\phi/dp_{s,s'}(p')$ is the distribution of momenta $p$ for members of species $s$ produced by the loss of a member of species $s'$ with momentum $p'$ (normalised so $\int (d\phi/dp)_{s,s'}(p')\, dp = 1$ for all $p'$), and $\tf_{s'}(p')$ is the CR distribution function for members of species $s'$ evaluated at momentum $p'$. In the discussion that follows, we will characterise the source functions for secondaries in terms of their values for the multiplicity $\xi$ and the momentum redistribution function $d\phi/dp$. We generically refer to members of species $s'$ as primaries, even if they were themselves produced by another, earlier collision. Similarly, we refer to members of species $s$ -- whose appearance is described by $\tilde{S}_s$ -- as secondaries, even if they are, in fact, just the initial CR particle after it has been scattered to lower energy.

In addition to computing the loss and source terms, it is also of interest to predict the observable emission from CRs. We write the rate of specific radiative emission per cosmic ray particle as $d\Psi/d\epsilon$, where
$\epsilon$ is the photon energy. We also calculate the CR ionisation rate via an analogous expression, since this is of interest for astrochemistry.

Different loss processes apply to different CR species, and neither our species list nor the set of processes we include are exhaustive; for example, at present \textsc{criptic} does not treat heavy CR nuclei or the spallation losses they suffer. Both the particle and process list may be expanded in future releases. At present the code tracks CR protons, electrons, and positrons, and it includes accurate treatments of all significant loss processes for those species at energies from $\approx 1$ MeV to $\approx 1$ PeV propagating through typical ISM conditions; on the low-energy end this limit is imposed by adopting the relativistic limit when computing electron radiative losses (our treatment of collisional processes is valid down to $\approx 0.1$ MeV), while on the high-energy side it is limited by the availability of tabulated or analytically-approximated cross sections. Below we describe all the processes we include, and the methods we use to compute them; we summarise our final expressions for all processes in \autoref{tab:loss_processes}. Note that we do not require an explicit additional term to represent streaming losses, as is required in hydrodynamic treatments of CRs, because in the FPE such losses are automatically included in the $\nabla\cdot(w\hat{t})$ term in the drift vector.

\begin{table*}
    \centering
    \begin{tabular}{lcccccc}
        \hline\hline
        Loss process & Affects & $\dot{p}_{\rm cts}$ & $L$ & $\xi$ & $d\phi/dp$ & $d\Psi/d\epsilon$ \\ \hline
        \multirow{2}{*}{Nuclear inelastic scattering} & \multirow{2}{*}{$p$} & 
        \multirow{2}{*}{-} & \multirow{2}{*}{\autoref{eq:loss_nuc}} & 1 ($p$)
        & \autoref{eq:phi_nuc_p} ($p$) & \multirow{2}{*}{\autoref{eq:Psi_nuc}} \\
        & & & & $\sigma_{\pi^\pm}/\sigma_{\rm nuc}$ ($e^\pm$) & \autoref{eq:phi_nuc} ($e^+,e^-$) \\
        \multirow{1}{*}{Ionisation} & $p,e^+,e^-$ & \autoref{eq:dpdt_ion} & - & - & - & - \\
        \multirow{1}{*}{Coulomb} & $p,e^+,e^-$ & \autoref{eq:dpdt_coul} & - & - & - & - \\
        Synchrotron & $e^+,e^-$ & \autoref{eq:dpdt_sync} & - & - & - & \autoref{eq:Psi_sync} \\
        Bremsstrahlung & $e^+,e^-$ & \autoref{eq:dpdt_brem} & \autoref{eq:loss_brem} & 1 ($e^+,e^-$) & \autoref{eq:phi_brem} & \autoref{eq:Psi_brem} \\
        Inverse Compton & $e^+,e^-$ & \autoref{eq:dpdt_IC} & \autoref{eq:loss_IC} & 1 ($e^+,e^-$) & \autoref{eq:phi_IC} & \autoref{eq:Psi_IC} \\
        Positron annihilation & $e^+$ & - & \autoref{eq:loss_pos} & - & - & \autoref{eq:Psi_pos} \\
        \hline\hline
    \end{tabular}
    \caption{Summary of loss processes included. Entries give the process name and the types of particles affected, followed by the equation or expression we use for the various terms -- $\dot{p}_{\rm cts}$, $L$, $\xi$, $\phi$, and $d\Psi/d\epsilon$ -- that describe the rate of continuous momentum loss, the rate of catastrophic loss, the secondary multiplicity, the secondary momentum distribution, and the specific power radiated per CR primary, respectively. For secondary multiplicities and momentum distributions, we indicate the type of secondary particle in parentheses. A blank entry indicates that the process does not produce the indicated effect, e.g., nuclear inelastic scattering does not produce continuous momentum loss.}
    \label{tab:loss_processes}
\end{table*}

\begin{figure*}
    \centering
    \includegraphics[width=0.8\textwidth]{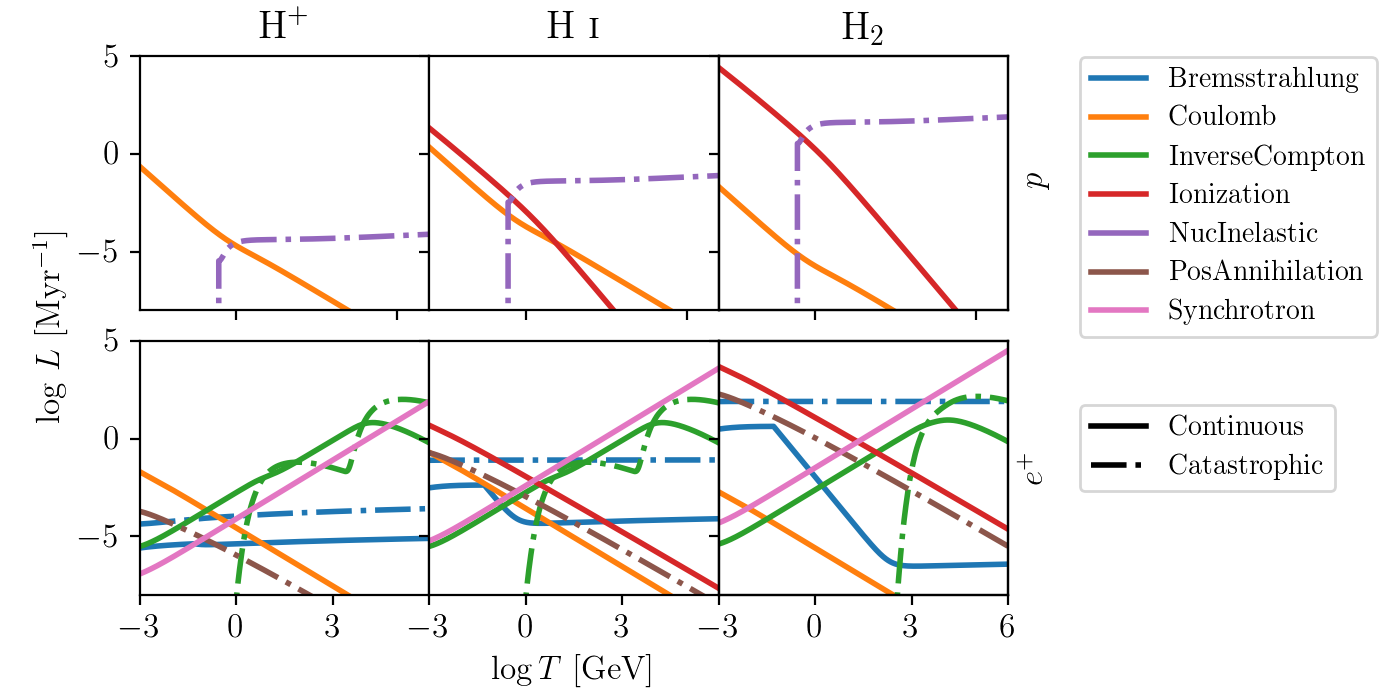}
    \caption{Loss rates due to all of the processes included in \criptic, as a function of CR kinetic energy $T$. For catastrophic losses (dashed lines) we show the catastrophic loss rate $L$, while for continuous processes (solid lines) the loss rate we how is computed as $L_{\rm cts} = \dot{p}_{\rm cts}/p$. Colours indicate different loss processes, as shown in the legend. The top row is for protons, the bottom for positrons; electrons are identical to positrons except that positron annihilation (brown dashed line) does not apply. Columns show, from left to right, loss rates in low-density ionised gas (H$^+$), moderate density atomic gas (H~\textsc{i}), and dense molecular gas (H$_2$); see main text for exact properties of the gas in each of these regions.}
    \label{fig:losses}
\end{figure*}

The remainder of this section describes the various loss processes we include, and how we treat them. To help provide intuition for this discussion, we plot  the various loss rates as a function of CR proton, electron, and positron energy for three example environments in \autoref{fig:losses}; here the catastrophic loss rate is $L$, and we define the equivalent continuous loss rate  as $\dot{p}_{\rm cts}/p$ for a CR packet of momentum $p$. Our example environments are representative of (i) a diffuse ionised region (H$^+$, first column), (ii) a medium density neutral atomic region (H~\textsc{i}, second column), and (iii) a dense molecular region (H$_2$, third column). These regions are characterised by, respectively, total densities of H nuclei $n_{\rm H} = 10^{-3}$, 1, and $10^3$ cm$^{-3}$, magnetic field strengths $B = 1$, 7, and 20 $\mu$G, and dilute blackbody radiation fields, in addition to the cosmic microwave background, with temperature $T_{\rm BB} = 5000$, 5000, and 10 K, and dilution factors $W_{\rm BB} = 10^{-13}$, $10^{-13}$, and 1; we explain $T_{\rm BB}$ and $W_{\rm BB}$ in more detail below. The H$^+$ region is composed of fully ionised H and singly-ionised He with \citet{Asplund09a} protosolar abundances; if we define $X_s$ as the number of members of a particular species $s$ per H nucleon in the background gas, the H$^+$ region has $X_{\rm H^+} = 1$,  $X_{\rm He^+} = 0.0955$, and $X_e = 1.0955$; the atomic region has $X_{\rm H^0} = 0.99$, $X_{\rm H^+} = 0.01$, $X_{\rm He^0} = 0.0955$, and $X_e = 0.01$; and the molecular region has $X_{\rm H_2} = 0.5$, $X_{\rm He^0} = 0.0955$, and $X_e = 10^{-7}$.

\subsubsection{Nuclear inelastic scattering}

CR protons with kinetic energies above a threshold value $T_\pi = (2 + m_\pi/2 m_p) m_\pi c^2$ can scatter inelastically off nuclei in the ISM, producing secondary particles, most commonly $\pi$ mesons; here $m_p = 0.9383$ GeV $c^{-2}$ and $m_\pi = 0.2797$ GeV $c^{-2}$ are the rest masses of the proton and $\pi^0$, respectively. In the process,  the CR proton retains, on average, only a fraction $\eta_{\mathrm{in}} \approx 1/2$ of its initial energy, which makes this a catastrophic loss process. We approximate the total cross section for nuclear inelastic scattering using the analytic fitting formula proposed by \citet{Kafexhiu14a}, which is well-calibrated against both collider experiments and particle physics simulations:
\begin{eqnarray}
    \lefteqn{\sigma_{\rm nuc} = \epsilon_{\rm nuc} \sigma_{pp} \approx \epsilon_{\rm nuc}\left[1 - \left(\frac{T}{T_\pi}\right)^{1.9}\right]^3 \times {}}
    \nonumber
    \\
    & & \left[30.7 - 0.96\, \log\left(\frac{T}{T_\pi}\right) + 0.18\, \log^2\left(\frac{T_\pi}{T}\right)\right]  \;\mathrm{mb},
\end{eqnarray}
where $\sigma_{pp}$ is the inelastic cross section for $pp$ collisions, $T = (\gamma - 1) m_p c^2$ is the kinetic energy of the incident proton in the frame of the ISM, $\gamma = (1 + p^2 / m_p^2 c^4)^{1/2}$ is the proton Lorentz factor, and $\epsilon_{\rm nuc}$ is the ``nuclear enhancement factor'' that represents the increase in the cross section relative to a pure H gas due to the presence of heavier nuclei. We compute the $\epsilon_{\rm nuc}$ from \citeauthor{Kafexhiu14a}'s equation 24, evaluated using the ISM elemental abundances provided by \citet[their Table 1.4]{Draine11a}, and for a beam of CR protons alone. Note that the cross section drops to zero nearly discontinuously near $T = T_\pi$, as is clear from the top row of \autoref{fig:losses}.

For protosolar abundances \citep{Asplund09a}, the mean gas mass per H nucleon is $\mu_{\rm H} m_{\rm H}$, where $m_{\rm H}$ is the hydrogen mass and $\mu_{\rm H} = 1.4$, and thus we can write the catastrophic loss rate for nuclear inelastic collisions as
\begin{equation}
    L_{\rm nuc} = \sigma_{\mathrm{nuc}} v \frac{\rho}{\mu_{\rm H} m_{\rm H}},
    \label{eq:loss_nuc}
\end{equation}
where $\rho$ is the total gas density, $m_{\rm H}$ is the hydrogen mass, and $v = c \sqrt{1 - \gamma^{-2}}$ is the proton velocity. Assuming each loss results in the CR proton surviving but having its kinetic energy reduced by a factor of $\eta_{\rm in} = 2$, we can write the corresponding source function (\autoref{eq:source_generic}) for ``secondary'' protons as having a multiplicity $\xi_{p,p} = 1$ and a momentum redistribution function 
\begin{equation}
    \phi_{\mathrm{nuc},p\to p}(p|p') = \delta\left[T(p) - \eta_{\rm in}T(p')\right],
    \label{eq:phi_nuc_p}
\end{equation}
where $T(p)$ is the kinetic energy of a proton with momentum $p$.

Inelastic collisions also produce both $\gamma$-rays (mainly via the decay of $\pi^0$ mesons) and secondary CR positrons and electrons (mainly via the decays of $\pi^+$ and $\pi^-$ mesons). We compute the $\gamma$-ray emission using the parameterisation provided by \citet{Kafexhiu14a}, which gives the differential cross-section for production of photons of energy $\epsilon$ by CR protons of kinetic energy $T$. This yields an emitted specific power per CR proton
\begin{equation}
    \frac{d\Psi}{d\epsilon} = \frac{f_{\rm nuc}}{\epsilon} A_{\rm max}(T) F(T,\epsilon) v \frac{\rho}{\mu_{\rm H} m_{\rm H}},
    \label{eq:Psi_nuc}
\end{equation}
where $\epsilon$ is the photon energy, and $A_{\rm max}(T)$ and $F(T,\epsilon)$ are parameterised functions of the CR kinetic energy $T$ given by equations 11 and 12 of \citet{Kafexhiu14a}.

To compute the source functions for CR electrons and positrons, we neglect the sub-dominant contribution from $\eta$ and more massive mesons, and focus on that from $\pi^{\pm}$. We compute the cross-sections for production of these products following \citet{Yang18a}; near the production threshold, and $T < 2$ GeV, we use an interpolated cross section taken from the tabulated values shown in \citeauthor{Yang18a}'s figure 4, while at larger energies we follow them in adopting\footnote{Note that in this expression we use $\sigma_{\rm nuc}$, rather that fully adopting \citeauthor{Yang18a}'s approach of using $\sigma_{\rm in}$ but then separately accounting for the contribution of heavier nuclei to pion production using the approach of \citet{Kafexhiu16a}; our approach amounts to assuming that heavier nuclei produce the same ratio of $\pi^+$ to $\pi^-$ and the same pion energy distribution as H. The error associated with this approximation is negligible except at energies $\lesssim 0.1$ GeV, where other processes are generally more important in any event.}
\begin{equation}
    \sigma_{\pi^{\pm}}(p) = \langle n_{\pi^\pm}\rangle \sigma_{\mathrm{nuc}}, \quad
    \left\langle n_{\pi^\pm}\right\rangle = 0.78 \,w^{-1/4}(w-2)^{3/4} -\frac{1}{6} \pm \frac{1}{3}.
    \label{eq:nuc_xi}
\end{equation}
Here $\langle n_{\pi^\pm}\rangle$ is the energy-dependent multiplicity of production of $\pi^{\pm}$ and $w = \sqrt{s}/m_p c^2 = [2 (1 + T/m_p c^2)]^{1/2}$ is the ratio of the center of mass energy to the proton rest mass. Decays of $\pi^{\pm}$ then produce final state $e^\pm$, with the distribution of electron / positron momenta given by
\begin{equation}
    \frac{d\phi_{\mathrm{nuc},p\to e^\pm}}{dp} \propto \frac{dE_{e^\pm}}{dp} \int f_e(x_{e^\pm}) f_\pi(x_{\pi^\pm}) \, dE_{\pi^\pm}.
    \label{eq:phi_nuc}
\end{equation}
Here $x_{\pi^\pm} = E_{\pi^{\pm}}/E_p$ is the energy of the pion created in the collision normalised to the initial proton energy, $x_{e^\pm} = E_{e^\pm}/E_{\pi^\pm}$ is the energy of the electron or positron normalised to the pion energy, and $f_\pi(x)$ and $f_e(x)$ are the distributions of the normalised energies given by equations (6) and (36) of \citet{Kelner06a}, respectively. Thus our final expression for the $e^\pm$ source functions, in terms of \autoref{eq:source_generic}, is that the multiplicity function $\xi_{e^\pm,p} = \sigma_{\pi^{\pm}}/\sigma_{\rm nuc}$, and the momentum redistribution function is given by \autoref{eq:phi_nuc}. Note that we do not at present follow final state neutrinos, although it would be a straightforward extension to the existing code to do so.

\subsubsection{Ionisation}
\label{sssec:ionisation}

At the energies with which we are concerned, ionisation of the background gas is a continuous loss mechanism for all CR particles.\footnote{Our assertion that electron ionisation losses can be treated as continuous differs from the conclusions of \citet{Ivlev21a}, who argue that ionisation losses for electrons must be treated as catastrophic. The difference is the energy range of interest: while their computations follow electrons down to $\sim$keV energies, we are limited to $\gtrsim 1$ MeV by our use of relativistic expressions for radiative loss rates. Calculations using the cross sections given below show that, for a 1 MeV electron, ionising collisions that change the energy of a CR electron by more than 10\% account for only $2.8\times 10^{-5}$ of all collisions, and are collectively responsible for only $8.9\%$ of the total energy loss rate; both these figures decrease as we go to higher electron energies. For this reason, we treat electron ionisation losses as continuous.} We can generically write the resulting rate of momentum loss for a CR of species $s'$ as
\begin{equation}
    \dot{p}_{\rm cts,ion} = \frac{v}{dT/dp} \frac{\rho}{\mu_{\rm H} m_{\rm H}} \sum_s X_s \mathcal{L}_{s',s,\rm ion},
    \label{eq:dpdt_ion}
\end{equation}
where $T$ and $p$ are the kinetic energy and momentum of the CR particle undergoing loss, the sum runs over all species $s$ with which the CR can interact, $X_s$ is the abundance of that species per H nucleon, and $\mathcal{L}_{s',s}$ is the loss function for that species, given by 
\begin{equation}
    \mathcal{L}_{s',s,\rm ion} = \int_0^{W_{s',s,\rm max}} \left(W + I_s\right) \frac{d\sigma_{s',s,\rm ion}}{dW} \, dW.
    \label{eq:ion_lossfunc}
\end{equation}
Here $I_s$ is the ionisation potential of species $s$, $W$ is the kinetic energy of the ejected electron, $W_{s',s,\rm max}$ is the maximum ejected electron kinetic energy allowed by kinematics, and $d\sigma_{s',s,\rm ion}/dW$ is the differential cross section for ejection of electrons with kinetic energy $W$ by collisions between CRs of species $s'$ and targets of species $s$. We include H, H$_2$ and He as target species, since these overwhelmingly dominate the ionisation losses; however, note that it is trivial to extend the formalism we describe below to include other targets.

We follow \citet{Ivlev21a} in taking our differential cross sections for ionisation by CR protons from the semi-analytic model of \citet{Rudd92a}, which gives
\begin{equation}
    \label{eq:dcs_ion_p}
    \frac{d\sigma_{p,s,\rm ion}}{dw} = \sigma_0 \left(\frac{R}{I_s}\right)^2
    \frac{F_{1,s}(T_p) + F_{2,s}(T_p) w}{(1 + w)^3},
\end{equation}
where $w = W/I_s$, $\sigma_0 = 4\pi a_0^2 N_s$, $a_0$ is the Bohr radius, $N_s$ is the number of electrons in the outer shell of the target species, $R = 13.6$ eV is the Rydberg energy, $T_p$ is the proton kinetic energy, and $F_{1,s}$ and $F_{2,s}$ are empirical fitting functions given by equations 43 and 44 of \citeauthor{Rudd92a}, which depend on $T_p$ and the properties of the target species, but not on $W$. The maximum energy allowed by kinematics is $W_{p,s,\rm max}= 4 (m_e/m_p) T_p - I_s$, and inserting this and \autoref{eq:dcs_ion_p} into \autoref{eq:ion_lossfunc}, the loss function is
\begin{eqnarray}
    \mathcal{L}_{p,s,\rm ion} & = & \sigma_0 I_s \left(\frac{R}{I_s}\right)^2
    \left[
    F_{2,s}(t) \log\left(1+ w_{p,s,\rm max}\right)
    +
    \vphantom{\left(F_{1,s}(t)-F_{2,s}(t)\right)\frac{w_{p,s,\rm max}}{1+w_{p,s,\rm max}}}
    {}
    \right.
    \nonumber \\
    & & \quad \left.
    \left(F_{1,s}(t)-F_{2,s}(t)\right)\frac{w_{p,s,\rm max}}{1+w_{p,s,\rm max}}
    \right],
\end{eqnarray}
where $w_{p,s,\rm max} = 4 t - 1$ and $t = (m_e/m_p) (T_p / I_s)$. The total ionisation cross section is
\begin{eqnarray}
\sigma_{p,s,\rm ion} & = & \sigma_0 \left(\frac{R}{I_s}\right)^2 
\left[
F_{2,s}(t)\frac{w_{p,s,\rm max}}{2(1+w_{p,s,\rm max})^2} + {}
\right.
\nonumber \\
& & \quad \left.
F_{1,s}(t) \frac{w_{p,s,\rm max}(2+w_{p,s,\rm max})}{2(1+w_{p,s,\rm max})^2}
\right].
\label{eq:sigma_ion_p}
\end{eqnarray}
Note from \autoref{fig:losses} that there is a clear inflection in the proton loss function near 1 GeV, where protons transition from sub-relativistic to relativistic.

Similarly, we take our differential cross sections for ejection of electrons of energy $W$ by CR electrons of energy $T_e$ from the RBEQ model of \citet[their equation 19, modified as per the description of the BEQ approximation in their text]{Kim00b},
\begin{eqnarray}
\label{eq:dcs_ion_e}
\lefteqn{\frac{d\sigma_{e,s,\rm ion}}{dw} = \sigma_0 \frac{\alpha^4}{2 i_s'\left(\beta_t^2 + \beta_{u_s}^2 + \beta_{i_s}^2\right)}
\times {} }
\\
& &
\left\{
\frac{Q_s-2}{t+1}\left(\frac{1}{1+w}+\frac{1}{t-w}\right)\frac{1+2t'}{(1+t'/2)^2}
+ {}
\phantom{\log\frac{\beta_t^2}{1-\beta_t^2}}
\right.
\nonumber \\
& &
\left(2-Q_s\right)\left(\frac{1}{(1+w)^2} + \frac{1}{(t-w)^2} + \frac{{i'_s}^2}{(1+t'/2)^2}\right)
+ {}
\nonumber \\
& &
\left.
\left(\frac{Q_s}{(1+w)^3}+\frac{Q_s}{(t-w)^3}\right)
\left[\log\frac{\beta_t^2}{1-\beta_t^2}-\beta_t^2 -\log(2i'_s)\right]
\right\},
\nonumber
\end{eqnarray}
where $\alpha$ is the fine structure constant and $Q_s$ and $U_s$ are the dimensionless dipole strength and mean outer shell electron kinetic energy for the target species; we take these latter two quantities from Table 1 of \citet{Kim94a}. In the expression above,
primes indicate quantities normalised to the electron rest energy (i.e., $t' = T_e/m_e c^2$, $i'_s = I_s/m_e c^2$), lower case indicates quantities normalised to the ionisation potential (i.e., $t = T_e / I_s$), and $\beta$'s indicate the $\beta$ factor corresponding to a particular energy (i.e., $\beta_t = [1 - 1/(t'+1)^2]^{1/2}$). The maximum ejected electron energy normalised to the ionisation potential is $w_{e,s,\rm max} = (t - 1)/2$, and inserting this and \autoref{eq:dcs_ion_e} into \autoref{eq:ion_lossfunc} gives
\begin{equation}
\mathcal{L}_{e,s,\rm ion} = I_s \sigma_0 f_{\rm ion} \left(g_{\rm ion,1} + g_{\rm ion,2}\right)
\end{equation}
where
\begin{equation}
f_{\rm ion} = \frac{\alpha^4}{2 i_s'\left(\beta_t^2 + \beta_{u_s}^2 + \beta_{i_s}^2\right)},
\end{equation}
\begin{eqnarray}
\lefteqn{g_{\rm ion,1} = }
\\
& & \frac{Q_s-2}{t+1}\left(\log\frac{2}{t+1} + t\log\frac{2t}{t+1}\right)\frac{1+2t'}{(1+t'/2)^2}
+ {}
\phantom{\log\frac{\beta_t^2}{1-\beta_t^2}}
\nonumber \\
& &
\left(2-Q_s\right)\left[\log\frac{(1+t)^2}{4t} + \frac{{i'_s}^2}{(1+t'/2)^2}\left(\frac{t-1}{2}\right)\right]
+ {}
\nonumber \\
& &
Q_s\frac{(t-1)^2}{2t(t+1)}
\left[\log\frac{\beta_t^2}{1-\beta_t^2} -\beta_t^2 -\log(2i'_s)\right],
\nonumber \\
\lefteqn{g_{\rm ion,2} = }
\\
& & \frac{Q_s}{2}\left(1-\frac{1}{t^2}\right)
\left[\log\frac{\beta_t^2}{1-\beta_t^2} -\beta_t^2 -\log(2i'_s)\right] + {}
\nonumber
\\
& &
(2-Q_s)\left[
1 - \frac{1}{t} - \frac{\ln t}{t+1}\frac{1+2t'}{(1+t'/2)^2} + \frac{{i'_s}^2}{(1+t'/2)^2}\frac{t-1}{2}
\right].
\nonumber
\end{eqnarray}
The total ionisation cross section is
\begin{equation}
    \label{eq:sigma_ion_e}
    \sigma_{e,s,\rm ion} = \sigma_0 f_{\rm ion} g_{\rm ion,2}
\end{equation}

At the $\gtrsim 1$ MeV energies with which we are concerned, positrons have almost exactly the same total ionisation cross sections as electrons \citep[e.g.,][]{Knudsen90a}, and it is therefore reasonable to assume similarly-identical differential cross sections. However, there is one subtlety: for electrons the kinematic limit $w_{e,s,\rm max} = (t - 1)/2$ is a result of the incident and ejected electrons being identical particles, so it is not possible to say which is the ``primary'' CR electron and which is the ``secondary'', ejected electron. The value of $w_{e,s,\rm max}$ for electrons amounts to treating whichever electron has lower energy as the secondary. For positrons, the incident and ejected particles are distinguishable, so it is not obvious what to choose for $w_{e^+,s,\rm max}$. Fortunately this choice makes relatively little difference for $\gtrsim 1$ MeV energies, since losses are dominated by collisions that eject electrons with $w \ll w_{\rm max}$; we therefore adopt the same kinematic limit for positrons as for electrons, and thus the same loss function, but warn that this approach would not be valid at lower energies.

In addition to the calculating losses, \criptic~reports the total ionisation rate for each target species. For a CR packet of species $s'$, the rate per primary CR at which ionisations of background species $s$ occurs is $\zeta_s = \sigma_{s',s,\rm ion} v_{\rm CR}$, where $v_{\rm CR}$ is the CR velocity, and $\sigma_{s',s,\rm ion}$ is the total ionisation cross section, given by \autoref{eq:sigma_ion_p} for protons and \autoref{eq:sigma_ion_e} for electrons and positrons.

\subsubsection{Coulomb losses}

In an ionised medium, all CRs lose energy via Coulomb interactions with the surrounding electrons. As with ionisation, this process is well-approximated as continuous at the energies with which we are concerned. We take our loss rates from \citet{Gould72a}, interpolating smoothly between the expressions provided for the classical limit, $\beta < \alpha$ (where $\beta=v/c$ is the CR velocity normalised to $c$ and $\alpha$ is the fine structure constant), the non-relativistic limit, $\alpha < \beta \ll 1$, and the ultrarelativistic limit, $1-\beta \ll 1$; these transitions are visible in the loss rates plotted in \autoref{fig:losses}, at least for protons. Our expressions are generically of the form
\begin{equation}
    \dot{p}_{\rm cts,Coul} = \left(\frac{dE_{\rm CR}}{dp}\right)^{-1} \frac{e^2 \omega_p^2}{v} B_{\rm stop},
    \label{eq:dpdt_coul}
\end{equation}
where $E_{\rm CR}$ is the CR energy,
\begin{equation}
    \omega_p = \sqrt{\frac{4 \pi n_e e^2}{m_e}}
\end{equation}
is the plasma frequency, $m_e$ is the electron mass, $n_e$ is free electron density, and $B_{\rm stop}$ is the stopping number, which is a function of $E_{\rm CR}$ with a different functional form for different CR particle types. For protons, interpolating between \citeauthor{Gould72a}'s three cases gives 
\begin{equation}
    B_{p,\rm stop} = \ln\left(\frac{2 \gamma m_e c^2 \beta^2}{\hbar \omega_p}\right) + \frac{1}{2} \ln\left[1 + \left(\frac{\Gamma_C\beta}{\alpha}\right)^2\right] - \frac{\beta^2}{2},
\end{equation}
where $\Gamma_C = 0.5615$ is a numerical constant. The equivalent expression for electrons is
\begin{eqnarray}
    B_{e^-,\rm stop} & = & \ln\left[\frac{\sqrt{2\delta(\gamma-1)} \beta m_e c^2}{\hbar\omega_p}\right] + \frac{1}{2}\ln\left[1+\left(\frac{\Gamma_C\beta}{\alpha}\right)^2\right]
    \nonumber \\
    & & {} + \frac{1}{2}\left[\left(1 + \frac{2\gamma-1}{\gamma^2}\right) \ln\left(1-\delta\right)
    + \frac{\delta}{1-\delta}\right]
    \nonumber \\
    & &
    {} + \frac{\delta^2}{4}\left(\frac{\gamma-1}{\gamma}\right)^2,
\end{eqnarray}
and for positrons is
\begin{eqnarray}
    \lefteqn{
    B_{e^+,\rm stop} = \ln\left[\frac{\sqrt{2\delta(\gamma-1)} \beta m_e c^2}{\hbar\omega_p}\right] + \frac{1}{2}\ln\left[1+\left(\frac{\Gamma_C\beta}{\alpha}\right)^2\right]
    }
    \nonumber \\
    & &
    {} + \frac{\delta^2}{4} \left(\frac{\gamma-1}{\gamma+1}\right)^2
    \left[
    \frac{1}{2}+\frac{1}{\gamma}+\frac{3}{2\gamma^2} -
    \left(\frac{\gamma-1}{\gamma}\right)^2\left(\frac{2\delta}{3}-\frac{\delta^2}{2}\right)
    \right]
    \nonumber \\
    & &
    {} + \delta\left[\frac{\delta}{8}\left(\frac{\gamma-1}{\gamma}\right)^2 - \frac{\gamma^2-1}{2\gamma^2}\right]
    \nonumber \\
    & &
    {} - \frac{\delta}{2}\left(\frac{\gamma-1}{\gamma+1}\right)
    \left[
    \frac{\gamma+2}{\gamma} - \frac{\gamma^2-1}{\gamma^2}\delta + \left(\frac{\gamma-1}{\gamma}\right)^2\frac{\delta^2}{3}
    \right].
\end{eqnarray}
Here $\gamma$ is the CR Lorentz factor, and $\delta$ is a fitting parameter for which we adopt the recommended value $\delta=1/2$.

\subsubsection{Synchrotron radiation}

Synchrotron radiation is a continuous loss mechanism for CR electrons and positrons.\footnote{In sufficiently strong magnetic fields synchrotron losses can be significant for protons as well, but at present we do not include these. It would be straightforward to do so in the future, however.} The rate of energy loss is given by the usual loss formula for an isotropic distribution of pitch angles \citep[e.g.,][]{Blumenthal70a},
\begin{equation}
    \left(\frac{dE}{dt}\right)_{\rm sync} = -\frac{4}{3}\sigma_T c \beta^2 \gamma^2 U_B,
\end{equation}
where $\beta=v/c$, $\sigma_T$ is the Thomson cross section, and $U_B$ is the magnetic energy density. We therefore have
\begin{equation}
    \dot{p}_{\rm cts,sync} = \frac{4}{3} \sigma_T c \beta^2 \gamma^2 U_B \left(\frac{dE_{e^\pm}}{dp}\right)^{-1}.
    \label{eq:dpdt_sync}
\end{equation}
The corresponding time- and direction-averaged specific power that an observer sees per CR electron or positron is
\begin{equation}
    \frac{d\Psi}{d\epsilon} = \frac{\sqrt{3} e^3 B h}{m_e c^2} \int_{0}^{\pi} \sin\alpha \frac{\nu}{\nu_{c,\perp}} \int_{\nu/(\nu_{c,\perp} \sin\alpha)}^\infty K_{5/3}(\xi)\, d\xi \, d\alpha,
    \label{eq:Psi_sync}
\end{equation}
where $\nu = \epsilon/h$ is the photon frequency and $\nu_{c,\perp} = 3 e B \gamma^2/4 \pi m_e c$ is the cutoff frequency for CRs with a pitch angle $\alpha=\pi/2$, and $K_{5/3}(\xi)$ is the modified Bessel function of order $5/3$. Note that synchrotron radiation is not isotropic with respect to the local magnetic field; the quantity we compute here is the mean power radiated over $4\pi$ sr. Similarly, we do not at present compute the polarised intensity that would be seen by an observer in a particular direction.

\subsubsection{Bremsstrahlung}

Bremsstrahlung is a catastrophic loss process for electrons and positrons, since, at least in the relativistic regime, energy loss is dominated by photons whose energies are a significant fraction of the CR energy. We adopt the bremsstrahlung differential cross sections given by \citet{Blumenthal70a},
\begin{equation}
    \frac{d\sigma_{e^\pm,s,\rm br}}{d\epsilon} = \frac{\alpha r_e^2}{\epsilon} \left\{
    \left[1 + \left(\frac{E_{e^\pm}-\epsilon}{E_{e^\pm}}\right)^2\right] \phi_1(\Delta) - \frac{2}{3} \frac{E_{e^\pm}-\epsilon}{E_{e^\pm}} \phi_2(\Delta), 
    \right\}
\end{equation}
where $\epsilon$ is the energy of the photon, $E_{e^\pm}$ is the total (rest plus kinetic) energy of the CR electron or positron, $\alpha$ is the fine structure constant, $r_e$ is the classical electron radius, 
\begin{equation}
    \Delta = \frac{\epsilon m_e c^2}{4 \alpha Z E_{e^\pm} (E_{e^\pm}-\epsilon)},
    \label{eq:delta}
\end{equation}
is the screening factor, $Z$ is the nuclear charge, and $\phi_1$ and $\phi_2$ are the screening functions. For species $s$ consisting of unshielded charges (free protons, electrons, and He nuclei),
\begin{equation}
    \phi_1 = \phi_2 = -4 Z^2 \left[\ln\left(2\alpha Z \Delta\right)+\frac{1}{2}\right],
\end{equation}
while for shielded nuclei (H, He, and He$^+$), we use the tabulated screening functions provided by \citeauthor{Blumenthal70a}.

The functional form of the differential cross section requires some care in our numerical treatment; the most obvious approach is to integrate the differential cross section to compute a total cross section, and set the catastrophic loss rate to be proportional to it. The problem is that, as a result of the $1/\epsilon$ dependence, the total cross section is logarithmically divergent, even though the total energy loss rate is finite. To handle this situation, we divide bremsstrahlung losses into a catastrophic component, representing losses due to photons with energies larger than $f_{\rm cat}=1/10$ of the CR kinetic energy, and a continuous part, accounting for losses due to lower-energy photons. For the catastrophic part the cross section is
\begin{equation}
    \sigma_{e^\pm,s,\rm br-cat} = \int_{f_{\rm cat} T_{e^\pm}}^{T_{e^\pm}} \frac{d\sigma_{e,s,\rm br}}{d\epsilon}\,d\epsilon,
\end{equation}
where $T_{e^\pm} = E_{e^\pm} - m_e c^2$ is the CR kinetic energy; the corresponding catastrophic loss rate
\begin{equation}
    L_{\rm br} = \frac{\rho}{\mu_{\rm H} m_{\rm H}} v \sum_s X_s \sigma_{e^\pm,s,\rm br-cat},
    \label{eq:loss_brem}
\end{equation}
where $v$ is the CR velocity, $\rho$ is the gas density, and $X_s$ is again the abundance of species $s$ per H nucleon. The momentum distribution function for the scattered CRs is
\begin{equation}
    \frac{d\phi_{\rm br}}{dp} \propto \frac{dT_{e^\pm}}{dp} \sum_s X_s \left(\frac{d\sigma_{e^\pm,s,\rm br}}{d\epsilon}\right)_{\epsilon=T_{e^\pm}(p')-T_{e^\pm}(p)},
    \label{eq:phi_brem}
\end{equation}
where the differential cross section $d\sigma_{e^{\pm},s,\rm br}/d\epsilon$ is evaluated at a photon energy $\epsilon$ corresponding to the different between the initial, $T_{e^\pm}(p')$, and final, $T_{e^\pm}(p)$, CR kinetic energies, and, following our catastrophic-continuous division, we set the catastrophic loss cross section to zero if $T_{e^\pm}(p) < f T_{e^\pm}(p')$. 

For the continuous part of the losses, we define a loss function by
\begin{equation}
    \mathcal{L}_{e^\pm,s,\rm br} = \int_0^{f_{\rm cat} T_{e^\pm}} \epsilon \frac{d\sigma_{e^\pm,s,\rm br}}{d\epsilon}\,d\epsilon,
\end{equation}
analogously to the ionisation loss function introduced in \autoref{sssec:ionisation}. The corresponding loss rate is
\begin{equation}
    \dot{p}_{\rm cts,br} = \frac{v}{dT_{e^\pm}/dp} \frac{\rho}{\mu_{\rm H} m_{\rm H}} \sum_s X_s \mathcal{L}_{s',s,\rm br}.
    \label{eq:dpdt_brem}
\end{equation}

We show the continuous and catastrophic loss rates in \autoref{fig:losses}. Note that the continuous loss rate drops sharply at energies $T_{e^\pm}\gtrsim 0.1$ GeV for the H~\textsc{i}- and H$_2$-dominated regions, but that a similar transition does not occur for the catastrophic loss rate. This is a direct result of the behaviour of atomic shielding, which is is significant when $\Delta \ll 1$. Examining \autoref{eq:delta}, we see that $\Delta \ll 1$ when $\alpha E_{e^\pm} \gg m_e c^2$ and $E_{e^\pm} \gg \epsilon$ are both satisfied. The former condition is met only when $T_{e^\pm} \gtrsim 0.1$ GeV, while the latter is met only for the continuous part of the loss rate, which is why we see shielding effects only for continuous losses at high energy. Also note that the bremsstrahlung loss rate, while it drops at $T_{e^\pm}\gtrsim 0.1$ GeV, eventually stabilises and becomes constant at yet higher energy. This is as direct result of the ionisation fraction being non-zero even in H~\textsc{i}- and H$_2$-dominated regions; at sufficiently high energy, the continuous loss rate becomes dominated by the residual population of free protons and electrons, which are unaffected by shielding.

Finally, the specific power radiated by bremsstrahlung photons per CR primary is
\begin{equation}
    \frac{d\Psi}{d\epsilon} = \frac{\rho}{\mu_{\rm H} m_{\rm H}} v \epsilon \sum_s X_s \frac{d\sigma_{e^\pm,s,\rm br}}{d\epsilon}.
    \label{eq:Psi_brem}
\end{equation}
Note that this includes photon emission at all energies; we separate catastrophic and continuous losses for the purposes of calculating CR propagation, but there is no need to separate them when computing photon emission.

\subsubsection{Inverse Compton scattering}

Inverse Compton (IC) scattering of electrons and positrons can be either a continuous or catastrophic loss process depending on the initial CR Lorentz factor $\gamma$ and initial photon energy $\epsilon'$. In the Thomson limit, which applies when
\begin{equation}
    \Gamma(\epsilon',\gamma) \equiv \frac{4\gamma \epsilon'}{m_e c^2} \ll 1,
\end{equation}
continuous loss is a good approximation, while in the Klein-Nishina regime, $\Gamma \gtrsim 1$, the CR typically loses a substantial fraction of its energy with each scattering \citep{Blumenthal70a}. \Criptic~must be able to operate in both regimes, since, for example, we wish to be able to model both $\sim$GeV CRs interacting with CMB photons ($\Gamma \sim 10^{-6}$) and $\sim 10$ TeV CRs interacting with visible or UV photons ($\Gamma \sim 10^3$). We therefore make use of the general Klein-Nishina expression for the cross section, rather than the simplified Thomson cross section. In this general case, the differential rate at which IC scattering produces photons of energy $\epsilon$ is \citep{Jones68a, Blumenthal70a}
\begin{equation}
    \frac{d\dot{N}}{d\epsilon} = \frac{2\pi r_e^2 c}{\gamma^2} \int_0^\infty \frac{1}{\epsilon'} \frac{dN'}{d\epsilon'} S(\epsilon, \epsilon',\gamma) \, d\epsilon',
    \label{eq:IC_scat_rate}
\end{equation}
where $r_e$ is the classical electron radius, $dN'/d\epsilon'$ is the specific number density of the photons being scattered, and
\begin{equation}
    S(\epsilon, \epsilon', \gamma) =
    \left\{
    \begin{array}{ll}
    2 q \ln q + (1+2q)(1-q) + {} \\
    \qquad \frac{1}{2} \frac{[\Gamma(\epsilon',\gamma) q]^2}{1 + \Gamma q}(1-q), & 0 < q < 1 \\
    0, & \mathrm{otherwise}
    \end{array}
    \right.
    \label{eq:klein-nishina}
\end{equation}
where $q = \epsilon/[\Gamma(\epsilon',\gamma) (\gamma m_e c^2 - \epsilon)]$. Here $S$ is a dimensionless function describing the shape of the Klein-Nishina cross section.

To proceed further, we assume that the radiation field with which CRs are interacting can be described as a sum of dilute blackbodies, characterised by a temperature $T_{\rm BB}$ and a dilution factor $W_{\rm BB}$; the CMB has $W_{\rm BB}=1$, $T_{\rm BB}=2.73$ K, while the starlight field of the Milky Way is well-approximated by three components with $(W_{\rm BB}, T_{\rm BB}/\mathrm{K}) = (7\times 10^{-13}, 3000), (1.7\times 10^{-13}, 4000), (1\times 10^{-14}, 7500)$ \citep{Mathis83a, Draine11a}. The corresponding photon number density for each component is
\begin{equation}
    \frac{dN'}{d\epsilon'} = W_{\rm BB}\frac{8\pi}{(hc)^3}\frac{\epsilon'^2}{e^{\epsilon'/k_B T_{\rm BB}}-1}.
\end{equation}
At this point it is convenient to define non-dimensional versions of the initial photon energy $\epsilon'$, the final photon energy $\epsilon$, and the blackbody temperature; we therefore define
\begin{equation}
    x = \frac{\epsilon'}{k_B T_{\rm BB}} \qquad y = \frac{\epsilon}{\gamma m_e c^2}
    \qquad \Gamma_{\rm BB} = \frac{4 \gamma k_B T_{\rm BB}}{m_e c^2}.
\end{equation}
With these definitions, we can rewrite \autoref{eq:IC_scat_rate} for a single component of the radiation field as
\begin{equation}
    \label{eq:ICscatrate}
    \frac{d\dot{N}}{dy} = \frac{\alpha^3}{8\pi \gamma^3} \left(\frac{c}{r_e}\right)  W_{\rm BB} \Gamma_{\rm BB}^2 \frac{dF_{\rm IC}}{dy},
\end{equation}
where $\alpha$ is the fine structure constant,
\begin{equation}
    \frac{dF_{\rm IC}}{dy} \equiv \int_{x_{\rm min}}^\infty \frac{x}{e^x-1} S(y,x,\gamma)\, dx,
    \label{eq:dFdy}
\end{equation}
describes the differential scattering rate in normalised photon energy units, $x_{\rm min} = y/[\Gamma_{\rm BB} (1-y)]$ is the minimum normalised initial photon energy that can produce a scattered photon with normalised energy $y$ (i.e., the minimum value of $x$ for which $q<1$), and $S(y,x,\gamma)$ is given by \autoref{eq:klein-nishina}, but with the substitution $\Gamma(\epsilon',\gamma) \to x \Gamma_{\rm BB}$ and $q \to y / [\Gamma_{\rm BB} x (1-y)]$. The total scattering rate and energy loss rate due to a single radiation field component are then
\begin{eqnarray}
\dot{N} & = & \frac{\alpha^3}{8\pi \gamma^3} \left(\frac{c}{r_e}\right)  W_{\rm BB} \Gamma_{e,\rm BB}^2 \int_0^1 \frac{dF_{\rm IC}}{dy}\, dy \\
\dot{E} & = & \frac{\alpha^3}{8\pi \gamma^2} \left(\frac{c}{r_e}\right)  W_{\rm BB} m_e c^2 \Gamma_{e,\rm BB}^2 \int_0^1 y \frac{dF_{\rm IC}}{dy}\, dy.
\end{eqnarray}

In the integrals above, the parts of the integrands at $y \ll 1$ are well-approximated as continuous loss, while those from $y$ near unity correspond to losses that should be treated catastrophically. Following our approach with bremmstrahlung, we handle this by somewhat arbitrarily placing the boundary between the continuous at catastrophic regimes at $y = f_{\rm cat} = 1/10$, and we therefore set the continuous loss rate to
\begin{equation}
    \dot{p}_{\rm cts,IC} = \left(\frac{dT_{e^\pm}}{dp}\right)^{-1} \frac{\alpha^3}{8\pi \gamma^2}\left(\frac{c}{r_e}\right) m_e c^2 \sum_i W_{i,\rm BB} \Gamma_{i,\rm BB}^2 G_{\rm IC}(\Gamma_{i,\rm BB}),
    \label{eq:dpdt_IC}
\end{equation}
where we have defined
\begin{equation}
    G_{\rm IC}(\Gamma_{i,\rm BB}) \equiv \int_0^{f_{\rm cat}} y \frac{dF_{\rm IC}}{dy}\, dy.
\end{equation}
The sum in \autoref{eq:dpdt_IC} runs over all the components of the radiation field, and $\Gamma_{i,\rm BB}$ and $W_{i,\rm BB}$ are the values that apply to the $i$th component. Similarly, the catastrophic loss rate is
\begin{equation}
    L_{\rm IC} = \frac{\alpha^3}{8\pi \gamma^3} \left(\frac{c}{r_e}\right) \sum_i W_{i,\rm BB} \Gamma_{i,\rm BB}^2 F_{\rm IC}(\Gamma_{i,\rm BB}),
    \label{eq:loss_IC}
\end{equation}
where
\begin{equation}
    F_{\rm IC}(\Gamma_{i,\rm BB}) \equiv \int_{f_{\rm cat}}^1 \frac{dF_{\rm IC}}{dy}\, dy.
\end{equation}
In the limit $\Gamma_{\rm BB}\to 0$, the functions $F_{\rm IC}$ and $G_{\rm IC}$ have the property that $G_{\rm IC}\to (\pi^4/135) \Gamma_{\rm BB}^2$ and $F_{\rm IC}/G_{\rm IC} \to 0$, so the continuous loss rate approaches the usual expression for the Thomson limit, and the catastrophic loss rate becomes negligible in comparison. On the other hand, in the limit $\Gamma_{\rm BB}\to \infty$, we have $G_{\rm IC}/F_{\rm IC}\to 0$, and catastrophic losses dominate. This behaviour is visible in \autoref{fig:losses}, where we see continuous losses being dominant at low CR energy and giving way to catastrophic losses at higher energy; also note that, in the H$^+$ and H~\textsc{i} regions, there are two distinct peaks of catastrophic loss, one at higher energy arising from the cosmic microwave background photon field with $T_{\rm BB} = 2.73$ K and one at lower energy from the starlight field with $T_{\rm BB} = 5000$ K.

For the part of IC losses that we treat as catastrophic, since electrons and positrons are conserved, the multiplicity $\xi_{e^{\pm},e^{\pm}} = 1$. Consistent with our division between continuous and catastrophic losses, the momentum redistribution function is
\begin{equation}
    \frac{d\phi_{\rm IC}}{dp} = \frac{1}{E_{e^\pm}} \frac{dE_{e^\pm}}{dp} \frac{\sum_i W_{i,\rm BB} \Gamma_{i,\rm BB}^2 \Theta(y - f_{\rm cat}) \frac{dF_{i,\rm IC}}{dy}}{\sum_i W_{i,\rm BB} \Gamma_{i,\rm BB}^2 F_{\rm IC}(\Gamma_{i,\rm BB})},
    \label{eq:phi_IC}
\end{equation}
where $y = 1 - E_{e^\pm}(p')/E_{e^\pm}(p)$, and the purpose of the Heaviside step function $\Theta(y - f_{\rm cat})$ is to enforce our approach that we only treat as catastrophic interactions that cause the CR energy to change by more than $f_{\rm cat} = 10\%$ at a time. Finally, the specific radiated power per CR primary is
\begin{equation}
    \frac{d\Psi}{d\epsilon} = 
    \frac{\alpha^3}{8\pi \gamma^3} \left(\frac{c}{r_e}\right) y
    \sum_i W_{i,\rm BB} \Gamma_{i,\rm BB}^2 \frac{dF_{i,\rm IC}}{dy},
    \label{eq:Psi_IC}
\end{equation}
where in this expression we do not include the step function because the observable emitted power includes both continuous and catastrophic losses.

\subsubsection{Positron annihilation}

The final (catastrophic) loss process we include is positron annihilation with electrons in the background gas. For the relativistic energies with which we are concerned, positronium formation is unimportant compared to direct annihilation, and Coulomb corrections to the annihilation cross section are small \citep[see the review by][]{Prantzos11a}, so the catastrophic loss rate is well-approximated by
\begin{equation}
    L_{\rm annih} = \sigma_d v \frac{\rho}{\mu_e m_{\rm H}},
\end{equation}
where $v$ is the CR positron velocity, $\mu_e m_{\rm H}$ is the mean mass per electron ($\mu_e = 1.17$ for \citealt{Asplund09a} abundances), and $\sigma_d$ is the \citet{Dirac30a} cross section, which for a CR positron with Lorentz factor $\gamma$ is given by
\begin{equation}
    \sigma_d = \frac{\pi r_e^2}{\gamma+1}\left[
    \frac{\gamma^2+4\gamma+1}{\gamma^2-1} \ln\left(\gamma+\sqrt{\gamma^2-1}\right) - \frac{\gamma+3}{\sqrt{\gamma^2-1}}
    \right].
\end{equation}
The energy-dependence of the cross section is visible in \autoref{fig:losses}.

We obtain the specific radiated power per CR positron by first transforming to the centre of mass frame of the electron-positron collision, which has Lorentz factor $\sqrt{\gamma}$ relative to the lab frame defined by the background gas. In this frame, the electron and positron both have energy $E_{\rm CM} = \sqrt{\gamma} m_e c^2$ and momentum $p_{\rm CM} = (\gamma-1) m_e c$, so each annihilation produces two photons with energy $\epsilon_{\rm CM} = \sqrt{\gamma} m_e c^2$, and an angular distribution \citep[equation 5.106]{Peskin95a}
\begin{equation}
    \frac{dp}{d\mu_{\rm CM}} \propto  
    \frac{2 + \gamma + (\gamma-1)\mu_{\rm CM}^2 - 2 \left[\gamma + (1-\gamma)\mu_{\rm CM}^2\right]^{-1}}{\gamma + (1-\gamma)\mu_{\rm CM}^2}
\end{equation}
where $\mu_{\rm CM}$ is the cosine of the angle between the direction of collision and the direction of the emitted photons. The corresponding angle measured in the lab frame is $\mu = (\mu_{\rm CM} + \beta_{\rm CM})/(1 + \beta_{\rm CM} \mu_{\rm CM})$, where $\beta_{\rm CM} = \sqrt{1-1/\gamma}$ is the normalised velocity of the centre of mass frame, and the corresponding photon energy measured in the lab frame is
\begin{equation}
    \epsilon = \frac{\epsilon_{\rm CM}}{\sqrt{\gamma}(1-\beta_{\rm CM}\mu)} = \left[\gamma + \sqrt{\gamma\left(\gamma-1\right)}\mu_{\rm CM}\right] m_e c^2
\end{equation}
Thus the energy distribution of the photons produced must be proportional to $(d\mu_{\rm CM}/d\epsilon)(dp/d\mu_{\rm CM})$.

While this functional form describes the distribution of CR energies that would be measured by observers in the lab frame isotropically distributed around the direction of the collision, we must of course obtain the same energy distribution for the situation of interest to us, where a single observer at rest measures the photon energy distribution emitted by an isotropic collection of annihilating positrons averaged over all $4\pi$ sr. We can therefore write the quantity of interest to us, the rate per unit energy per CR positron emitted by an isotropic positron population as measured in the observer frame, as
\begin{equation}
    \frac{d\dot{N}}{d\epsilon} = \frac{2L}{m_e c^2} \frac{dF_{\rm pos}}{dy},
    \label{eq:loss_pos}
\end{equation}
where we have defined $y = \epsilon/m_e c^2$ as the ratio of photon energy to electron rest energy, the function $dF_{\rm pos}/dy$ describes the energy distribution of the received photons, and we adopt the normalisation $\int (dF_{\rm pos}/dy) \, dy = 1$, so that the total photon production rate integrated over all energies is $2L$, i.e., every positron that annihilates produces two photons. Since the energy distribution function $dF_{\rm pos}/dy \propto (d\mu_{\rm CM}/d\epsilon)(dp/d\mu_{\rm CM})$, it is straightforward to work out its functional form:
\begin{equation}
    \frac{dF_{\rm pos}}{dy} = \mathcal{N}^{-1} \left[\frac{y^4-4 y^3 + 2 \gamma (1+3\gamma) y^2 - 4 \gamma^2(1+\gamma) y + 2\gamma^2}{y^2(y-2\gamma)^2}\right],
\end{equation}
where $dF_{\rm pos}/dy$ is non-zero only for energies $y\in (y_-,y_+)$ with $y_\pm = \gamma \pm g$ and $g \equiv \sqrt{\gamma(\gamma-1)}$. The normalisation factor $\mathcal{N}$ required to ensure unit integral over this energy range is
\begin{eqnarray}
    \lefteqn{\mathcal{N} = \frac{2}{\gamma}\left(3 - \frac{g}{\gamma}-4\gamma + 4 g\right) \cdot{}
    } 
    \nonumber \\
    & &
    \left\{
    3g\gamma-g\gamma^{2}-4g\gamma^{3} - \gamma\left[1+\gamma\left(4\gamma^2-\gamma-4\right)\right] + {}
    \right.
    \nonumber \\
    & &
    \left.
    \,\left[(3-10\gamma)\gamma + g-6g\gamma+6g\gamma^{2}+8g\gamma^{3}\right]\tanh^{-1}
    \frac{g}{\gamma}
    + {}
    \right.
    \nonumber \\
    & & 
    \left.
    \,2\gamma^3(1+4\gamma)\sinh^{-1}\frac{g}{\sqrt{\gamma}} 
    \right\}.
\end{eqnarray}

We can now write down our final expression for the specific power per CR primary,
\begin{equation}
    \frac{d\Psi}{d\epsilon} = 2 \sigma_d v \left(\frac{\rho}{\mu_e m_{\rm H}}\right) y \frac{dF_{\rm pos}}{dy}.
    \label{eq:Psi_pos}
\end{equation}
However, we caution that this expression only includes emission from positron losses ``in flight'', which likely represent only a minority of total positron annihilations, with the balance occurring due to the formation of positronium after positrons have dropped to near-thermal energies via other loss processes \citep{Prantzos11a}. We do not include a treatment of positronium formation or the resulting emission in \criptic, though it would be straightforward to apply such a model to the output of a \criptic~calculation, since \criptic~records the location and time at which each CR packet drops below the minimum momentum threshold at which we cease to follow it.

\section{Numerical method}
\label{sec:numerics}

We can obtain solutions to the FPE, \autoref{eq:fp_spherical}, by solving the corresponding SDE, \autoref{eq:sde}, to obtain the trajectories through phase space $\mathbf{q}(t)$ for a large number of sample CR packets, including extra steps to account for losses and sources. The phase space distribution of those packets at any time $t$ then provides an estimate of the phase space density $\tf(\mathbf{q})$ at that time. Each sample packet is characterised by a phase-space position $(\mathbf{x}, p)$, a weight $\Upsilon$ indicating the number of individual particles it represents, and the mass $m$ and charge $Ze$ of the particles that comprise it; each packet represents only a single species, but a computation may include an arbitrary number of species, each with its own distribution function $\tilde{f}$ and corresponding sample packets.

\Criptic~advances the sample packets through a series of time steps $\Delta t$. The procedure for updating from time $t^{(n)}$ at the end of the $n$th time step to time $t^{(n+1)} = t^{(n)}+\Delta t^{(n)}$ has four parts, which we describe in detail in the subsequent sections:
\begin{enumerate}
    \item If the diffusion vector, drift vector, or loss rate depends on the CR distribution function $\tilde{f}$ (i.e., if the problem is non-linear), estimate the required functions of $\tilde{f}$ at the position of each packet (\autoref{ssec:reconstruction}).
    \item If any CR sources are present, inject new packets (\autoref{ssec:injection}).
    \item Advance all packets to time $t^{(n+1)}$ using an Euler-Maruyama (EM) update; in the process determine the next time step $\Delta t^{(n+1)}$ (\autoref{ssec:advance}).
    \item Check for production of secondaries during the time step, which are treated stochastically; if any secondary packets are produced during the time step, update them to time $t^{(n+1)}$ as well, iterating until no packets remain (\autoref{ssec:secondaries}).
\end{enumerate}
We first describe the operations of each of these steps in serial, and then how we modify the procedure for parallel computation in \autoref{ssec:parallel}. We describe some general features of the \criptic~implementation in \autoref{ssec:implementation_notes}.

\subsection{Step 1: reconstructing the distribution function}
\label{ssec:reconstruction}

In a non-linear problem, the drift vector $\mathbf{A}$ or diffusion vector $\bm{\kappa}$ depend on the CR distribution function $\tilde{f}$ itself, so the first step in an advance is to reconstruct the distribution function seen by each sample packet so that these non-linear dependencies can be evaluated; in a linear problem we skip this step, as it is computationally expensive. As we discuss below, \criptic~is largely agnostic about the particular form of the non-linearity, and can accommodate a wide range of CR propagation models. However, it is not computationally practical to allow arbitrary functional dependence on $\tilde{f}$. We therefore limit the type of non-linearity we allow to what is by far the most common case.
This, namely, is that the agent responsible for generating the non-linearity is resonant interactions between CRs and waves in the background plasma, and the waves are themselves generated by the CRs via the streaming instability. This constrains the functional form of the non-linear dependence on $\tf$, because CRs with a particular momentum $p$ can only resonantly interact with waves whose wavelength is smaller than the CR gyroradius $r_g$. When dealing with multiple CR species, this condition is most conveniently expressed in terms of the CR rigidity $R = p c/ |Z| e$, where $Z$ is the CR charge in units of the elementary charge $e$. The gyroradius $r_g = R/B$, where $B$ is the local magnetic field strength and, since this is fixed at any given position, the condition for resonant interaction then implies that CRs of rigidity $R$ can experience non-linear interactions with other CRs whose rigidity satisfies $R' > R$. Given this consideration, we restrict \criptic~to computing non-linear effects that can be described in terms of a dependence of the propagation or loss rates for a CR of rigidity $R$ only on integrals of $\tilde{f}$ over particles with rigidities $R' > R$. Other dependenies will be considered in future expansions. 

With this physical picture in mind, \criptic~estimates the CR number density $n_{R'>R}$, pressure $P_{R'>R}$, and (kinetic) energy density $U_{R'>R}$, and the gradients of these quantities at the position of each packet, considering only the contributions from CRs with rigidity greater than or equal to that of the packet being considered; we will drop the $R'>R$ subscript from this point forward for brevity. For a packet with momentum $p$ and charge $Ze$, these quantities are \citep{Zweibel17a}
\begin{equation}
\left(\begin{array}{c}
n\\ P\\ U
\end{array}\right) = \sum_s \int_{p Z_s/Z}^\infty 
\left(
\begin{array}{c}
1\\ v_s p' \\ T_s
\end{array}
\right) \tilde{f}_s  \, dp' 
\label{eq:quantities}
\end{equation}
where the sum runs over all CR species, $Z_s e$ is the charge on species $s$, $\tilde{f}_s$ is distribution function for species $s$ evaluated at position $\mathbf{x}$, and $v_s$ and $T_s$ are the velocity and kinetic energy of a CR of species $s$ with momentum $p'$; note that these functions depend on species $s$ because they depend on the particle mass. Thus $n$, $P$, and $U$ are simply integrals of $\tilde{f}$ over momentum, evaluated with different weights -- 1 for $n$, $v_s p$ for $P$, and $T_s$ for $U$.\footnote{Extensions of \criptic~to compute other quantities that are defined by similar weighted integrals over $\tilde{f}$ are trivial to implement if required, and simply require supplying the weight function for that quantity.} The expressions for the gradients are completely analogous, simply replacing $\tf$ with $\nabla\tf$ in \autoref{eq:quantities}.

We evaluate these integrals by approximating them with Gaussian kernel density estimates; for each packet we define a bandwidth tensor $\tensor{H}$ (computed as we describe below), and approximate the integrals above as
\begin{equation}
    \label{eq:kde}
    \left(\begin{array}{c} n\\ P\\ U\end{array}\right) =
    \sum_{s,i} K_{\tensor{H}}(\mathbf{x}-\mathbf{x}_{si})\Upsilon_{si} \Theta_{si}
    \left(\begin{array}{c} 1\\ v_{si} p_{si}\\ T_{si}\end{array}\right)
\end{equation}
where the sum runs over all species $s$ and all packets $i$ belonging to that species, $\Upsilon_{si}$, $\mathbf{x}_{si}$, $p_{si}$, $v_{si}$, and $T_{si}$ are the weight, position,  momentum, velocity, and kinetic energy of packet $i$ of species $s$, $\Theta_{si}$ is unity for $p_{si} > p (Z_s/Z)$ and zero otherwise, and
\begin{equation}
    K_{\tensor{H}}(\mathbf{x}) = \sqrt{\frac{1}{8\pi^3\,\mathrm{det}\,\tensor{H}}} \exp\left(-\frac{1}{2}\mathbf{x}^{\mathbf{T}} \tensor{H}^{-1} \mathbf{x}\right)
\end{equation}
is the usual three-dimensional Gaussian kernel. The analogous expression for the gradients of these quantities are
\begin{equation}
    \label{eq:kdegrad}
    \left(\begin{array}{c} \nabla n\\ \nabla P\\ \nabla U\end{array}\right) =
    -\tensor{H}_\nabla^{-1} \sum_{s,i} (\mathbf{x}-\mathbf{x}_{si}) K_{\tensor{H}_\nabla}(\mathbf{x}-\mathbf{x}_{si})\Upsilon_{si} \Theta_{si}
    \left(\begin{array}{c} 1\\ v_{si} p_{si}\\ T_{si}\end{array}\right).
\end{equation}
Note that the bandwidth $\tensor{H}_\nabla$ used to estimate the gradient is not the same as that used to estimate the quantities themselves, as discussed below.

We evaluate the sums in \autoref{eq:kde} using an order $N\ln N$ algorithm based on a kd-tree decomposition. Our procedure is as follows. First, we sort the packets into a balanced kd-tree, and for each node in the tree we record the sum of the weights $\sum \Upsilon$ and squared weights $\sum \Upsilon^2$ for all packets contained in that node. Once the tree has been constructed, the next step is to determine the bandwidth tensor $\tensor{H}$ for each packet. There is a vast body of literature on optimal methods for bandwidth selection, but the overriding constraint for us is that we require a method that operates quickly and without requiring global communication (for distributed memory calculations); the latter constraint rules out methods such as cross-validation or multi-stage plug-in selectors.

Instead, we make use of the tree structure itself to make an estimate of the local bandwidth, by choosing the bandwidth 
that brings
a target number of neighbours $N_{\rm ngb,target}$ within the kernel; our default value for this parameter is 1024, but users can choose alternate values. We define the effective neighbour number for each node of the tree as $N_{\rm ngb,node} = (\sum \Upsilon)^2 / \sum \Upsilon^2$, and for each packet we start at the leaf of the tree that contains it, and climb the tree until we reach a node for which $N_{\rm ngb,node} \geq N_{\rm ngb,target}$ (or the root of the tree). At this point, we set the bandwidths $\tensor{H}$ and $\tensor{H}_\nabla$ for the packet by applying the optimal normal scale bandwidth selectors \citep[equations 3.17 and 3.18]{Garcia-Portugues22a},
\begin{eqnarray}
    \tensor{H} & = & \left(\frac{4}{5}\right)^{2/7} N_{\rm ngb,node}^{-2/7} \mathbf{\Sigma} \\
    \tensor{H}_\nabla & = & \left(\frac{4}{7}\right)^{2/9} N_{\rm ngb,node}^{-2/9} \mathbf{\Sigma},
\end{eqnarray}
where $\mathbf{\Sigma}$ is the covariance matrix for the points in the node. 

Once we have selected a bandwidth for each packet, we use the tree to evaluate \autoref{eq:kde} and \autoref{eq:kdegrad}. We defer details of the algorithm to \aref{app:kdtree}. 

\subsection{Step 2: injecting packets}
\label{ssec:injection}

The second step in our algorithm is that each CR source present in the simulation volume adds new CR packets; in terms of \autoref{eq:fp_spherical}, this represents the source term $\tilde{S}$. A source is characterised by its position $\mathbf{x}_s$ at the start of the time step, the first two derivatives of its position $\dot{\mathbf{x}}_s$ and $\ddot{\mathbf{x}}_s$,
the species of CRs it produces, the total rate $\dot{N}$ at which it produces CRs (measured as particles injected per unit time), and the momentum distribution $df/dp$ of those particles, where we normalise $df/dp$ to have unit integral. Our current implementation uses sources with truncated powerlaw distributions in momentum
$df/dp \propto p^q$ over the interval $(p_0,p_1)$, but extension to alternative functional forms of the momentum distribution is trivial.

When we inject packets, we assign each packet an injection time $t_i$ drawn from a uniform distribution from $t^{(n)}$ to $t^{(n)}+\Delta t^{(n)}$, and an injection position $\mathbf{x}_i = \mathbf{x}_s + \dot{\mathbf{x}}_s (t_i - t^{(n)}) + \ddot{\mathbf{x}}_s (t_i-t^{(n)})^2 / 2$. Assigning packet momenta requires some subtlety, because the naive approach -- drawing momenta from $df/dp$ -- is very computationally inefficient. For most realistic sources the momentum distribution is very steep, $df/dp \sim p^{-2.2}$, and so if every injected packet represents an equal fraction of the CR population, then a very large number of packets are needed to capture the behaviour at high momenta.

For this reason, we do not draw momenta from $df/dp$, but instead from an alternative distribution $df_{\rm samp}/dp$ that is generally flatter; our default choice is $df_{\rm samp}/dp \propto p^{-1}$, corresponding to a uniform distribution in $\log p$, but users can alter this. To compensate for under-sampling low-$p$ packets compared to their true numbers, we increase the weight of those packets we do draw. Specifically, during a time step in which we draw a total of $N_{\rm packet}$ CR packets to represent the CRs injected by the given source, we set the weight of each packet we inject to
\begin{equation}
    \Upsilon = \frac{\dot{N}\, \Delta t^{(n)}}{N_{\rm packet}} \left(\frac{df/dp}{df_{\rm samp}/dp}\right),
\end{equation}
where $df/dp$ and $df_{\rm samp}/dp$ are both evaluated at the momentum $p$ of the newly-drawn packet. The factor in parentheses ensures that the momentum distribution function of the injected packets, weighted by the packet weights $\Upsilon$, follows $df/dp$.

The number of packets $N_{\rm packet}$ injected by each source is set by a user-specified packet injection rate $\dot{N}_{\rm packet}$, which specifies the number of primary CR packets injected per unit time by all sources in the calculation; this choice determines the trade-off between computational cost and fidelity in sampling the distribution function, and the optimal choice is necessarily problem-dependent. When only a single source is present, we trivially have $N_{\rm packet} = \dot{N}\, dt$. In the more general case where multiple sources are present, we assign each source a weight
\begin{equation}
    \Upsilon_s = \dot{N} \int \frac{df_{\rm samp}}{dp} \, dp,
\end{equation}
and then
\begin{equation}
    N_{{\rm packet},i} = \frac{\Upsilon_{s,i}}{\sum_j \Upsilon_{s,j}} \dot{N}_{\rm packet}\, \Delta t^{(n)}
\end{equation}
packets for source $i$, where the sum runs over all sources present. This ensures that the total number of packets injected is $\dot{N}_{\rm packet}\, dt$, and that the total momentum distribution of all packets injected is distributed as $df_{\rm samp}/dp$.

Once packets have been injected, we reconstruct the number density $n$, pressure $P$, and energy density $U$ at their phase space locations using the procedure described in \autoref{ssec:reconstruction}, exactly as for the packets that already exist at the start of the time step.

\subsection{Step 3: advancing packets}
\label{ssec:advance}

Consider a sample CR packet that starts a time step with phase space position $\mathbf{q}^{(*)}$ and weight $\Upsilon^{(*)}$, and let $t^{*}$ be the time at which the packet starts the step; for packets that existed at the start of the time step $t^* = t^{(n)}$ and similarly for $\mathbf{q}^{(*)}$ and weight $\Upsilon^{(*)}$, while for newly-created packets $t^{(*)} = t_i$, where $t_i$ is the time at which that packet was injected, and the phase space position $\mathbf{q}$ and weight $\Upsilon$ correspond to those with which the packet was injected. We must advance the packet to time $t^{(n+1)}$ through a series of sub-steps. We begin each sub-step by computing the drift and diffusion vectors $\mathbf{A}$ and $\bm{\kappa}$, and the sum of the catastrophic loss rates due to all processes $L$, given the current properties and phase space position of the packet; if the distribution function $\tf$ or quantities derived from it are required, we use the reconstructed value obtained in Step 1. From these, we compute a series of time step constraints associated with spatial drift, spatial diffusion, momentum drift, momentum diffusion, and catastrophic loss:
\begin{eqnarray}
\Delta t_{\rm x-drift} & = & \frac{\Delta x}{\sqrt{A_1^2+A_2^2+A_3^2}} \\
\Delta t_{\rm x-diff} & = & \frac{\Delta x^2}{\max(K_\parallel,K_\perp)} \\
\Delta t_{\rm p-drift} & = & \frac{p}{|A_4|} \\
\Delta t_{\rm p-diff} & = & \frac{p^2}{K_{\rm pp}} \\
\Delta t_{\rm loss} & = & \frac{1}{L}
\end{eqnarray}
Here $\Delta x$ is the smaller of (i) a user-specified length scale that describes the structure of the background gas through which the CR packets move (and thus the size scale on which $\mathbf{A}$ and $\bm{\kappa}$ might be expected to vary) and (ii) the square root of the geometric mean of the eigenvalues of the bandwidth tensor $\tensor{H}$, which describes the typical size of the kernel used to estimate $\tf$. We then set the overall time step to
\begin{eqnarray}
    \lefteqn{\Delta t = \min\left(t^{(n+1)}-t^{(*)}, \phantom{\frac{C}{\Delta t_{\rm x-drift}^{-1}}}
    \right.}
     \\
    & &
    \left.
    \frac{C}{\Delta t_{\rm x-drift}^{-1} + \Delta t_{\rm x-diff}^{-1} + 
    \Delta t_{\rm p-drift}^{-1} + \Delta t_{\rm p-diff}^{-1} + \Delta t_{\rm loss}^{-1}}\right), \nonumber
\end{eqnarray}
where $C$ is a user-specified tolerance with a default value of 0.25; note that the sub-step size $\Delta t \leq \Delta t^{(n)}$.

Once the time step has been set, we first update the packet weight via
\begin{equation}
    \Upsilon^{(\dagger)} = \Upsilon^{(*)} e^{-L\, \Delta t},
\end{equation}
where $^{(\dagger)}$ indicates the state after the update, and we next update the momentum through a standard EM step \citep{Gardiner09a},
\begin{equation}
    p^{(\dagger)} = p^{(*)} + A_4 \Delta t + \eta \sqrt{K_{\rm pp} \Delta t},
\end{equation}
where $\eta$ is a random variable drawn from a distribution with unit mean and zero variance. At this point we check for creation of secondaries, a procedure we describe in \autoref{ssec:secondaries}, and we delete packets for which $p$ or $\Upsilon/\Upsilon_i$, where $\Upsilon_i$ is the packet weight at injection, fall below user-specified tolerances; this is to limit the use of computational resources following packets that represent either a negligible number of particles or that have fallen to energies below those in which we are interested.

The final step is to compute the new spatial position, which we also do through an EM update,
\begin{equation}
    x_i^{(\dagger)} = x_i^{(*)} + A_i \Delta t + \eta_i \sqrt{\kappa_i \Delta t},
\end{equation}
where $\bm{\eta}$ is a vector of three independent random variables with unit mean and zero variance. However, there is a subtle difficulty in implementing this expression: the positions here are written in the local TNB basis, which is not the same at every position in space. In principle we could handle this issue simply by evaluating the TNB basis vectors in our fixed coordinate system at the start of every sub-step, using the equation above to compute the displacement of the CR packet along these basis vectors, and then transforming these displacements into our fixed coordinate system. However, due to the first-order nature of the EM scheme,\footnote{Higher-order schemes such as the Milstein algorithm \citep{Gardiner09a} are unfortunately not usable for our problem, due to the potentially complex dependence of the diffusion coefficients on position.} this approach leads to considerable numerical drift of packets across field lines. We therefore follow \citet{Merten17a} in circumventing this problem by dividing each EM step into a series of sub-steps during each of which we recompute the TNB basis vectors, carrying out those sub-steps using an adaptive, higher-order method that provides vastly better field-line tracing. 

Specifically, let $\bm{\xi}$ be the position in the fixed frame of the simulations, and define an effective ``velocity'' in the local TNB frame by
\begin{equation}
    \dot{x}_i = A^{(v=0)}_i \Delta t + \eta_i \sqrt{\frac{\kappa_i}{\Delta t}},
\end{equation}
where the superscript $(v=0)$ indicates the drift vector $\mathbf{A}$ evaluated with the advection velocity $\mathbf{v}$ set to zero; we separate out the advection velocity because it is \textit{not} defined relative to the TNB basis. We then carry out a Runge-Kutta-Fehlberg 4th-5th order (RKF45; \citealt{Fehlberg70a}) update of the position in the fixed frame $\bm{\xi}$ by setting the derivative at each stage of the RKF45 update to
\begin{equation}
    \dot{\bm{\xi}} = \dot{x}_1 \That + \dot{x}_2 \Nhat + \dot{x}_3 \Bhat + \mathbf{v},
    \label{eq:rkf}
\end{equation}
where the basis vectors $\That$, $\Nhat$, and $\Bhat$ are recomputed at every stage, taking into account how the TNB basis vectors change as the packet moves. If the field does not vary in space, then the basis does not change, and \autoref{eq:rkf} reduces to the standard EM update. As is usual with the RKF45 update, at the same time we compute the update, we can also compute an error estimate; if that error estimate exceeds some specified tolerance, we divide the RKF45 time step $\Delta t$ into two sub-steps of size $\Delta t/2$, repeating the subdivision recursively as necessary until the error estimate drops below a specified tolerance. In this way we ensure that packets follow field lines accurately.

At this point we have updated all packet variables for the sub-step $\Delta t$. We repeat this process until each packet is advanced to the final time $t^{(n+1)}$.

\subsection{Step 4: catastrophic losses and secondaries}
\label{ssec:secondaries}

One of the steps in the advance procedure described in \autoref{ssec:advance} is to create secondary packets, where secondary here is used to mean any packet created from another packet, rather than from a source or present in the initial conditions. As described in \autoref{ssec:microphysics}, the source function describing the secondaries created by some catastrophic loss process $i$ can be characterised in terms of the loss rate $L_i$ for the process, the multiplicity function $\xi_{i,s}$ that describes the multiplicity of secondaries of species $s$ for that process, and the momentum distribution function $\phi_{i,s}$ for those secondaries. Note that $L_i$, $\xi_{i,s}$, and $\phi_{i,s}$ are all functions of the momentum $p'$ of the CRs undergoing loss, but in the discussion below we do not write out this functional dependence explicitly for compactness.

Now consider a time step of length $\Delta t$ during which a CR packet of statistical weight $\Upsilon$ experiences a total loss rate from all processes $L = \sum_i L_i$. As a result, the weight of the primary packet is reduced to $\Upsilon e^{-L \Delta t}$. The total number of secondaries of species $s$ created by loss process $i$ in the course of this evolution is 
\begin{equation}
    \Upsilon_{i,s,\rm sec} = \xi_{i,s} \frac{L_i}{L} \left(1 - e^{-L\Delta t}\right) \Upsilon.
\end{equation}

A naive implementation of secondary production would be to create secondaries with this weight every time step. However, doing so would quickly make the calculation impossibly expensive due to the rapid proliferation of packets. We therefore instead set a probability $p_{i,s,\rm sec} \leq 1$ that each packet spawns a secondary of species $s$ via loss process $i$ during each sub-step of its advance (\autoref{ssec:advance}), and increase the statistical weight of the secondaries by a factor $1/p_{i,s,\rm sec}$ to compensate, so that the expected statistical weight has the correct value. To be precise, for any secondaries we do create, we assign them a weight
\begin{equation}
    \Upsilon_{i,s} = \frac{\xi_{i,s}}{p_{i,s,\rm sec}} \frac{L_i}{L} \left(1 - e^{-L\Delta t}\right) \Upsilon
\end{equation}
If a secondary is created, we assign its initial position to be the same as that of its parent at the start of the sub-step, and we assign its initial momentum by drawing from the momentum redistribution function $\phi_{i,s}$.

We set the secondary creation probability $p_{i,s,\rm sec}$ for each process and species to a value proportional to the expected secondary creation rate $\xi_{i,s} L_i$. We then normalise the probability as follows:
\begin{equation}
    p_{i,s,\rm sec} = f_{\rm sec} \frac{L_i \xi_{i,s}}{\sum_{j} L_j \sum_{s'}\xi_{j,s'}} \left(1 - e^{-L\Delta t}\right),
\end{equation}
where the sums in the denominator run over all processes $j$ and secondary species $s'$, and $f_{\rm sec}$ is a user-settable dimensionless parameter that functions analogously to $\dot{N}_{\rm packet}$ in that it parameterises the trade-off between fidelity and computational cost in tracking secondaries. With this choice, the weight assigned to each secondary created then reduces to
\begin{equation}
    \Upsilon_{i,s} = \frac{\sum_{j} L_j \sum_{s'} \xi_{j,s'}}{f_{\rm sec} L} \Upsilon.
\end{equation}

To understand the meaning of the parameter $f_{\rm sec}$, first note that the expected number of secondaries created during a time step, summing over all loss processes and all species, is
\begin{equation}
    \langle N_{\rm sec} \rangle_1 = \sum_{i,s} p_{i,s,\rm sec} = f_{\rm sec} \left(1 - e^{-L\Delta t}\right).
\end{equation}
After $N$ such steps the expected number of secondaries created is $\langle N_{\rm sec}\rangle = N\langle N_{\rm sec}\rangle_1$, and the weight of the original packet will have been reduced by a factor $e^{-NL\Delta t}$. Thus if we let $\Upsilon_i$ and $\Upsilon_f$ be the initial and final weights of the packet undergoing losses, and adopt the limit of small time steps, 
$L \Delta t \ll 1$, then we can write the expected number of secondaries created as
\begin{equation}
    \langle N_{\rm sec}\rangle = f_{\rm sec} \ln\frac{\Upsilon_i}{\Upsilon_f}.
\end{equation}
Thus we see that $f_{\rm sec}$ determines the 
mean
number of secondary packets created per $e$-folding of the primary packet weight; a value of $f_{\rm sec} = 1$ corresponds to creating an average of one secondary per $e$-folding, while a value of $f_{\rm sec} = 1/\ln 10 \approx 0.434$ corresponds to creating one secondary per factor of 10 reduction in the weight of the primary. We choose the latter as our default, but the optimal choice likely varies from problem to problem, depending on how much loss the primary particles suffer.

\subsection{Parallelism}
\label{ssec:parallel}

\Criptic~is parallelised using a hybrid MPI-openMP model. Within a single MPI rank, it uses openMP threads to carry out the advance for all the packets owned by that rank. This involves two synchronisation points: one at the end of the step when we compute the next time step, and one after calculating any non-linear terms in the drift and diffusion vectors prior, since this calculation must be completed prior to moving any packets.

The MPI parallelism has two parts. The first is a step to distribute the packets across ranks to maintain load balance. For this purpose we use the PANDA algorithm described by \citet{Patwary16a}; briefly summarising here, the algorithm constructs a global kd-tree and uses it to partition packets between ranks based on their spatial position. Construction of a kd-tree requires two steps at each level of partition: choosing a dimension along which to divide the packets, and then choosing a point in that dimension to partition the data. The first step is straightforward to carry out in parallel: the ranks compute the variance of positions in each dimension, and since this involves only summation operations, parallelising is straightforward. We choose to split along the dimension of maximum variance. The second step, choosing a partition point, requires more care, since it is  expensive to find the median point -- the traditional choice for the partition -- in a distributed memory parallel calculation. Instead, we find the median approximately by selecting a set of sample points from all ranks to mark the edges of a histogram, and counting how many points on each rank fall into the various histogram bins. The counts can then be reduced in parallel, allowing us to approximate the value of the median. This process repeats recursively at each level of the tree, until the desired level of leaves are created. Each leaf is assigned to an MPI rank, and the packets within each leaf are then sent to that rank. In practice, since the tree changes little from time step to time step, during most time steps the boundaries of leaves move little, and thus little data need be communicated to maintain load balance. Once the leaves of the global tree have been assigned to MPI ranks, we construct local kd-trees below those leaves, exactly as we do in a non-parallel calculation.

The second part of MPI parallelism is to carry out kernel density estimation in parallel when packets are distributed across ranks. As with the remainder of the tree algorithm we use for the kernel density computation, we defer details to \aref{app:kdtree}.

\subsection{Implementation notes}
\label{ssec:implementation_notes}

\Criptic~is written in C++, based on the C++17 standard; we have avoided C++20 features to ensure compatibility with somewhat older compilers. A key aspect of the design is to maximise user flexibility in specifying (1) the initial conditions, (2) the properties of the background gas and radiation field through which CRs propagate, and (3) the underlying plasma physical model that describes CR propagation. While flexibility in initial conditions is standard in simulation codes, flexibility in the background gas state and propagation model is more challenging.

In \criptic, we achieve this using C++ classes. The state of the background gas is defined by a pure virtual interface class, which can be specialised by a user to describe an arbitrary gas distribution. We provide specialisations for some standard cases, for example where the gas distribution is specified in terms of an analytic function, a static Cartesian grid, or a series of snapshots in time that are each stored on Cartesian grids but, given this flexibility, users can implement their own classes to describe arbitrary time- and position-dependent magnetic fields, gas densities, ionisation states, compositions, and background radiation fields.

We take a similar approach to CR propagation. In practice, CR propagation in \criptic~is defined using a pure virtual interface class \texttt{Propagation}, which defines the call operator \texttt{Propagation::()} as a pure virtual function that takes as input the spatial position $\mathbf{x}$, the time $t$, the properties of the CR packet (type of particle, momentum, etc.), the properties of the background gas (total density, ion density, composition, magnetic field, etc.), and the CR field quantities $n$, $P$, and $U$ and their gradients. This function must return all of the quantities that appear in the drift vector $\mathbf{A}$ (\autoref{eq:drift}) and the diffusion tensor $\tensor{D}$ (\autoref{eq:diffusion}): the parallel diffusion coefficient $K_\parallel$ and its spatial gradient $\nabla K_\parallel$, the perpendicular diffusion coefficient $K_\perp$ and its spatial gradient $\nabla K_\perp$, the momentum diffusion coefficient $K_{pp}$ and its derivative with respect to momentum $\partial K_{pp}/\partial p$, and the streaming speed $w$ and its spatial gradient $\nabla w$ and derivative with respect to momentum $\partial w/\partial p$.

To define a CR propagation model, the user defines a class derived from \texttt{Propagation} that provides an implementation of the call operator and computes the required outputs from the provided inputs; the implementation of this function is entirely up to the user, and thus, for example, CR propagation can include arbitrary combinations of streaming and diffusion, which can depend in arbitrary ways on position, time, CR properties, gas properties, and the field quantities $n$, $P$, and $U$ and their gradients. As with the gas properties, we provide implementations for some standard cases -- for example, a model where the CR diffusion coefficient is a powerlaw function of CR momentum, and where CRs stream down field lines at the ion Alfv\'en speed -- but users are not limited to these choices. The only restrictions are that, in its current form, \criptic~cannot capture CR propagation coefficients that depend on something other than the provided inputs listed above, or where CR propagation is not describable by the pitch angle-averaged Fokker-Planck equation (e.g., because the pitch angle distribution is not close to isotropic).

One implication of this flexibility is that \criptic~can be run using exactly the same interstellar gas and radiation field distributions, and CR propagation models, as standard CR propagation codes such as \textsc{Galprop} \citep{Strong07a}.

\section{Code tests}
\label{sec:tests}

Here we describe the various validation tests to which we have subjected \criptic.

\subsection{Transport tests}

Our first batch of tests evaluates \criptic's performance in simulating CR transport, including the step of reconstructing the CR field where necessary because transport rates depend non-linearly on it. For all the tests in this section, we disable all catastrophic and continuous loss terms, so we are testing the transport parts of the code only. In these tests we will characterise the performance of the code in terms of its $L^1$ error; we do not experiment with varying the number of packets explicitly, but below we show that the errors we obtain are generally consistent with Poisson noise, and thus in general we expect the error to depend on number of packets used in a given simulation as $L^1_{\rm err} \propto N_{\rm packet}^{-1/2}$.

\subsubsection{Anisotropic diffusion}
\label{sssec:diffusion}

Our first test is to validate \criptic's treatment of diffusion, including anistropy and momentum-dependence of the diffusion coefficients. We consider a uniform region containing gas at rest, threaded by a uniform magnetic field in the $\hat{x}$ direction. The CR diffusion coefficients parallel and perpendicular to the field have a fixed ratio $\chi$ and vary as powerlaws in the CR momentum, i.e.,
\begin{equation}
    K_{(\parallel,\perp)} = (\chi, 1) \cdot K_0 \left(\frac{p}{p_0}\right)^q,
\end{equation}
where $\chi$, $K_0$, $p_0$, and $q$ are constants. There is no momentum diffusion or streaming. The domain is initially empty of CRs, but at $t=0$ a point source of CR protons turns on at the origin. The source is monochromatic, and is characterised by its luminosity $\mathcal{L}$ and by the momentum $p_0$ of the protons it injects. For this problem setup, the Fokker-Planck equation reduces to
\begin{equation}
    \frac{\partial \tf}{\partial t} = K \left(\chi \frac{\partial^2 \tf}{\partial x^2} + \frac{\partial^2 \tf}{\partial y^2} + \frac{\partial^2 \tf}{\partial z^2}\right) + \frac{\mathcal{L}}{T} \delta\left(\mathbf{r}\right)\delta\left(p - p_0\right) \Theta(t),
\end{equation}
where $\mathbf{r} = (x,y,z)$, $K \equiv K_0 (p/m_p c)^q$, $\Theta(x)$ is the Heaviside step function, and
\begin{equation}
    T = m_p c^2 \left[\sqrt{1+\left(\frac{p_0}{m_p c}\right)^2}-1\right]
\end{equation}
is the kinetic energy of the injected protons. 

Making the change of variable $x = \sqrt{\chi} x'$ reduces this problem to a constant-coefficient diffusion equation
\begin{equation}
    \frac{\partial \tf}{\partial t} = K \left(\frac{\partial^2 \tf}{\partial x'^2} + \frac{\partial^2 \tf}{\partial y^2} + \frac{\partial^2 \tf}{\partial z^2}\right) + \frac{1}{\sqrt{\chi}}\frac{\mathcal{L}}{T} \delta\left(\mathbf{r'}\right)\delta(p-p_0)\Theta(t),
\end{equation}
where $\mathbf{r'} = (x',y,z)$; note the extra factor of $1/\sqrt{\chi}$ in the final term, which arises from the change of variable. The Green's function for the spatial distribution of the CRs is then
\begin{equation}
    G(r', t) = \frac{1}{\left(4\pi K t\right)^{3/2}} e^{-r'^2/4Kt},
    \label{eq:greens_aniso_diff}
\end{equation}
where $r' = |\mathbf{r'}|$, so for the case our case of a source that turns on at $t=0$, the solution for the spatial distribution for $t>0$ is
\begin{eqnarray}
    \tf(r',t) & = & \int_0^t \frac{1}{\sqrt{\chi}} \frac{\mathcal{L}}{T} \delta(p-p_0) G(r',t')\, dt' 
    \nonumber \\
    & = & \frac{1}{\sqrt{\chi}}\frac{\mathcal{L}}{T} \frac{1}{4\pi K r'} \mbox{erfc}\left(\frac{r'}{2\sqrt{K t}}\right)\delta\left(p - p_0\right).
    \label{eq:diffusion_analytic}
\end{eqnarray}
The corresponding CR energy density is simply $U_{\rm CR} = T \int f(r',t) \, dp$.

\begin{figure}
\includegraphics[width=\columnwidth]{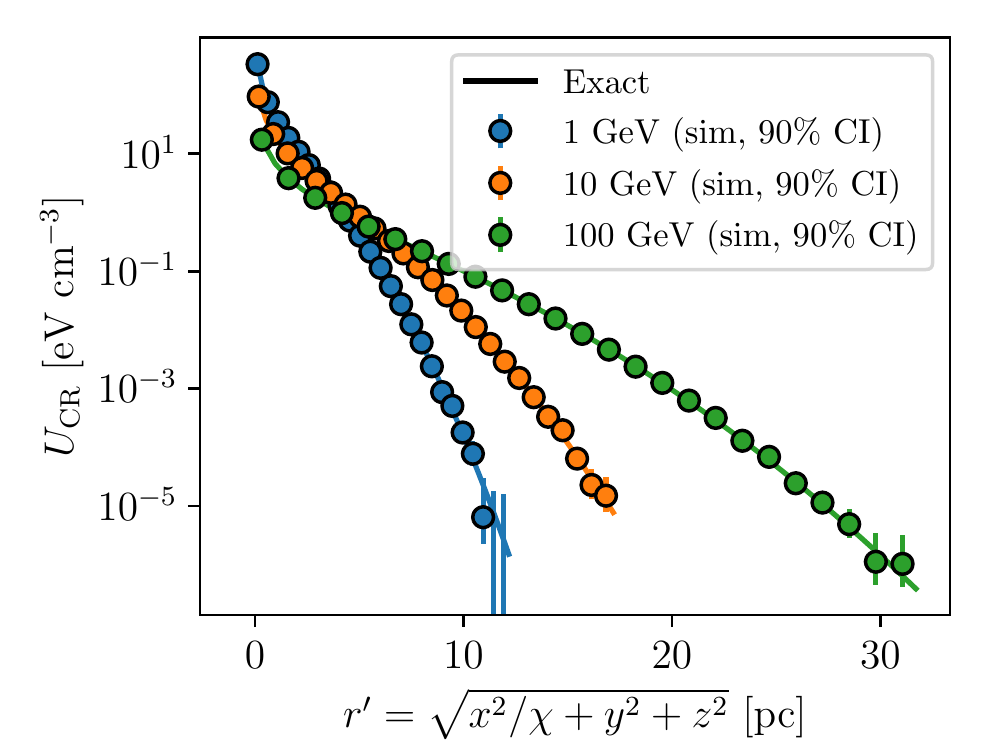}
\caption{
Energy density $U_{\rm CR}$ as a function of effective radius $r'$ in the anisotropic diffusion test (\autoref{sssec:diffusion}). Solid lines show the exact solution computed from \autoref{eq:diffusion_analytic} for the three sources producing CRs of energy ($T=1,10$, and 100 GeV; blue, orange, and green), while circles show the \criptic~result; for clarity we plot only every other bin. Error bars indicate the 90\% confidence interval derived from the number of CRs in the bin, assuming the CRs number counts are Poisson-distributed.
}
\label{fig:AnisoDiff1}
\end{figure}

\begin{figure}
\includegraphics[width=\columnwidth]{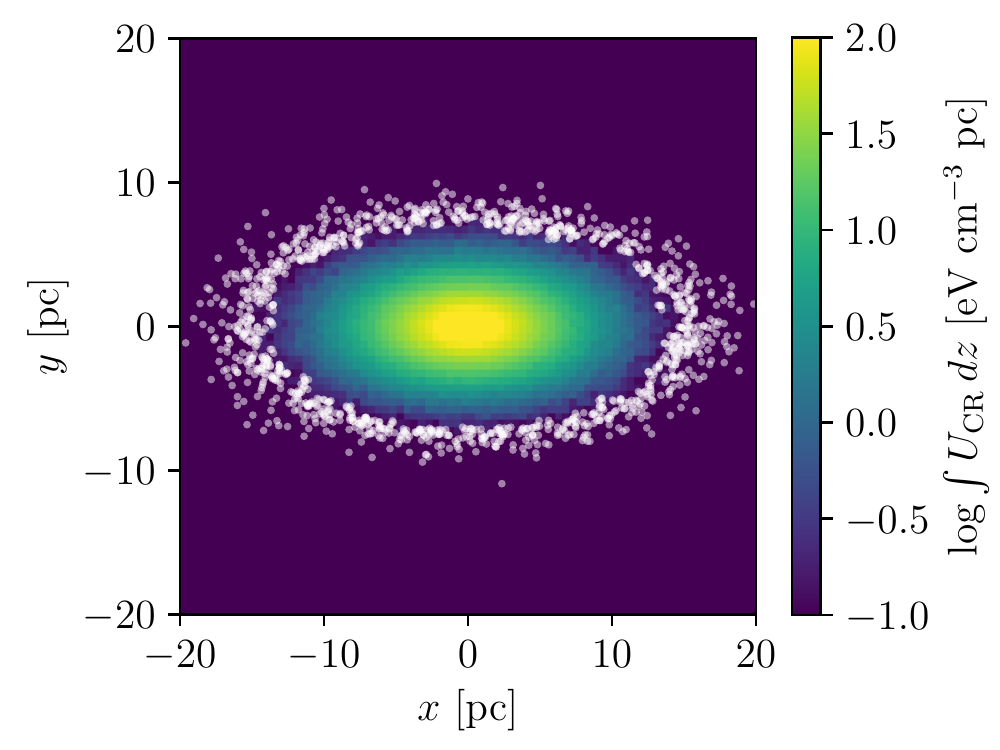}
\caption{
Projected energy density $\int U_{\rm CR}\, dz$ in the anisotropic diffusion test (\autoref{sssec:diffusion}), where energy density is computed only for CRs with energies $T=1$ GeV. Colour shows the projected energy density from $10^{-1} - 10^2$ eV cm$^{-3}$ pc, and white points show the positions of individual CR packets in regions where the CR energy density falls below this level.
}
\label{fig:AnisoDiff2}
\end{figure}

We test \criptic~by simulating this problem using $K_\perp = 10^{28}(p/m_p c)^{1/2}$ cm$^2$ s$^{-1}$ and $\chi = 4$, with three sources producing CRs with kinetic energy $T = 1$, $10$, and 100 GeV; each source has luminosity $\mathcal{L} = 10^{38}$ erg s$^{-1}$; since the CRs from the different sources do not interact, the solution \autoref{eq:diffusion_analytic} applies independently to the CRs produced by each source. We run the simulation for $t = 2\times 10^9$ s using a packet injection rate $10^{-3}$ s$^{-1}$, so that there are $2\times 10^6$ CR packets at the end of the simulation. We show the results in \autoref{fig:AnisoDiff1} and \autoref{fig:AnisoDiff2}.

As the plots show, \criptic~recovers the exact solution to very high precision, covering $\approx 8$ orders of magnitude in CR energy density; errors in the solution are consistent with expectations from Poisson statistics given the finite number of CR packets used in the simulation. The $L^1$ error, defined by
\begin{equation}
    L^1_{\rm err} = \frac{4\pi }{Lt}\int | U_{\rm CR,exact} - U_{\rm CR,sim} | r'^2 \, dr',
\end{equation}
where $U_{\rm CR,exact}$ and $U_{\rm CR,sim}$ correspond to the exact and simulation solutions shown in \autoref{fig:AnisoDiff1} (using the bins shown to compute $U_{\rm CR,sim}$), is below 1\% for all three sources. Visual inspection of the projected CR distribution also demonstrates that it is anistropic by a $4:1$ ratio, exactly as expected.

\subsubsection{Variable diffusion}
\label{sssec:vardiff}

Our next test investigates \criptic's performance when the diffusion coefficient is non-constant. We simulate the transport of CRs with an isotropic diffusion coefficient, but one that varies as a powerlaw in both space and time:
\begin{equation}
    K = K_\parallel = K_\perp = K_0 \left(\frac{r}{r_0}\right)^{q_r} \left(\frac{t}{t_0}\right)^{q_t}.
\end{equation}
There is no streaming in this test, and the background gas is at rest. No sources are present, and all CRs have momentum $p_0$. For this setup the FPE is
\begin{equation}
    \frac{\partial \tf}{\partial t} = K \left(\frac{\partial^2 f}{\partial x^2} + \frac{\partial^2 \tf}{\partial y^2} + \frac{\partial^2 \tf}{\partial z^2}\right),
\end{equation}
and one can verify by substitution that the system admits a similarity solution for the spatial distribution
\begin{eqnarray}
    \lefteqn{\tf(r,t) = N_{\rm CR} \delta(p-p_0) \frac{3(2-q_r)^{\frac{6}{q_r-2}}}{4\pi r_0^3 \Gamma\left(\frac{q_r-5}{q_r-2}\right)} 
    \left[(q_t+1)\eta\right]^{3/(2-q_r)}
    }
    \nonumber \\
    & &
    \left(\frac{t}{t_0}\right)^{3\frac{q_t+1}{q_r-2}}\exp\left[-\frac{(q_t+1)\eta}{(q_r-2)^2} \left(\frac{r}{r_0}\right)^{2-q_r}\left(\frac{t}{t_0}\right)^{-q_t-1}
    \right],
    \label{eq:vardiffexact}
\end{eqnarray}
where $\Gamma(x)$ is the $\Gamma$ function, $\eta\equiv r_0^2/K_0 t_0$, and $N_{\rm CR}$ is the total number of CRs. The energy density $U_{\rm CR} = T \int \tf \, dp$, where $T$ is the kinetic energy corresponding to CR momentum $p_0$.

For our test we set $K_0 = 4\times 10^{28}$ cm$^2$ s$^{-1}$, $r_0 = 10^{18}$ cm, $t_0 = 10^8$ s, $q_r=-1/2$, and $q_t=1$. We initialise the simulation with $2.5\times 10^5$ CR packets with a total energy of $E_0 = N_{\rm CR} T = 10^{48} $ erg; the initial radial distribution of the packets follow the exact solution, \autoref{eq:vardiffexact}, evaluated at $t = t_0$. We then use \criptic~to advance the system to time $t=2t_0$.

\begin{figure}
    \centering
    \includegraphics[width=\columnwidth]{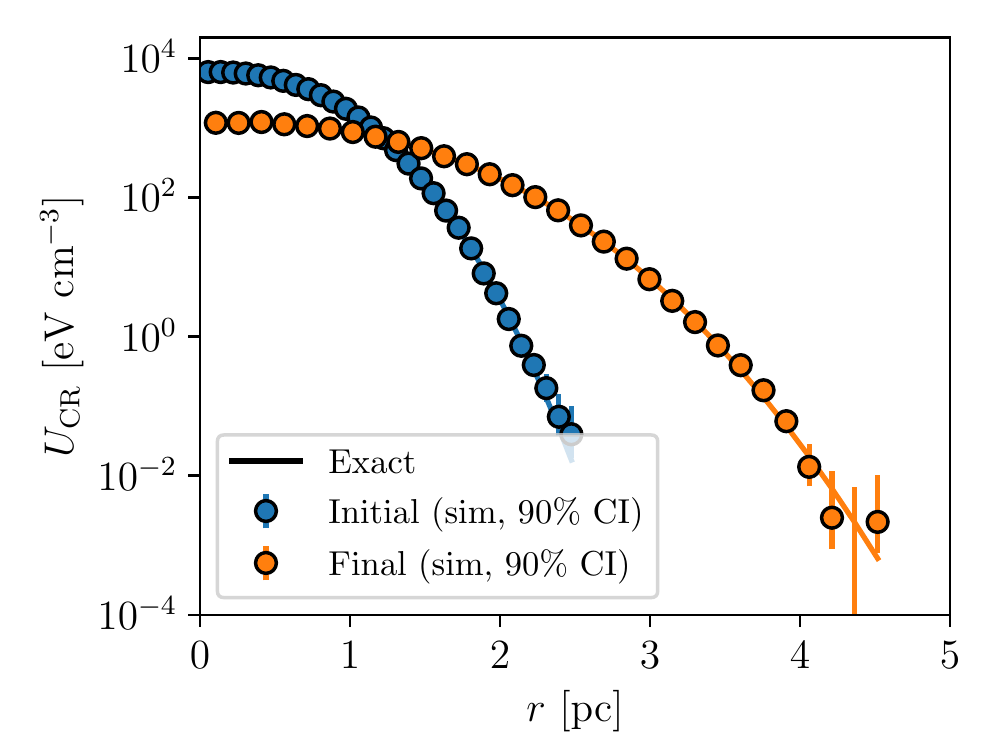}
    \caption{Results of the variable diffusion coefficient test described in \autoref{sssec:vardiff}. We show the CR energy density $U_{\rm CR}$ in the initial condition (blue) at $t = t_0$ and in the final simulation snapshots at $t=2t_0$ (orange). Solid lines show the exact solution given by \autoref{eq:vardiffexact}, while circles show the \criptic~results, with error bars to indicate the 90\% confidence interval assuming the number of CR packets in each radial bin is Poisson-distributed. 
    }
    \label{fig:VarDiff1}
\end{figure}

We show the results in \autoref{fig:VarDiff1}. The figure shows that \criptic~recovers the exact solution to very high accuracy; errors are consistent with Poisson sampling, and are no larger in the final time step than in the initial setup. The $L^1$ error at the final time, defined by
\begin{equation}
    L^1_{\rm err} = \frac{4\pi }{E_0}\int | U_{\rm CR,exact} - U_{\rm CR,sim} | \, r^2 dr,
\end{equation}
is below 1\%.

\subsubsection{Oscillating field loop}
\label{sssec:loopdiff}

Our next test checks the ability of the code to trace diffusion along curved, moving field lines. In this test, we place a single CR source with CR luminosity $\mathcal{L} = 10^{38}$ s$^{-1}$ at $x=r_0 = 1$ pc, $y=z=0$ at time $t=0$. The magnetic field consists of loops around the $z$-axis, so $\mathbf{B} = B_0 \hat{\phi}$, where $\hat{\phi}$ is the $\phi$ unit vector in an $(r,\phi,\theta)$ cylindrical coordinate system. CRs diffuse with zero perpendicular diffusion, and parallel diffusion described by a coefficient $K_\parallel = 10^{28}$ cm$^{2}$ s$^{-1}$. The FPE governing the system then reduces to
\begin{equation}
    \frac{\partial \tf}{\partial t} = \frac{K_\parallel}{r_0^2} \frac{\partial^2 f}{\partial s^2} + \frac{\mathcal{L}}{T}\delta(s)\delta(p-p_0),
\end{equation}
where $T = 1$ GeV is the energy of a single CR, $p_0$ is the corresponding CR momentum, and $s = 2\pi r_0 \phi$ is the position along the current loop, with the source located at $s = \phi = 0$.

The FPE in this case is equivalent to one-dimensional diffusion in a periodic domain $(-\pi r_0, \pi r_0)$. This problem may be solved by standard Fourier methods, and the exact solution for a point source located at $\phi = 0$ that begins injecting CRs at $t=0$ is that the CR energy per unit angle along the loop is
\begin{equation}
    \frac{dE_{\rm CR}}{d\phi} = \frac{\mathcal{L}t}{2 \pi} \left[1 - 2 \frac{t_{\rm diff}}{t} \sum_{n=1}^\infty \frac{\cos(n\phi)}{n^2} \left(e^{-n^2 t/t_{\rm diff}} - 1\right)\right],
    \label{eq:loopexact}
\end{equation}
where $t_{\rm diff} = r_0^2 / K_\parallel = 9.55 \times 10^8$ s is the characteristic diffusion timescale. To add a complication to this test, both the source and the background gas perform simple harmonic oscillation in the $x$ direction, with amplitude $r_0$ and angular frequency $\varpi = 4\pi/t_{\rm diff}$; this allows us to test how well CRs follow field lines when the field lines are attached to a fluid that is accelerating, and where the direction of the acceleration can be both perpendicular and parallel to the field line. The exact solution is still given by \autoref{eq:loopexact}, provided that we define $\phi$ relative to the time-dependent centre of the loop, since advection should move all the CR packets together.

We simulate the system for a time $t = 4 t_{\rm diff}$ using a packet injection rate $3\times 10^{-4}$ s$^{-1}$, so over the course of the simulation $3.81\times 10^5$ packets are injected into the domain, and the loop performs eight full periods of oscillation.  This test therefore evaluates not only how well CRs follow curved field lines, but how well they do so when the field lines and the gas to which they are anchored are moving at arbitrary, time-variable angles relative to the field direction.

\begin{figure}
    \centering
    \includegraphics[width=\columnwidth]{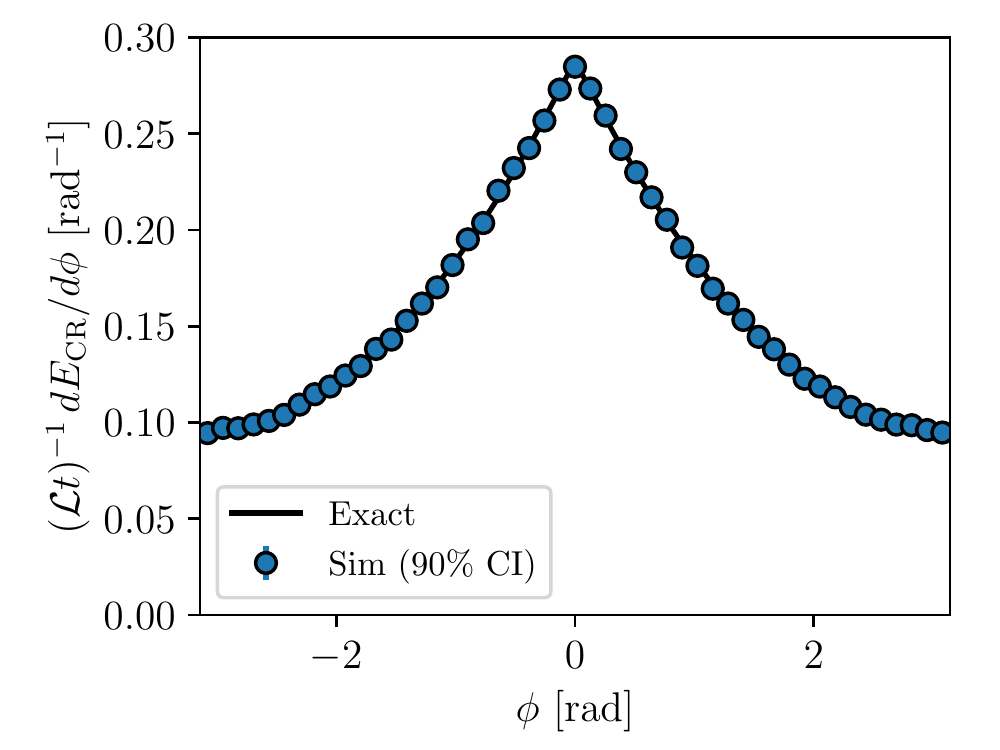}
    \caption{Results for the oscillating loop diffusion test (\autoref{sssec:loopdiff}). The figure shows the CR energy density per unit angle $dE_{\rm CR}/d\phi$, normalised by the total CR energy injected $\mathcal{L}t$. The solid line is the exact solution, and points show the numerical solution computed with \criptic; as in previous tests, the points include error bars showing the 90\% confidence interval from Poisson statistics, but for this test the error bars are for the most part comparable to the sizes of the plot markers and thus are hidden.}
    \label{fig:LoopDiff1}
\end{figure}

\autoref{fig:LoopDiff1} shows the comparison between the \criptic~numerical solution and the exact solution given by \autoref{eq:loopexact}. Clearly the agreement is very good. Defining the $L^1$ error by
\begin{equation}
    L^1_{\rm err} = \frac{1}{\mathcal{L}t}\int \left| \frac{dE_{\rm CR,exact}}{d\phi} - \frac{dE_{\rm CR,sim}}{d\phi} \right| \, d\phi,
\end{equation}
we find that the error is $<1\%$. We also check quantitatively how well the CR packets remain confined to the field line to which they are attached, by examining the standard deviation of CR packet radial coordinates at the end of the simulation, $\sigma_r$. We find that $\sigma_r / r_0 \approx 5.7\times 10^{-6}$, so over 4 diffusion times and 8 oscillations periods, numerical diffusion causes CR packets to drift across field lines by less than 1 part in $10^5$.

\subsubsection{Momentum diffusion}
\label{sssec:momdiff}

Our next test evaluates the ability of the code to handle momentum diffusion, or, equivalently, second-order Fermi acceleration. For this test we turn on a single source of CRs at $t=0$ at the origin, which injects CRs with luminosity $\mathcal{L} = 10^{38}$ erg s$^{-1}$. All injected CRs have a momentum of exactly $p_0=1$ GeV$/c$. There is no spatial diffusion or streaming, but CRs diffuse in momentum with a diffusion coefficient $K_{\rm pp}$, which we set implicitly by setting the diffusion time at momentum $p_0$ to $t_{\rm diff} = 1$ Myr. The corresponding diffusion coefficient is $K_{\rm pp} = p_0^2 / t_{\rm diff}$, and the corresponding FPE is
\begin{equation}
\frac{\partial\tf}{\partial t} = K_{\rm pp} \left[\frac{\partial^2 \tf}{\partial p^2} - \frac{\partial}{\partial p}\left(\frac{2}{p} \tf\right) \right] + \frac{\mathcal{L}}{T}\delta(\mathbf{r})\delta(p-p_0),
\end{equation}
where $T$ is the kinetic energy corresponding to momentum $p_0$. Using the change of variables $\tf'=p \tf$ reduces the problem to a 1D diffusion equation, subject to the boundary condition $\tf' = 0$ at $p = 0$, which can be solved by standard Green's function methods. The exact solution for the CR momentum distribution is
\begin{eqnarray}
    \lefteqn{\frac{dn_{\rm CR}}{dp} = \frac{\mathcal{L}}{T} \frac{t}{\sigma_p} \frac{p}{p_0}
    }
    \nonumber \\
    & &
    \left[
    \sqrt{\frac{2}{\pi}}\left(e^{-\left(\frac{p-p_0}{2\sigma_p}\right)^2} -
    e^{-\left(\frac{p+p_0}{2\sigma_p}\right)^2}\right) +
    \frac{p+p_0}{\sigma_p}\mbox{erfc}\left(\frac{p+p_0}{\sqrt{2}\sigma_p}\right) + {}
    \right.
    \nonumber \\
    & &
    \qquad
    \left.
    \frac{\left|p-p_0\right|}{\sigma_p} \mbox{erfc}\left(\frac{|p-p_0|}{\sqrt{2}\sigma_p}\right) - 1
    \right],
    \label{eq:momdiffexact}
\end{eqnarray}
where $\sigma_p^2 = 2 K_{\rm pp} t$.

\begin{figure}
    \centering
    \includegraphics[width=\columnwidth]{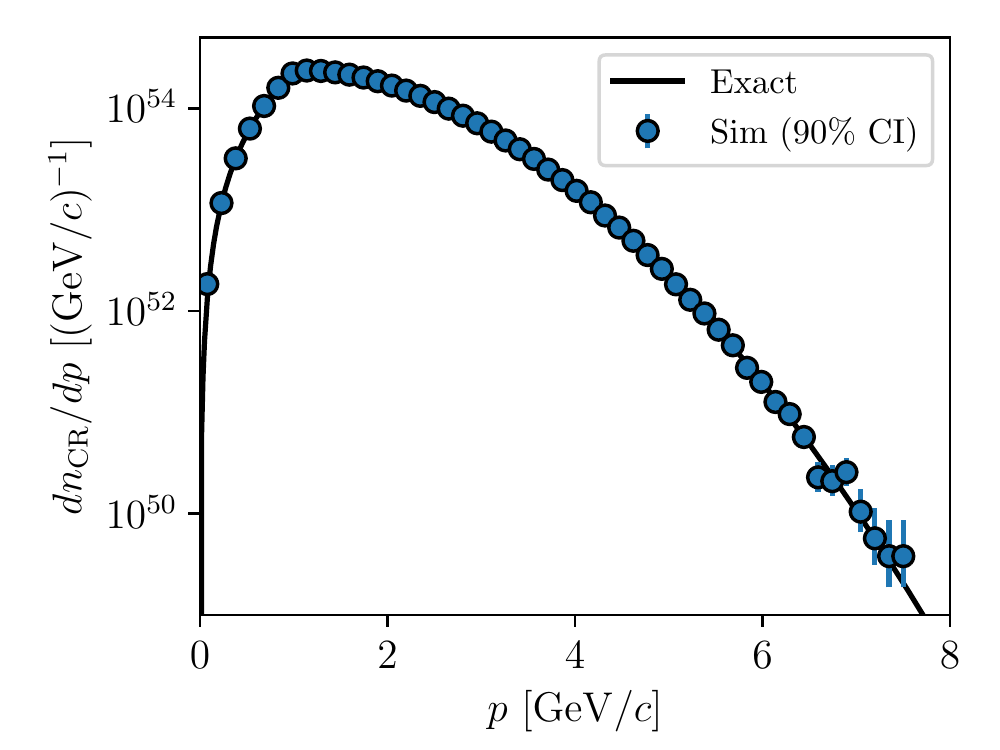}
    \caption{Results for the momentum diffusion test (\autoref{sssec:momdiff}). The plot show the CR momentum distribution $dn_{\rm CR}/dp$ at the end of the simulation. The solid line shows the exact solution, \autoref{eq:momdiffexact}, while points show the \criptic~result. Error bars on the points show the 90\% confidence interval, computed assuming the number of CRs in each bin is Poisson-distributed.}
    \label{fig:MomDiff1}
\end{figure}

We simulate this problem with \criptic~using a packet injection rate of $10^{-7}$ s$^{-1}$, running to time $t=t_{\rm diff} = 1$ Myr, so there are $3.16\times 10^6$ CR packets present at the end of the simulation. We compare the exact and numerical solutions in \autoref{fig:MomDiff1}. The Figure shows very good agreement between the exact and numerical results. Quantitatively, the $L^1$ error, defined for this problem as
\begin{equation}
    L^1_{\rm err} = \frac{1}{(\mathcal{L}/T)t}\int \left| \frac{dn_{\rm CR,exact}}{dp} - \frac{dn_{\rm CR,sim}}{dp} \right| \, dp,
\end{equation}
is $<1\%$.

\subsubsection{Non-linear diffusion}
\label{sssec:nonlindiff}

All of our tests thus far have been for problems where the diffusion coefficient does not depend on the CR field, and thus which can be solved without the reconstruction step in our algorithm. We now consider a problem where reconstruction is required. We consider a system where diffusion is isotropic, and the diffusion coefficient is a powerlaw function of the CR energy density:
\begin{equation}
    K_\parallel = K_\perp = K = K_0 \left(\frac{U_{\rm CR}}{U_{\rm CR,0}}\right)^q.
\end{equation}
No sources are present. Such a system admits a similarity solution \citep{Pattle59a}
\begin{equation}
    U_{\rm CR} = \frac{E_{\rm CR,tot}}{\sqrt{\pi} r_0^3} \frac{\Gamma\left(\frac{1}{q}+\frac{5}{2}\right)}{\Gamma\left(\frac{1}{q}+1\right)} \left[1 - \left(\frac{r}{r_{\rm out}}\right)^2\right] \left(\frac{t}{t_0}\right)^{-3/(3q+2)}, 
    \label{eq:nonlinexact}
\end{equation}
where $E_{\rm CR,tot}$ is the total CR energy in the system,
\begin{equation}
    r_{\rm out} = r_0 \left(\frac{t}{t_0}\right)^{1/(3q+2)},
\end{equation}
and $r_0$ and $t_0$ are given by
\begin{eqnarray}
r_0^3 & = & \frac{E_{\rm CR,tot}}{U_{\rm CR,0}} \pi^{-3/2} \frac{\Gamma\left(\frac{1}{q}+\frac{5}{2}\right)}{\Gamma\left(\frac{1}{q}+1\right)} \\
t_0 & = & \frac{q r_0^2}{2(3q+2) K_0}.
\end{eqnarray}
Note that the system has the property that $U_{\rm CR} = 0$ exactly at $r \geq r_{\rm out}$, so this problem represents a severe test of our method, since the exact solution has a sharp cutoff in the CR energy density.

\begin{figure}
    \centering
    \includegraphics[width=\columnwidth]{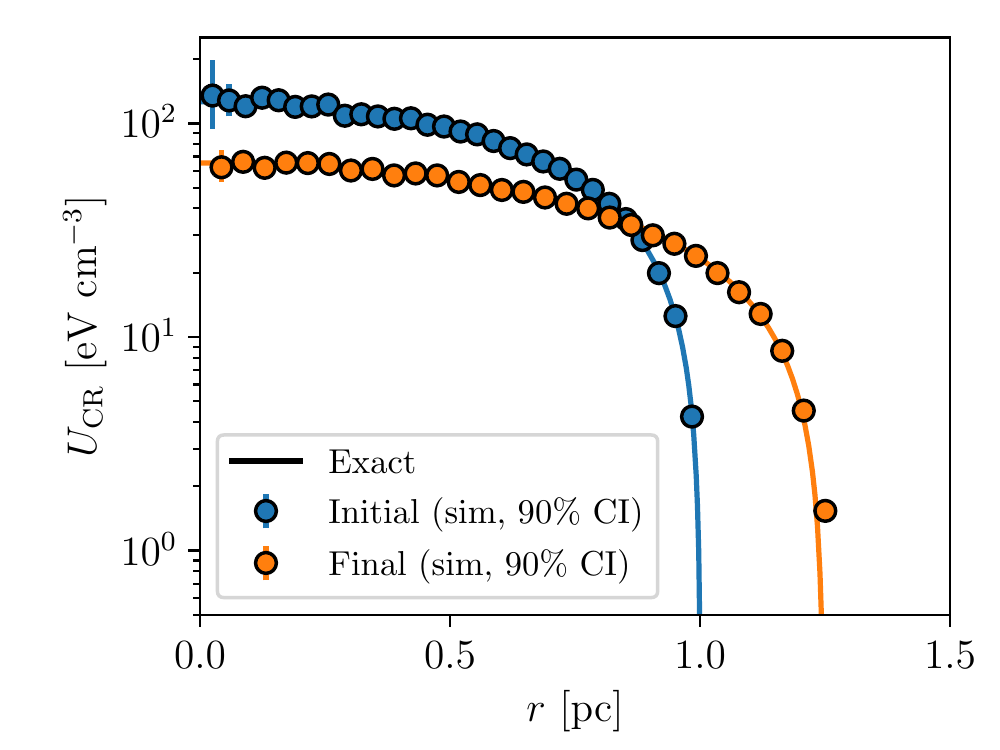}
    \caption{Results for the non-linear diffusion test (\autoref{sssec:nonlindiff}). The plot shows the CR energy density $U_{\rm CR}$ at times $t=t_0$ (``Initial'', blue) and $t=3t_0$ (``Final'', orange). Lines show the exact solution given by \autoref{eq:nonlinexact}, while circles with error bars show the solution computed by \criptic; errors indicate the 90\% confidence interval derived by assuming the number of CR packets in each bin is drawn from a Poisson distribution.}
    \label{fig:NonLinDiff1}
\end{figure}

For our test with \criptic, we set the total CR energy to $E_{\rm CR,tot} = 10^{46}$ erg, adopt index $q=1$, and specify $U_{\rm CR,0}$ and $K_0$ implicitly by setting $r_0 = 1$ pc and $t_0 = 10^7$ s. We initialise the distribution of CR packets to the analytic solution at $t=t_0$, using a total of $10^5$ CR packets, and evolve the system to $t=3t_0$. We show the results of this test in \autoref{fig:NonLinDiff1}. We find that the agreement between the numerical and exact solutions is very good. At smaller radii the match is almost perfect, and \criptic~recovers the location of the sharp edge of the exact solution with only a very small amount of numerical blurring. Defining the $L^1$ error for this test as
\begin{equation}
    L^1_{\rm err} = \frac{4\pi }{E_{\rm CR,tot}}\int | U_{\rm CR,exact} - U_{\rm CR,sim} | r'^2 \, dr',
\end{equation}
we find $L^1_{\rm err} = 2.4\%$.

\begin{figure}
    \centering
    \includegraphics[width=\columnwidth]{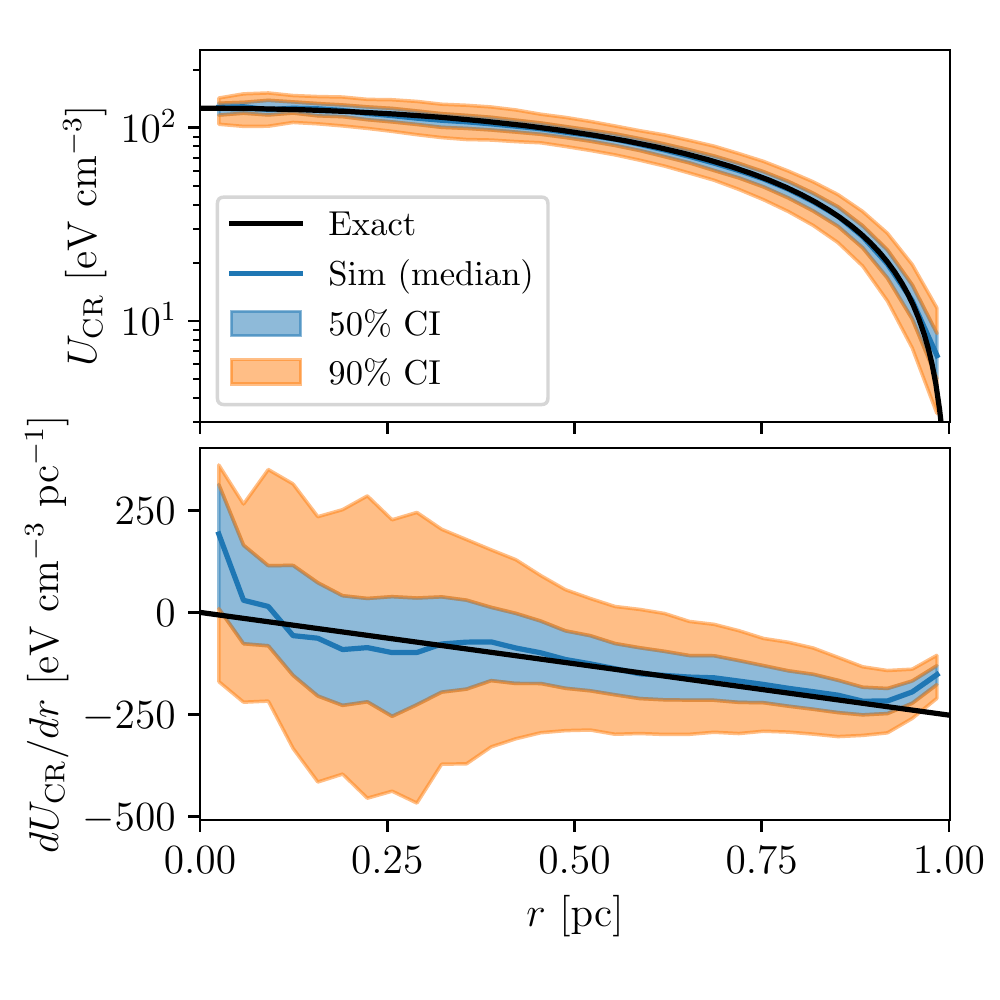}
    \caption{Reconstruction of the CR energy density $U_{\rm CR}$ (top) and its radial derivative $dU_{\rm CR}/dr$ (bottom) at the initial time in the non-linear diffusion test (\autoref{sssec:nonlindiff}). Black lines show the exact solution as a function of radius $r$, solid blue lines show the median value of the \criptic~CR packets, computed in 30 radial radial bins, and blue and orange bands show the 25th - 75th percentile range (50\% confidence interval) and 5th - 95th percentile range (90\% CI), respectively, in the same bins.}
    \label{fig:NonLinDiff2}
\end{figure}

To understand the quality of the solution, it is helpful to examine the reconstructed CR energy density and its derivative, which are used to construct the diffusion coefficient and its spatial derivative. We show these reconstructions for the initial time in \autoref{fig:NonLinDiff2}; results are qualitatively similar at later times. We see that our kernel density estimation algorithm reconstructs the CR energy density with very high accuracy; the median is so close to the exact solution that the line showing it is nearly invisible in the figure, hidden behind the exact solution, and the 50th and 90th percentile ranges lie within $\approx 10\%$ and $\approx 20\%$ of the exact solution except at large radii.

The derivative, which is inherently harder to reconstruct, shows a larger range, and deviates from the exact solution at both small radii, where the number of CR packets is small due to the small volume, and at the edge of the distribution; nonetheless, the median agrees very well with the exact value over most of the radial range. Errors are most significant at the largest radii, where the exact solution goes to zero exactly, a feature that is necessarily blurred somewhat in \criptic's reconstruction due to the finite size of the kernel. It is this effect that is responsible for the sharp edge of the exact solution expanding very slightly too quickly in the \criptic~simulation. Nonetheless, the overall error is very small.

\subsubsection{Streaming and streaming losses}
\label{sssec:streaming}

Our next test checks the ability of the code to handle streaming down CR pressure gradients, together with the associated adiabatic changes in CR momentum when the divergence of the streaming velocity is non-zero. For this test we consider a fully ionised medium at rest with a powerlaw distribution of density and a ``split monopole'' magnetic field. The density and magnetic field as a function of position are
\begin{eqnarray}
\rho & = & \rho_0 \left(\frac{r}{r_0}\right)^{k_\rho} \\
\mathbf{B} & = & B_0 \left(\frac{r_0}{r}\right)^2 \mbox{sgn}(z) \hat{r},
\end{eqnarray}
where $r$ is the distance from the origin. For this test we use $r_0 = 1$ pc, $\rho_0 = 2.34\times 10^{-24}$ g cm$^{-3}$, $k_{\rho} = -2$, and $B_0 = 10$ $\mu$G. The simulation begins with no CRs, but we place a point source of CRs with luminosity $\mathcal{L} = 10^{38}$ erg s$^{-1}$ at the origin, where it injects CRs with initial energy $T = 1$ GeV.\footnote{To avoid CRs near the origin requiring infinitely small time steps, we flatten the density and magnetic field profiles at very small radii, and make the CR source slightly extended. Specifically, we adopt a flattening radius $r_{\rm \rm flat} = 10^{-3} r_0$ and evaluate the density and magnetic field using $\max(r, r_{\rm flat})$ rather than simply $r$; we likewise inject CRs at a random radius uniformly distributed from 0 to $r_{\rm flat}$, rather than exactly at $r=0$.} For the CR transport model in this set, we let CRs stream down the CR pressure gradient at the ion Alfv\'en velocity,
\begin{equation}
    \mathbf{w} = \frac{\mathbf{B}}{\sqrt{4\pi \rho}} \mbox{sgn}\left(-\nabla P_{\rm CR} \cdot \mathbf{B}\right).
\end{equation}

In this test the CR pressure gradient always points to the origin, and thus CRs should always stream away from the origin at the streaming speed
\begin{equation}
    w = v_{A0} \left(\frac{r}{r_0}\right)^{-2-k_\rho/2},
\end{equation}
where $v_{A0} = B_0 /\sqrt{4\pi \rho_0}$. Individual CRs therefore obey an equation of motion $dr/dt = w$ which, for a CR injected at $t = t_{\rm inj}$, has the solution
\begin{equation}
    r(t - t_{\rm inj}) = r_0 \left[\left(3 + \frac{k_\rho}{2}\right) \frac{v_{A0} t}{r_0}\right]^{1/(3+k_\rho/2)}
    \label{eq:rstreamexact}
\end{equation}
for times $t \gg t_{\rm inj}$. Moreover, the momentum of individual CRs evolves with radius due to adiabatic cooling as
\begin{equation}
    \frac{dp}{dr} = \left(\frac{dr}{dt}\right)^{-1} \frac{dp}{dt} = -\frac{p}{3 w} \nabla \cdot \mathbf{w},
\end{equation}
which has the solution that a CR injected with momentum $p_{\rm inj}$ at radius $r_{\rm inj}$ has momentum
\begin{equation}
    p = p_{\rm inj} \left(\frac{r}{r_{\rm inj}}\right)^{-k_\rho/6}
    \label{eq:pstreamexact}
\end{equation}
once it reaches radius $r$.

\begin{figure}
    \centering
    \includegraphics[width=\columnwidth]{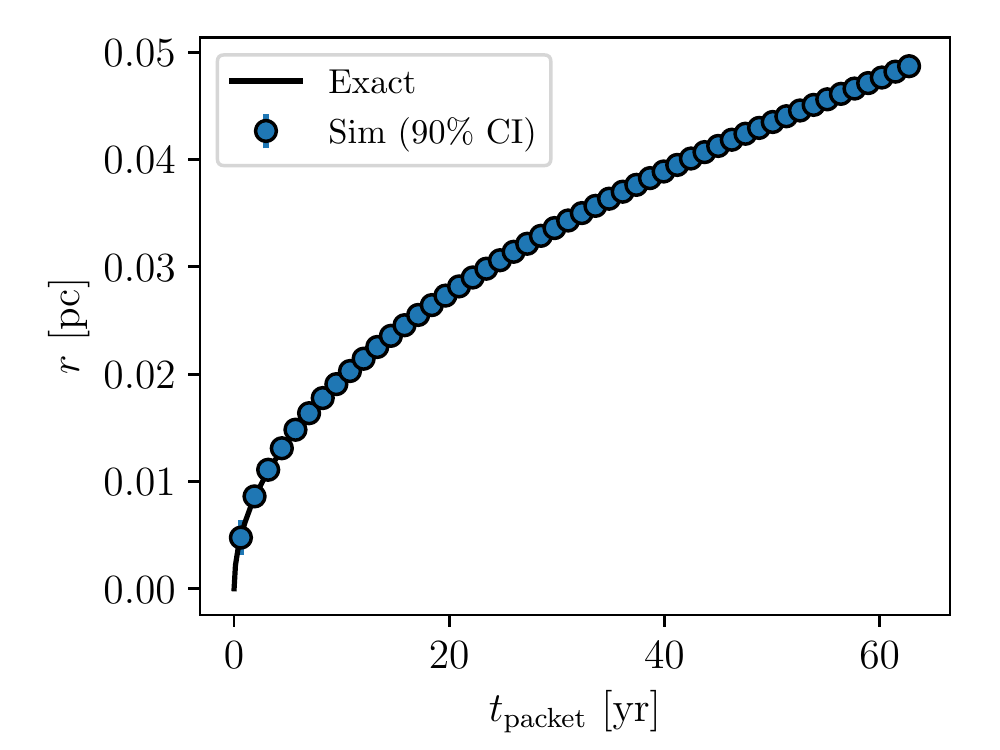}
    \includegraphics[width=\columnwidth]{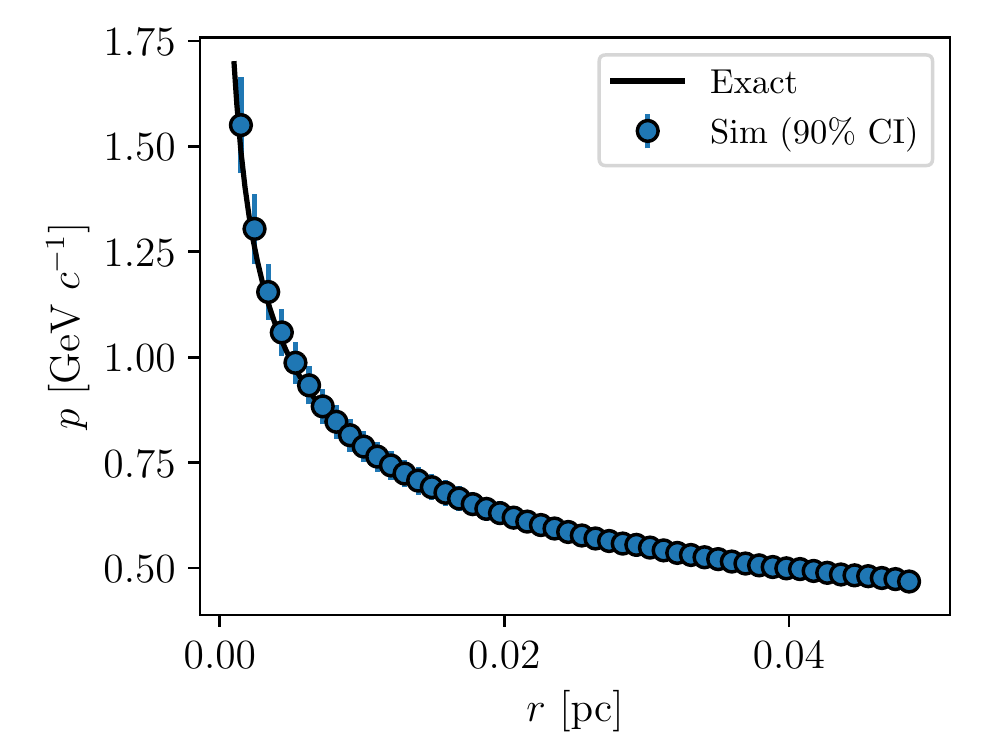}
    \caption{Solutions for the streaming problem (\autoref{sssec:streaming}). In the top panel we show CR packet radial position $r$ versus packet age $t_{\rm packet} = t - t_{\rm inj}$, and in the bottom we show CR packet momentum $p$ as a function of radius. In both panels, solid lines are the exact solutions given by \autoref{eq:rstreamexact} and \autoref{eq:pstreamexact}, respectively, and circles with error bars are the numerical solutions computed by \criptic, averaged over bins in age (top) and radius (bottom). Error bars show the 5th to 95th percentile range in each bin.}
    \label{fig:RadStream}
\end{figure}

We simulate the system for $10^9$ s using a packet injection rate of $10^{-4}$ s$^{-1}$, so that by the end of the simulation there are $10^5$ packets. We show the \criptic~results in comparison to the exact solutions for CR packet position versus age and momentum versus radius in \autoref{fig:RadStream}. We again see excellent agreement between \criptic~and the exact solution, indicating the code successfully reconstructs the direction of the CR pressure gradient and correctly computes the rate of adiabatic cooling caused by the non-uniform streaming speed. Quantitatively, we define the radial position and momentum error for each CR packet by
\begin{equation}
    e_r = \frac{r}{r_{\rm exact}} - 1 \qquad e_p = \frac{p}{p_{\rm exact}} - 1,
\end{equation}
where $r_{\rm exact}$ and $p_{\rm exact}$ are the exact solutions given by \autoref{eq:rstreamexact} and \autoref{eq:pstreamexact}. We find that the mean values of $e_r$ and $e_p$ are $0.37\%$ and $0.40\%$, respectively, with variances $1.5\%$ and $4.6\%$; the errors are therefore small, and in fact the variance in $e_p$ is an overestimate of the true error because it is mostly a result of the dependence of $p$ on the initial injection radius $r_{\rm inj}$, which is randomly varied by a small amount for numerical reasons as discussed above.

\subsubsection{Number density computation}
\label{sssec:nden}

Our final transport test evaluates \criptic's ability to reconstruct the CR field in a more realistic problem where there is a continuous distribution of CR positions and momenta. In this problem we place a source of CR protons at the origin, which injects CRs at a specific rate $d\dot{n}_{\rm src}/dp \propto p^{q}$ over a momentum range from $p_0$ to $p_1$. The CRs then diffuse isotropically away from the source with a constant diffusion coefficient $K$. The exact solution for the CR distribution is then just given by \autoref{eq:diffusion_analytic} with the $(\mathcal{L}/T)\delta(p-p_0)$ replaced by $d\dot{n}_{\rm src}/dp$, and we can immediately write down the expected number density of CRs at any given radius $r$ and time $t$ with momentum $>p$ for any $p \in (p_0,p_1)$,
\begin{equation}
    n_{\rm CR}(>p) = \frac{1}{4\pi K r} \mbox{erfc}\left(\frac{r}{2\sqrt{Kt}}\right) \dot{n}_{\rm src} \frac{1-(p/p_1)^{q+1}}{1-(p_0/p_1)^{q+1}},
    \label{eq:nden_exact}
\end{equation}
where $\dot{n}_{\rm src}$ is the total CR injection rate integrated over all momenta. The test is to compare this exact value for the CR number density to the value reconstructed by \criptic, given by \autoref{eq:kde}. We run the test using a diffusion coefficient $K=10^{28}$ cm$^2$ s$^{-1}$ and a source with total energy injection rate $\mathcal{L} = 10^{38}$ erg s$^{-1}$, with $p_0 = 1$ GeV$/c$, $p_1 = 10^3$ GeV$/c$, and $q=-2.2$, for a time, $t=10^9$ s. We use a packet injection rate $10^{-4}$ s$^{-1}$ for the test.

\begin{figure}
    \centering
    \includegraphics[width=\columnwidth]{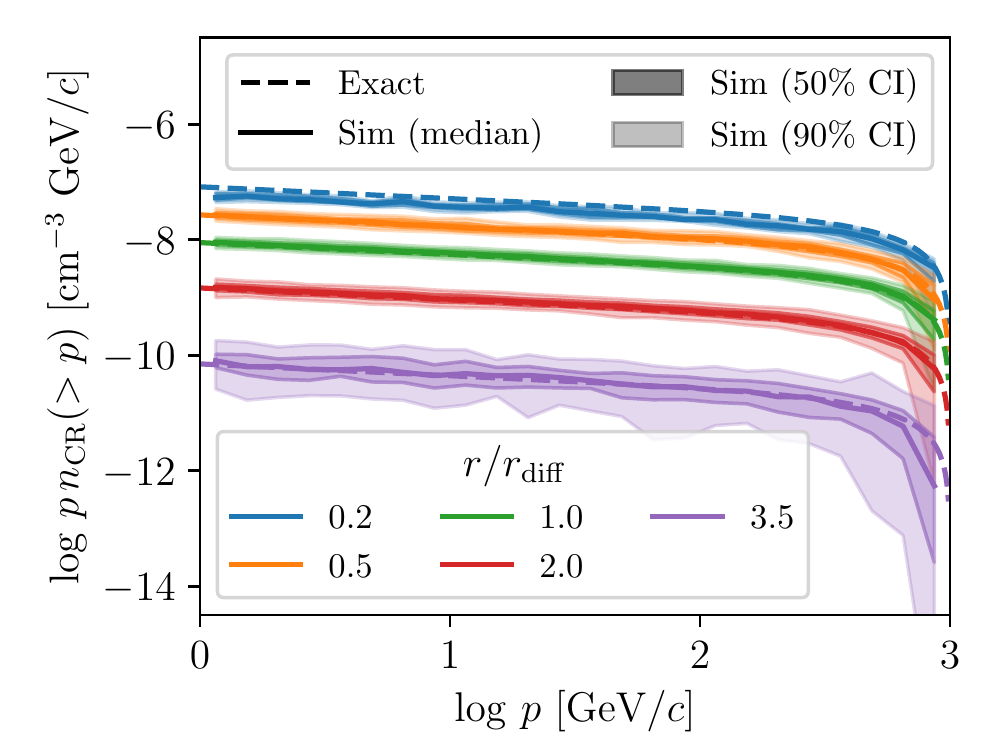}
    \caption{Number density of CRs with momentum $>p$ in the reconstruction test described in \autoref{sssec:nden}. Solid lines show the median value of $n_{\rm CR}(>p)$ computed by \criptic~in 25 logarithmically-spaced bins of $p$, evaluated at 5 different radii (as indicated by the colours). Shaded bands around the medians show the ranges within which 50\% and 90\% of the reconstructed values fall. Dashed lines show the exact solution (\autoref{eq:nden_exact}).}
    \label{fig:NdenComp1}
\end{figure}

\begin{figure}
    \centering
    \includegraphics[width=\columnwidth]{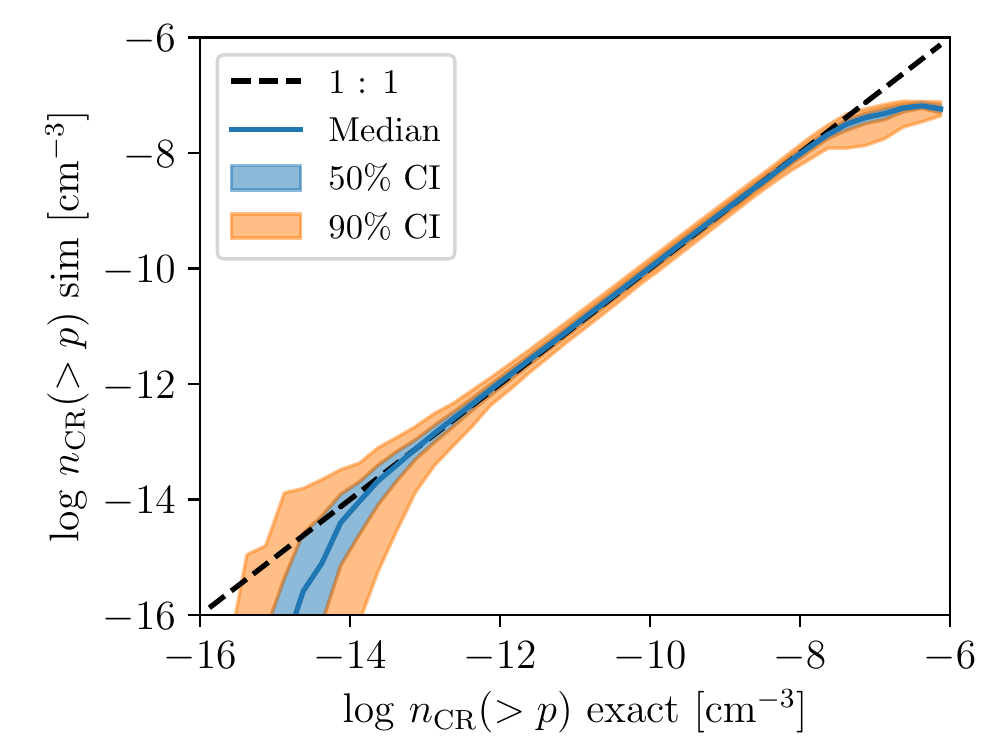}
    \caption{Distribution of exact (horizontal axis) and \criptic-computed (vertical axis) values of $n_{\rm CR}(>p)$, the number density of CRs with momentum $>p$. The blue solid line shows the median, while shaded regions indicate the range into which 50\% and 90\% of the reconstructed values fall; the black dashed line is the $1:1$ line, corresponding to perfect agreement.}
    \label{fig:NdenComp2}
\end{figure}

We compare the exact and reconstructed CR number densities in \autoref{fig:NdenComp1}, which shows $n_{\rm CR}(>p)$ computed in 25 logarithmically-spaced bins in $p$, evaluated at radii $r/r_{\rm diff} = 0.2, 0.5, 1.0, 2.0$, and $3.5$, where $r_{\rm diff} = \sqrt{Kt}$, and for each radial bin we consider CR packets whose radii are within 5\% of the target value. Solid lines show the median reconstructed value of $n_{\rm CR}(>p)$ for packets in that bin, and shaded bands show the ranges within which 50\% and 90\% of the packet values fall. We see that the median reconstructed value of $n_{\rm CR}(>p)$ is in almost perfect agreement with the exact result except at the highest values of $p$, where the finite number of sample packets causes deviation, and at $r/r_{\rm diff} = 0.2$, where the simulation median is $\sim 10\%$ below the true value due to the finite size of the smoothing kernel, which blurs out the sharp peak at small radii visible in, e.g., \autoref{fig:AnisoDiff1}. The other clear trend is that the 50 and 90 percent ranges expand at larger radii, simply because our finite number of packets leads to larger Poisson errors.

To make a quantitative analysis of the error, for each CR packet we evaluate the exact value of $n_{\rm CR,exact}(>p)$ at its position from \autoref{eq:nden_exact}, and we place the packets in 40 bins of $n_{\rm CR, exact}(>p)$. For each bin, we examine the distribution of \criptic-reconstructed number densities, $n_{\rm CR,sim}(>p)$, in that bin, and compute the median and the 50\% and 90\% confidence intervals of this distribution. We plot the median and confidence intervals as a function of $n_{\rm CR,exact}(>p)$ in \autoref{fig:NdenComp2}. We see the same patterns that were visible in \autoref{fig:NdenComp1}, i.e., agreement is excellent over most of the range of $n_{\rm CR,exact}(>p)$, but there are deviations at both the highest values, where the smoothing kernel blurs out the very sharp peak around the source, and the lowest values, where the finite number of CR packets in the simulation leads to errors in reconstructing very low density regions. However, over 5 decades in $n_{\rm CR}(>p)$, from $10^{-13} - 10^{-8}$ cm$^{-3}$, the error is very small: averaged over this range, the median value differs from the exact one by only $0.005$ dex, and the 50\% range is only $0.16$ dex wide.

\subsection{Microphysics tests}

Our second set of tests evaluates the performance of our code in simulating the microphysical processes that govern CR loss and secondary production, and the radiation spectra produced thereby. In only some of these cases is an exact analytic solution available, and, where it is not, we compare to expected physical behaviour and limiting cases.

\subsubsection{Proton diffusion with collisional loss}
\label{sssec:AnisoDiffLoss}

Our first test of microphysics evaluates our treatment of proton catastrophic losses, and the coupling between them and transport. To this end, we repeat the anisotropic diffusion experiment described in \autoref{sssec:diffusion}, but including losses due to CR proton inelastic scattering due to a background gas of constant density $\rho$, and using a source that injects a continuous momentum distribution of CR protons at a rate per unit momentum $d\dot{n}_{\rm src}/dp$; we disable all other loss mechanisms for the purposes of this test. Considering only primary protons (i.e., those that have not yet been scattered), the FPE that governs this system is
\begin{equation}
    \frac{\partial \tf}{\partial t} = K \left(\chi\frac{\partial^2\tf}{\partial x^2} + \frac{\partial^2 \tf}{\partial y^2} + \frac{\partial^2 \tf}{\partial z^2}\right) + \frac{d\dot{n}_{\rm src}}{dp} \delta(\mathbf{r}) - \sigma_{\rm nuc} v \frac{\rho}{\mu_{\rm H} m_{\rm H}},
\end{equation}
where $K$ is the perpendicular diffusion coefficient, $\chi$ is the ratio of parallel and perpendicular coefficients (with the magnetic field oriented in the $\hat{x}$ direction), and $\sigma_{\rm nuc}$ and $v$ are the nuclear inelastic cross section and particle velocity as a function of particle momentum $p$. The system can be solved exactly by making the same change of variable $x=\sqrt{\chi}x'$ and $r'^2 = x'^2 + y^2 + z^2$ as in \autoref{sssec:diffusion} to transform the problem to a standard diffusion equation with a loss term, and then writing down the Green's function including the loss term. Since the loss rate is independent of position and time, this is simply
\begin{equation}
    G(r',t) = \frac{1}{(4\pi K t)^{3/2}} \exp\left(-\frac{r'^2}{4 K t} - \frac{t}{t_{\rm loss}}\right),
\end{equation}
where $t_{\rm loss} = \mu_{\rm H} m_{\rm H} / \sigma_{\rm nuc} v \rho$ is the loss timescale. The exact solution (again, considering only the primary proton population), is
\begin{eqnarray}
\lefteqn{
\tf(r',p,t) = \int_0^t \sqrt{\chi} \frac{d\dot{n}_{\rm src}}{dp} G(r',t') \, dt' 
}
\nonumber
\\
& = &\frac{d\dot{n}_{\rm src}}{dp} \frac{1}{8\pi \sqrt{\chi} K r'} \left[
e^{-r'/r_{\rm loss}}\mbox{erfc}\left(\frac{r'}{2r_{\rm diff}} - \frac{r_{\rm diff}}{r_{\rm loss}}\right) + {}
\right.
\nonumber \\
& &
\qquad
\left.
e^{r'/r_{\rm loss}}\mbox{erfc}\left(\frac{r'}{2r_{\rm diff}} + \frac{r_{\rm diff}}{r_{\rm loss}}\right)
\right],
\label{eq:AnisoDiffLossExact}
\end{eqnarray}
where we have defined $r_{\rm loss}^2 = K t_{\rm loss}$ and $r_{\rm diff}^2 = K t$.

We run the test using a momentum-dependent diffusion coefficient $K = 10^{28}(p/m_p c)^{1/2}$ cm$^2$ s$^{-1}$ and anisotropy parameter $\chi=4$. The central CR source produces CR protons with a momentum distribution $d\dot{n}_{\rm src}/dp \propto p^{q}$ with $q=-2.2$ over a momentum range from $p_0 = 0.1$ GeV/$c$ to $p_1 = 10^5$ GeV/$c$, with a total luminosity $L = 10^{38}$ erg s$^{-1}$. The CRs propagate through a uniform background gas of density $\rho = 2.34\times 10^{-21}$ g cm$^{-3}$, and we run the simulation for $t = 2\times 10^{12}$ s, using a packet injection rate $\Gamma=10^{-6}$ s$^{-1}$, so there are $2\times 10^6$ packets present at the end of the simulation. Given this setup, we have $r_{\rm diff} = 45.8 (p/m_p c)^{1/2}$ pc, and $r_{\rm loss}/r_{\rm diff} \approx 0.5$ at momenta far above the pion production threshold, $p_{\rm th} = 0.78$ GeV/$c$; thus we have selected parameters so that losses are relatively important over most of the momentum range of the test, but become unimportant ($r_{\rm loss}\to\infty)$ at the lowest CR momenta.

\begin{figure}
    \centering
    \includegraphics[width=\columnwidth]{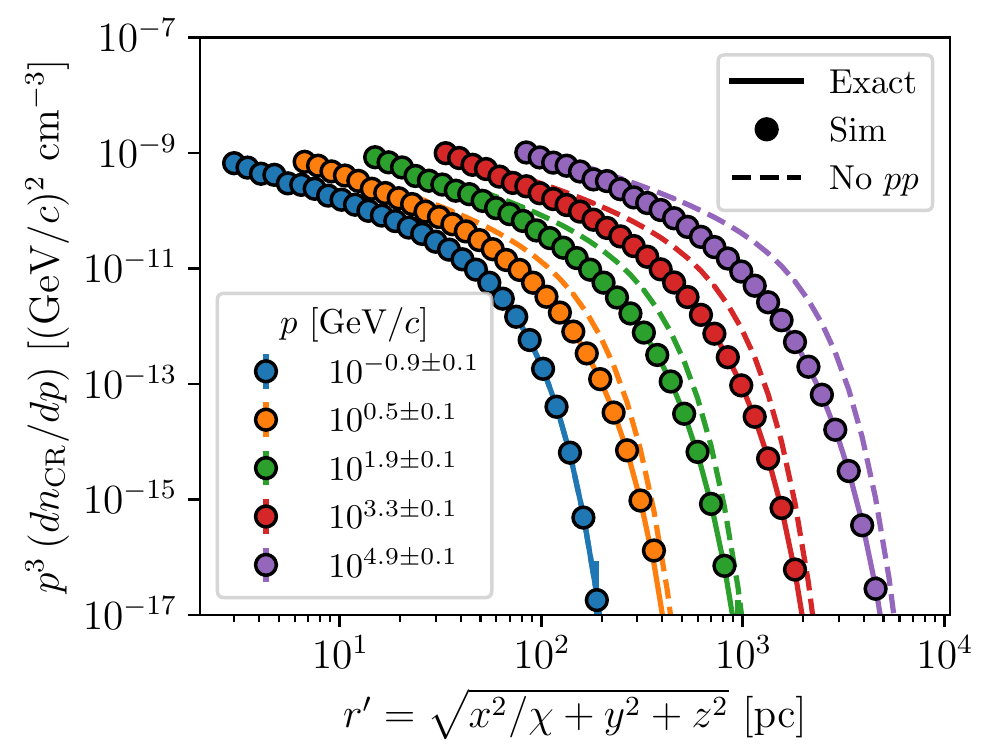}
    \caption{CR proton number density per unit momentum, $dn_{\rm CR}/dp$, as a function of effective radius $r'$ in the anisotropic diffusion with loss test problem (\autoref{sssec:AnisoDiffLoss}). We show $dn_{\rm CR}/dp$ evaluated in five sample momentum bins, each $0.2$ dex wide, spanning from the minimum to the maximum momenta present in the problem; note that our plotted values of $dn_{\rm CR}/dp$ have been scaled by $p^3$ in order to make it easier to display such a wide range of momenta on the same plot. Points with error bars show \criptic~simulation results, with the error bar indicating the Poisson uncertainty on the mean value in each bin due to the finite number of packets. Solid lines show the exact solutions given by \autoref{eq:AnisoDiffLossExact}, while dashed lines show the solutions we would expect to find if we were to disable losses (\autoref{eq:AnisoDiffLossExact} with $r_{\rm loss}\to\infty$).
    }
    \label{fig:AnisoDiffLoss1}
\end{figure}

We show the radial distribution of CR number density for a range of sample momenta in \autoref{fig:AnisoDiffLoss1}, comparing the simulation results to the exact solution given by \autoref{eq:AnisoDiffLossExact}; for comparison we also show the solution that would be expected in the absence of losses, i.e., setting $r_{\rm loss} = \infty$ in \autoref{eq:AnisoDiffLossExact}. We find that \criptic~recovers the correct exact solution, including momentum-dependent loss rates, to the level expected from Poisson statistics. We define the $L^1$ error norm for this problem by 
\begin{equation}
    L^1_{\rm err} = \frac{
    \int_0^\infty \int_{p_0}^{p_1} |\tf_{\rm sim} - \tf_{\rm exact}| \, 4\pi r'^2 p^{q} \, dp \, dr'
    }{
    t\int_{p_0}^{p_1} \frac{d\dot{n}_{\rm src}}{dp} p^{q} \, dp
    },
\end{equation}
where $\tf_{\rm sim}$ are the simulation results and $\tf_{\rm exact}$ is the exact solution. Note the weight factor of $p^{q}$ in this integral is included to ensure that all momenta are weighted equally in computing the error estimate, so the integrand in the denominator is independent of $p$. Defined this way, we find that the $L^1$ error in our \criptic~solution is $2.0\%$.

\subsubsection{Electron streaming with synchrotron and inverse Compton loss}

Our second test evaluates our implementation of synchrotron and inverse Compton losses for electrons. We place a point source of CR electrons with a powerlaw momentum distribution $d\dot{n}_{\rm src}/dp = (\dot{n}_0/p_0) (p/m_e c)^q$ over the range $(p_0, p_1)$ in a medium with a uniform magnetic field $\mathbf{B} = B\hat{x}$ and a uniform radiation field with dilution factor $W_{\rm BB}$ and blackbody temperature $T_{\rm BB}$. The electrons stream in the $+x$ direction at constant speed $w$, while suffering synchrotron and inverse Compton losses. All other loss processes are disabled, as is diffusion. The Fokker-Planck equation for this system is
\begin{equation}
    \frac{\partial \tf}{\partial t} = -\frac{\partial}{\partial x}\left(w \tf\right) + \frac{\partial}{\partial p}\left(\dot{p}_{\rm cts}\tf\right) + \frac{d\dot{n}_{\rm src}}{dp},
\end{equation}
where $\dot{p}_{\rm cts}$ is the rate of momentum loss due to synchrotron and inverse Compton radiation. Since the loss rate is independent of position, the spatial and momentum parts of the system are separable, and there is a one-to-one relationship between the position $x$ and the time $t = x/w$ for which CRs have been subject to loss processes. Over this time, a CR injected with momentum $p_i$ will have been reduced to momentum $p$ given implicitly by $t = -\int_{p_i}^{p} dp'/\dot{p}_{\rm cts}(p')$. If we further adopt the ultrarelativistic limit $\gamma\gg 1$ and $p \approx \gamma m_e c$ and assume that we are far from the Klein-Nishina regime, then $\dot{p}_{\rm cts} \propto p^2$, and we can evaluate the integral analytically; from \autoref{eq:dpdt_sync} and \autoref{eq:dpdt_IC}, we have
\begin{equation}
    \gamma = \frac{\gamma_i}{1 + \gamma_i (t/t_{\rm loss})},
\end{equation}
where for convenience we have expressed the momentum in terms of the Lorentz factor $\gamma$, and
\begin{equation}
    t_{\rm loss} = \frac{3 m_e c}{4 \sigma_T \left(U_B + U_R\right)},
\end{equation}
where $U_B$ and $U_R$ are the energy densities of the magnetic field and radiation field, respectively. This in turn allows us to write down the solution to the Fokker-Planck equation,
\begin{equation}
    \tf(x,\gamma,t) = \frac{\dot{n}_0}{\gamma_0} \left(\frac{1}{1 - \xi \gamma}\right)^2 \left(\frac{\gamma}{1-\xi \gamma}\right)^q
    \label{eq:dist_streamloss}
\end{equation}
for $\xi < t/t_{\rm loss}$ and $\gamma \in [\gamma_0/(1+\xi\gamma_0), \gamma_1/(1+\xi\gamma_1)]$, and 0 otherwise, where here $\gamma = p / m_e c$ is the Lorentz factor in the ultrarelativistic limit and $\xi \equiv x/w t_{\rm loss}$ is the dimensionless distance. Integrating over $x$, the momentum distribution of all electrons with $\gamma > \gamma_0$ is
\begin{equation}
    \frac{dn_{\rm CR}}{d\gamma} = \frac{\dot{n}_0 t_{\rm loss}}{\gamma_0}
    \left(\frac{\gamma^{q-1}}{q+1}\right)
    \left[
    \left(1-\xi_{\rm max}\gamma\right)^{-q-1} - 1
    \right],
    \label{eq:dndp_streamloss}
\end{equation}
where $\xi_{\rm max} = \min(t/t_{\rm loss}, \gamma^{-1}-\gamma_1^{-1})$; for $1\ll \gamma \ll \gamma_1$ and $t\to\infty$, this gives the classical ``cooled synchrotron'' $dn/dp\propto p^{q-1}$, i.e., the spectral index is the injection index minus one.

\begin{figure}
    \centering
    \includegraphics[width=\columnwidth]{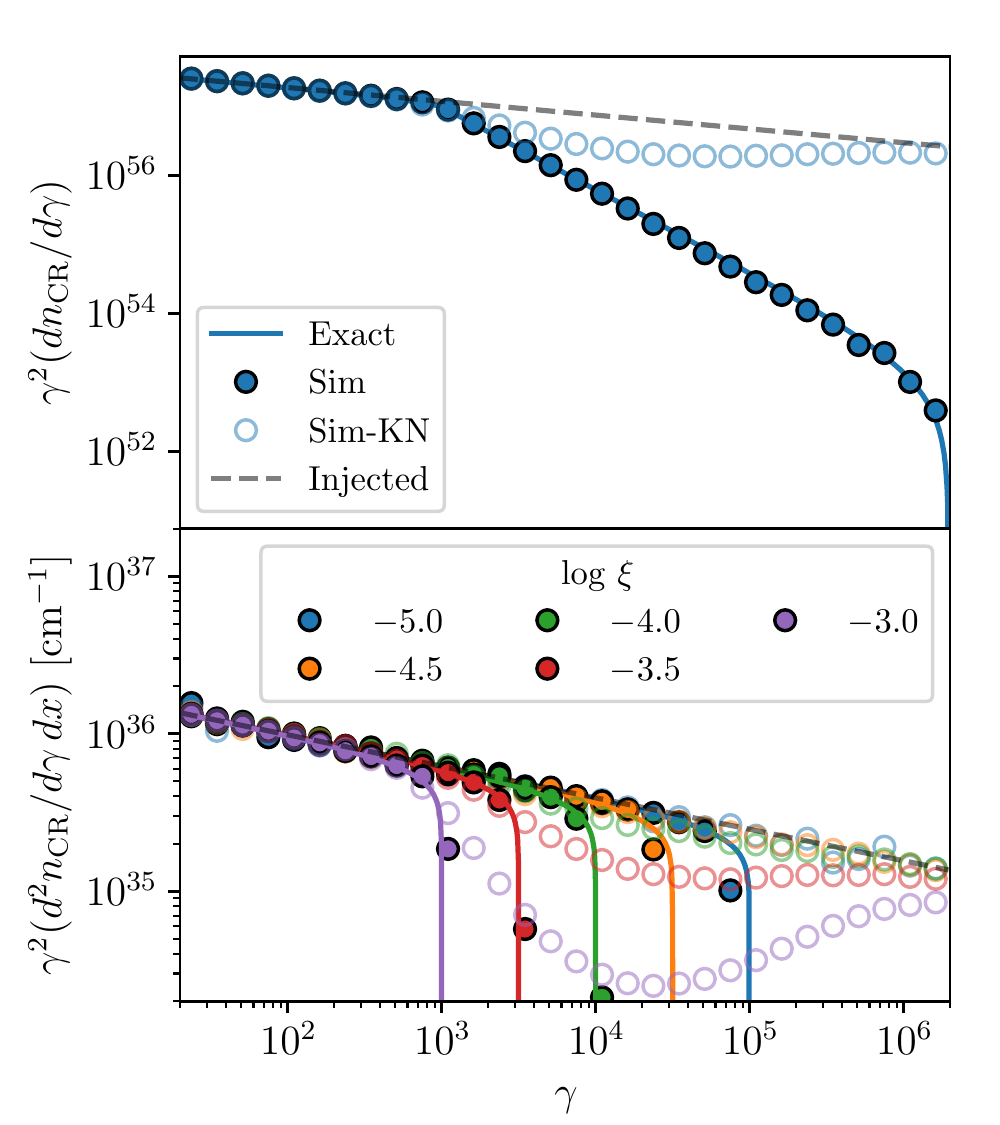}
    \caption{Results of the electron streaming with losses test. In the top panel we show the compensated CR spectrum integrated over all space, $\gamma^2 (dn_{\rm CR}/d\gamma)$ at the end of the simulation. Filled circles show the numerical results evaluated in 30 logarithmically-spaced bins of Lorentz factor $\gamma$, the solid line shows the exact solution (\autoref{eq:dndp_streamloss}), the dashed grey line shows the distribution of CRs injected by the source, and the open circles are the results of our test in the Klein-Nishina regime (see main text). In the lower panel, we show momentum distributions evaluated at five sample positions given by the dimensionless position variable $\xi$ indicated in the legend. As in the top panel, filled circles are the \criptic~numerical solution, solid lines are exact solutions (\autoref{eq:dist_streamloss}), the grey dashed line is the injected spectrum, and the open circles are the results of running the simulation in the Klein-Nishina regime rather than the Thomson regime.}
    \label{fig:ElectronStreamLoss1}
\end{figure}

We carry out this test with a source with total luminosity $10^{38}$ erg s$^{-1}$, which injects CR electrons with a powerlaw momentum distribution characterised by $q=-2.2$ at momenta from $p_0 = 10^{-2}$ GeV$/c$ to $p_1 = 10^3$ GeV$/c$. The electrons stream at $w = 100$ km s$^{-1}$, and we set the energy densities in the background magnetic field and radiation field to $U_B = U_R = 50$ eV cm$^{-3}$; this corresponds to a magnetic field $B=44.87$ $\mu$G, a blackbody radiation temperature $T_{\rm BB} = 10.144$ K (for dilution factor $W_{\rm BB} = 1$), and a loss time $t_{\rm loss} = 6.09$ Gyr. We run the simulation for $10^{-3}t_{\rm loss}$, using a packet injection rate $10^{-8}$ s$^{-1}$. We show the results in \autoref{fig:ElectronStreamLoss1}; the upper panel shows the CR spectrum integrated over all positions compared to the analytic solution given by \autoref{eq:dndp_streamloss}, while the lower panel shows the distribution function evaluated at selected positions, compared to the analytic solution given by \autoref{eq:dist_streamloss}.

As \autoref{fig:ElectronStreamLoss1} shows, \criptic~reproduces the exact solutions very well. Quantitatively, we define the $L^1$ error for the integrated spectrum for this problem by 
\begin{equation}
    L_{\rm err}^1 = \frac{1}{\ln(\gamma_1/\gamma_0)} \int_{\ln \gamma_0}^{\ln \gamma_1} \left|\log_{10} 
    \frac{\left(dn_{\rm CR}/d\gamma\right)_{\rm sim}}{\left(dn_{\rm CR}/d\gamma\right)_{\rm exact}}
    \right| \, d\ln \gamma,
\end{equation}
so the error is the mean logarithmic deviation between the exact and simulated spectra; note that we use this definition because the steep nature of the spectrum means that, if we do not measure the deviation logarithmically, the error norm is dominated by the parts of the spectrum at low $\gamma$, where we simply have the original spectrum. Using this definition, we find $L^1_{\rm err} = 0.012$ dex for the solution shown in \autoref{fig:ElectronStreamLoss1}. We can also define the $L^1$ error at a given dimensionless position $\xi$ analogously, simply by replacing $dn_{\rm CR}/d\gamma$ by $\tf$, and replacing $\gamma_1$ with $\gamma_1/(1 + \xi \gamma_1)$ in the upper integration limit; doing so we find $L^1_{\rm err} = 0.026$ dex at $\xi = 10^{-5}$ to $0.01$ dex at $\xi = 10^{-3}$; the error is largest at $10^{-5}$ because this is sampled by the fewest packets. Nonetheless, even at this small value of $\xi$, the agreement with the exact solution is clearly very good.

Although we do not have exact solution for it, we can also slightly modify this test to verify that \criptic~behaves as expected qualitatively in the Klein-Nishina regime. To do so, we change the background magnetic field to $B = 4.487\times 10^{-9}$ $\mu$G and the radiation field to $T_{\rm BB} = 1.0144\times 10^5$ K with $W_{\rm BB} = 2\times 10^{-16}$; this has the effect of setting $U_R = 100$ eV cm$^{-3}$ and $U_B = 5\times 10^{-19}\mbox{ eV cm}^{-3}\approx 0$. Thus the total magnetic plus radiation energy density is unchanged from the original version of the test, and in the Thomson limit we should recover exactly the same solution. However, while for our previous value of $T_{\rm BB}$ the momentum range $(p_0,p_1)$ for the injected CRs corresponded to $\log\Gamma_{\rm BB} = -6.9$ to $-1.9$, with the higher value of $T_{\rm BB}$ we now have $\log\Gamma_{\rm BB} = -2.9$ to $2.1$. Consequently $\Gamma_{\rm BB} = 1$ occurs in the middle of the injected momentum range, at $p = 7.5$ GeV$/c$ ($\gamma = 1.5\times 10^4$), and we therefore expect Klein-Nishina effects to become significant at momenta approaching this value. 

The open circles in \autoref{fig:ElectronStreamLoss1} show the results of the test in the Klein-Nishina regime. The qualitative result is as expected: for $\gamma \ll 1.5\times 10^4$, we are in the limit $\Gamma_{\rm BB} \ll 1$, and the solution matches the Thomson case. For $\gamma \gg 1.5\times 10^4$, on the other hand, the rate of energy loss scales as $dE/dt\propto \ln \gamma$ \citep{Blumenthal70a}, so the loss timescale obeys $t_{\rm loss}\propto \gamma/\ln\gamma$, rather than $t_{\rm loss}\propto 1/\gamma$ as in the Thomson regime. Thus losses become increasingly unimportant at high $\gamma$, and the spectrum approaches the injected spectrum rather than the Thomson limit solution.

\subsubsection{A thick target}

Our final microphysical test is to simulate a thick target with all microphysical processes enabled, and with a source injecting both primary protons and electrons. In this test \criptic~is performing a calculation similar to that carried out by other authors who treat CR microphysics but do not include transport, or include it only in a simplified parameterised way such as by analytically specifying a loss time or calorimetry fraction \citep[e.g.,][]{Yoast-Hull14a, Peretti19a, Roth21a}. Our goal is to show that \criptic, though not optimised for this type of calculation (since it is possible to obtain the answer much more efficiently if one is uninterested in a detailed treatment of spatial transport), nonetheless recovers CR and emitted $\gamma$-ray spectra similar to those that have been reported in the literature.

For this test we disable spatial transport and momentum diffusion, and we consider an environment such as might be found in a starburst galaxy: a uniform medium of molecular hydrogen characterised by number density $n_{\rm H} = 10^3$ H nuclei cm$^{-3}$, an ionisation fraction by mass $\chi = 10^{-4}$, a magnetic field strength $B = 0.3$ mG, and two radiation fields both with $W_{\rm BB} = 1$, one with $T_{\rm BB} = 2.73$ K (the CMB) and one with $T_{\rm BB} = 20$ K (representing a reprocessed dust radiation field). We place two sources in the medium. One injects protons with a momentum distribution $dn/dp \propto p^{-2.2}$ over a range in kinetic energy $T = [10^{-3}, 10^6]$ GeV, and has total luminosity $L = 3\times 10^{42}$ erg s$^{-1}$, corresponding roughly to the luminosity expected for a galaxy with a star formation rate of $\approx 100$ M$_\odot$ yr$^{-1}$, assuming 1 SN per 100 M$_\odot$ of stars formed, a total energy of $10^{51}$ erg per SN, with $\approx 10\%$ of that taking the form of CRs. The other source injects electrons with the same spectrum, but a total luminosity a factor of 10 smaller. We run the simulation for 0.5 Myr using a primary packet injection rate $\Gamma_{\rm inj} = 2\times 10^{-7}$ s$^{-1}$; this is longer than the loss time at all energies for the background environment, so by this time the system settles to steady state. At the end of the simulation, there are approximately $4.3\times 10^6$ CR packets present.

\begin{figure}
    \centering
    \includegraphics[width=\columnwidth]{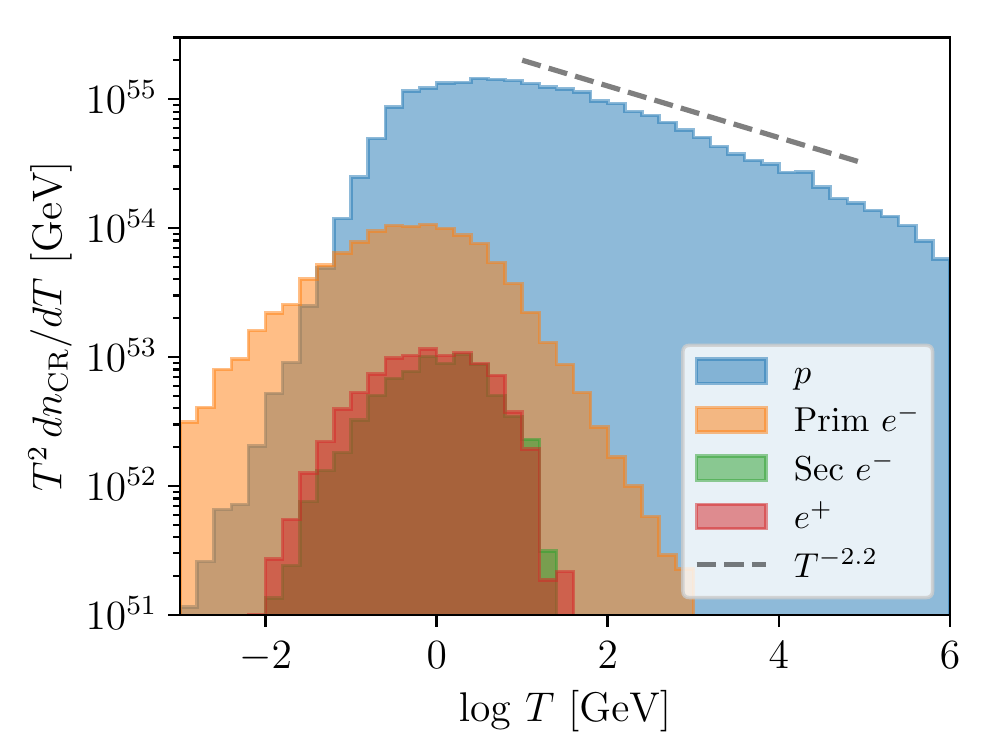}
    \caption{Steady-state CR spectra in the thick target problem. For each species we plot $T^2\, dn_{\rm CR}/dT$, where $dn_{\rm CR}/dT$ is the number of individual CR particles in a particular energy bin. Different colours indicate primary protons, primary electrons, secondary electrons, and (secondary) positrons, as indicated in the legend. The dashed black line labelled $T^{-2.2}$ shows the shape of the injection spectrum.}
    \label{fig:ThickTarget1}
\end{figure}

We show the resulting steady-state CR spectra in \autoref{fig:ThickTarget1}. The result is in accord with what we would expect: the proton spectrum has a slope that is very slightly shallower than the injection spectrum at high energies, reflecting the slight increase in $pp$ cross section with energy. This continues to $\approx 1$ GeV, and below this energy the spectrum dies off quickly, reflecting the strong ionisation losses that low-energy CRs suffer in a starburst environment (c.f.~\autoref{fig:losses}); indeed, each second the CR protons ionise a total of $6.0\times 10^{50}$ H$_2$ molecules and $7.2\times 10^{49}$ He atoms.\footnote{Note that these should be understood as primary ionisations, since we do not track electrons with energies $\lesssim 1$ MeV, which overwhelmingly dominate secondary ionisations.} The primary electron spectrum is both lower in absolute value and substantially steeper, with a slope closer to $-3.2$ at high energy, as a result of the quadratic dependence of synchrotron and inverse Compton losses on CR energy. It too falls off below $\approx 1$ GeV due to ionisation losses -- the total ionisation rate produced by electrons and positrons is $6.8\times 10^{50}$ s$^{-1}$ and $8.7\times 10^{49}$ s$^{-1}$ for H$_2$ and He, respectively. Finally, the secondary electron and positron spectra are nearly identical, and are lower still, reflecting the relatively large value we have chosen for the primary electron / proton ratio in this test.

\begin{figure*}
    \centering
    \includegraphics[width=\textwidth]{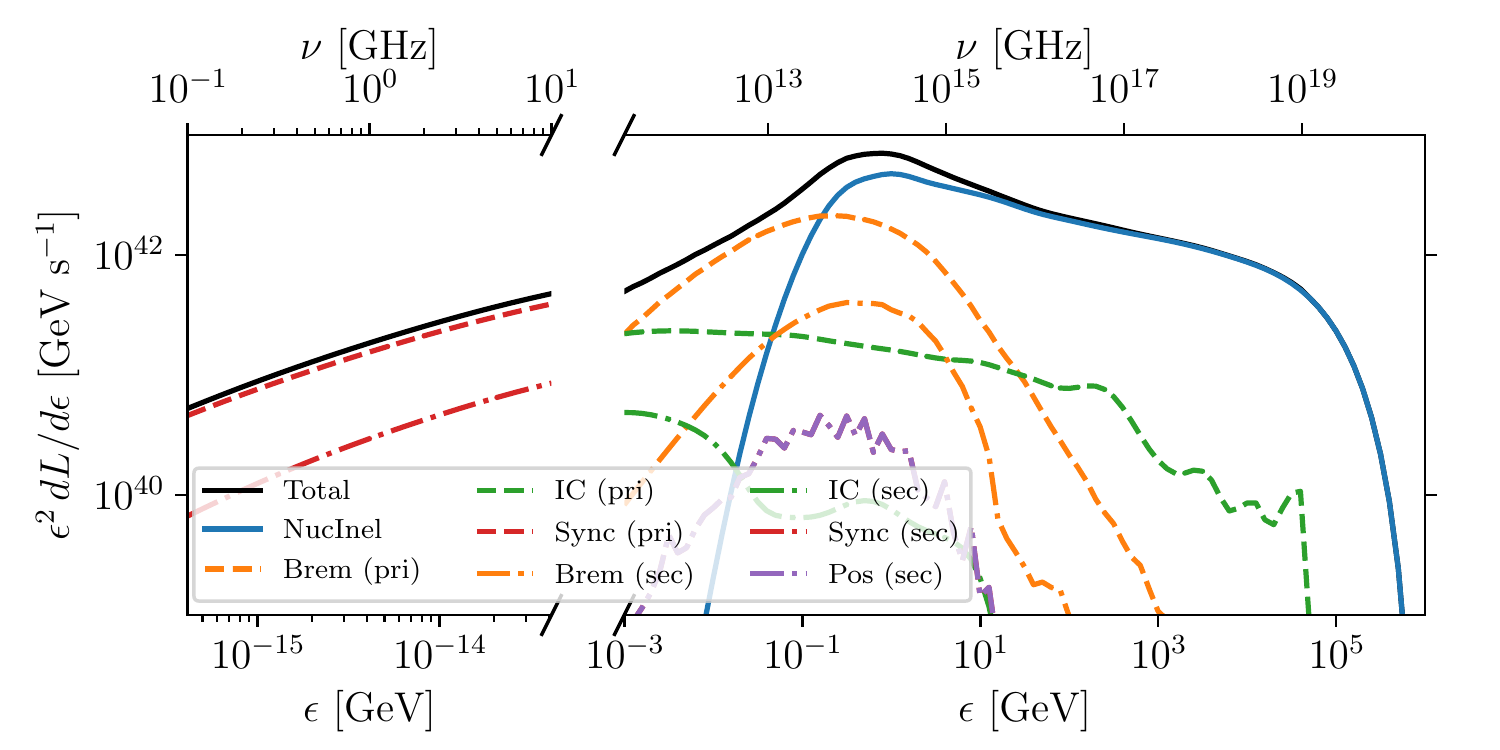}
    \caption{Emitted $\gamma$-ray spectra produced in the thick target test. The black solid line shows the total spectrum produced by all particles and processes, and the blue solid line shows the contribution from nuclear inelastic scattering. Dashed and dot-dashed lines in different colours show emission by primary and secondary electrons and positrons, respectively, with the colour indicating the emission process -- bremsstrahlung, inverse Compton, and synchrotron.}
    \label{fig:ThickTarget2}
\end{figure*}

We show the $\gamma$-ray emission that \criptic~predicts for this system in \autoref{fig:ThickTarget2}. We find that nuclear inelastic scattering dominates at high energies, with a sharp cutoff just below $10^6$ GeV reflecting the cutoff in the injection spectrum that we apply there. This gives rise to the usual bump at $\approx 1$ GeV
in the spectral energy distribution.  Bremsstrahlung and inverse Compton emission from primary electrons significantly contribute to the total
emission
at energies below $\approx 1$ GeV. In the radio, we see a dominant contribution from primary electron synchrotron, with a $\approx 10\%$ additional contribution from secondary electrons. All other processes are subdominant. Note that the wavy structure in the inverse Compton and bremmstrahlung spectra at high energy is real, and reflects the modulation imposed by the fact that we have two blackbody radiation fields at different temperatures present. On the other hand, the somewhat spiky contribution from positron annihilation (which includes only the Doppler-boosted 511 keV photons, not the inverse Compton, bremmstrahlung, or synchrotron contributions from positrons) is a result of the relatively small number of positron packets present in the calculation, which causes some numerical noise. However, since this component is subdominant by $\approx 2$ dex, this has no noticeable effect on the total spectrum. In general, our result is qualitatively consistent with the results of other thick target calculations of starburst $\gamma$-ray and radio spectra.

\section{Discussion and conclusion}
\label{sec:discussion}

We conclude by discussing applications of \criptic, as well as the limitations of the current code and our plans for future expansions. 

\subsection{Applications}

\Criptic~can be used to solve a wide range of problems in CR transport. In \citet{Sampson22a}, we have already applied it to the problem of determining an effective transport theory for CRs that stream through a turbulent plasma. This application exploits \criptic's ability to model transport through an arbitrary, time-dependent background -- in this case the output of an MHD turbulence simulation.

In future work we intend to use \criptic~to post-process MHD simulations of Milky Way-like galactic discs \citep[e.g.,][]{Wibking22b}, in order to compare the results produced by different CR transport models with observable quantities such as the $\gamma$-ray spectral index as a function of height above the galactic plane. We also intend to post-process CR hydrodynamics simulations in order to predict detailed observables from them.

Astrochemistry and the link between it and high-energy phenomena represents another immediate application. There has been considerable debate about the CR ionisation rate in starburst galaxies \citep[e.g.,][]{Papadopoulos10a, Bisbas15a, Narayanan17a} and in the Milky Way Central Molecular Zone \citep{Ginsburg16a, Oka19a, Tanaka21a}, but thus far this discussion has focused on molecular indicators, and has not taken advantage of constraints offered by high-energy tracers such as $\gamma$-ray emission. However, the CRs that drive ionisation and those that produce $\gamma$-rays are ultimately part of the same population, albeit at somewhat different energies. Efforts to combine constraints on the low-energy population that drives ionisation and the high-energy population that drives $\gamma$-ray emission have thus far been very limited \citep[e.g.,][]{Armillotta22a}, but \criptic's ability to simulate the full CR energy range should greatly facilitate these efforts.

\subsection{Limitations and expansion plans}

No simulation code can ever capture all of physics, and that is certainly true of \criptic. It is therefore worth concluding pointing out some limitations of the current code, and some plans for future expansion to remedy at least some of these limitations. One limitation is that \criptic~is a post-processing code, and therefore does not allow for calculation of the back-reaction on
the flow due to CR pressure forces or heating. In this regard, it is analogous to Monte Carlo post-processing radiative transfer codes, which can be used to predict detailed spectra, but do not allow self-consistent calculation of how a flow reacts to radiation forces. To the extent that a calculation using criptic predicts a CR field where CR forces are important (or, analogously, a Monte Carlo calculation predicts a dynamically-important radiation pressure force), the correct way to proceed is to evolve the system with a code that self-consistently includes those forces, and then use a post-processing tool like \criptic~to calculate the observable emission at much higher resolution that would be possible from the self-consistent calculation alone. Beyond this limit to the overall \criptic~methodology, though, we here identify three other limitations that are ripe for improvement in future releases.

First, \criptic~does not yet include all of the CR species or loss processes for which observational constraints exist. Since we have focused on radiative and astrochemical signatures from galaxies, and particularly radiative signatures that are observable from beyond the Milky way, in this first release we have included the species that dominate these. However, direct \textit{in situ} measurements exist for a range of heavier CR nuclei, most prominently He, B, and C. We intend to include these species in a future release. We will also add anti-protons and the process of positron annihilation via positronium formation, which do produce radiative signatures that are observable in the Galaxy, if not from external galaxies. Finally, we have focused on CRs in the $\sim 1\mbox{ MeV} - 1\mbox{ PeV}$ range in typical galactic environments, and have not included loss processes that become dominant outside this range or in highly-magnetised environments such as around active galactic nuclei. At low energies the main omitted process is charge exchange \citep[e.g.,][]{Schultz08a}, while at the high energy end it is photo-hadronic interactions \citep[e.g.,][]{Mucke99a}; photon-photon scattering can also become a significant loss process for high-energy photons in strong radiation environments, and is not yet included. In highly-magnetised environment, we should also include synchrotron losses for protons. The modular nature of the code makes addition of such processes straightforward, and these too may be included in future releases.

Second, at present \criptic~solves the Fokker-Planck equation in the spatial and momentum directions, but not in pitch angle; it is therefore valid only on scales large enough that the local CR pitch angle distribution has become approximately isotropic. This limits applications to ultra-high energy CRs, and potentially to environments where CR scattering is very strongly suppressed since, in both these regimes, the mean free path that CRs travel before becoming isotropised can be large. Fortunately our It\^o calculus-based formulation of the problem is readily extensible to include the pitch angle dimension as well, and we intend to include an option to solve problems in the anisotropic regime in future releases.

Third, \criptic~does not yet predict neutrino emission. This is a straightforward extension to the existing code, since the main process responsible for producing neutrinos -- nuclear inelastic collisions leading to charged pion production -- is already included via calculation of secondary electrons and positrons. The limitation is rather that existing neutrino observatories are limited to very high energy neutrinos, where the point raised above applies, i.e., it is unclear that one can assume pitch angle isotropisation for the CRs that drive the observable neutrino emission. Extension to neutrinos is therefore likely to wait until after the extension to follow diffusion in pitch angle is complete.

Finally, we note that \criptic~is released under an open source license. Users are encouraged to contribute their own expansions, which can be incorporated into future releases.

\section*{Data availability}

\Criptic~is available from \url{https://bitbucket.org/krumholz/criptic/src/master/} under an open source license. The setup files and analysis scripts required to reproduce all the test problems described in this paper are included in the repository. An introduction and users' guide is available at \url{https://criptic.readthedocs.io/}.

\section*{Acknowledgements}

We thank I.~Kafexhiu for assistance with cross section calculations, and C.~Federrath for assistance in developing an interface for \criptic~to read \textsc{flash} simulations. \Criptic~and its associated support software use the following software packages: Astropy\footnote{http://www.astropy.org} \citep{Astropy-Collaboration13a, Astropy-Collaboration18a}, Numpy \citep{Numpy20a}, Matplotlib \citep{Hunter07a}, GNU Scientific Library \citep{Galassi09a}, PCG \citep{ONeill14a}. This research was undertaken with the assistance of resources and services from the National Computational Infrastructure (NCI), award jh2, which is supported by the Australian Government. This research was funded by the Australian Research Council through its Discovery Projects funding scheme, award DP190101258.




\bibliographystyle{mnras}
\bibliography{refs} 

\begin{thebibliography}{}
\makeatletter
\relax
\def\mn@urlcharsother{\let\do\@makeother \do\$\do\&\do\#\do\^\do\_\do\%\do\~}
\def\mn@doi{\begingroup\mn@urlcharsother \@ifnextchar [ {\mn@doi@}
  {\mn@doi@[]}}
\def\mn@doi@[#1]#2{\def\@tempa{#1}\ifx\@tempa\@empty \href
  {http://dx.doi.org/#2} {doi:#2}\else \href {http://dx.doi.org/#2} {#1}\fi
  \endgroup}
\def\mn@eprint#1#2{\mn@eprint@#1:#2::\@nil}
\def\mn@eprint@arXiv#1{\href {http://arxiv.org/abs/#1} {{\tt arXiv:#1}}}
\def\mn@eprint@dblp#1{\href {http://dblp.uni-trier.de/rec/bibtex/#1.xml}
  {dblp:#1}}
\def\mn@eprint@#1:#2:#3:#4\@nil{\def\@tempa {#1}\def\@tempb {#2}\def\@tempc
  {#3}\ifx \@tempc \@empty \let \@tempc \@tempb \let \@tempb \@tempa \fi \ifx
  \@tempb \@empty \def\@tempb {arXiv}\fi \@ifundefined
  {mn@eprint@\@tempb}{\@tempb:\@tempc}{\expandafter \expandafter \csname
  mn@eprint@\@tempb\endcsname \expandafter{\@tempc}}}

\bibitem[\protect\citeauthoryear{{Ambrosone}, {Chianese}, {Fiorillo},
  {Marinelli}  \& {Miele}}{{Ambrosone} et~al.}{2022}]{Ambrosone22a}
{Ambrosone} A.,  {Chianese} M.,  {Fiorillo} D.~F.~G.,  {Marinelli} A.,
  {Miele} G.,  2022, arXiv e-prints, \href
  {https://ui.adsabs.harvard.edu/abs/2022arXiv220303642A} {p. arXiv:2203.03642}

\bibitem[\protect\citeauthoryear{{Armillotta}, {Ostriker}  \&
  {Jiang}}{{Armillotta} et~al.}{2021}]{Armillotta21a}
{Armillotta} L.,  {Ostriker} E.~C.,   {Jiang} Y.-F.,  2021, \mn@doi [\apj]
  {10.3847/1538-4357/ac1db2}, \href
  {https://ui.adsabs.harvard.edu/abs/2021ApJ...922...11A} {922, 11}

\bibitem[\protect\citeauthoryear{{Armillotta}, {Ostriker}  \&
  {Jiang}}{{Armillotta} et~al.}{2022}]{Armillotta22a}
{Armillotta} L.,  {Ostriker} E.~C.,   {Jiang} Y.-F.,  2022, \mn@doi [\apj]
  {10.3847/1538-4357/ac5fa9}, \href
  {https://ui.adsabs.harvard.edu/abs/2022ApJ...929..170A} {929, 170}

\bibitem[\protect\citeauthoryear{{Asplund}, {Grevesse}, {Sauval}  \&
  {Scott}}{{Asplund} et~al.}{2009}]{Asplund09a}
{Asplund} M.,  {Grevesse} N.,  {Sauval} A.~J.,   {Scott} P.,  2009, \mn@doi
  [\araa] {10.1146/annurev.astro.46.060407.145222}, \href
  {http://adsabs.harvard.edu/abs/2009ARA%26A..47..481A} {47, 481}

\bibitem[\protect\citeauthoryear{{Astropy Collaboration} et~al.,}{{Astropy
  Collaboration} et~al.}{2013}]{Astropy-Collaboration13a}
{Astropy Collaboration} et~al., 2013, \mn@doi [\aap]
  {10.1051/0004-6361/201322068}, \href
  {http://adsabs.harvard.edu/abs/2013A%26A...558A..33A} {558, A33}

\bibitem[\protect\citeauthoryear{{Astropy Collaboration} et~al.,}{{Astropy
  Collaboration} et~al.}{2018}]{Astropy-Collaboration18a}
{Astropy Collaboration} et~al., 2018, \mn@doi [\aj] {10.3847/1538-3881/aabc4f},
  \href {https://ui.adsabs.harvard.edu/abs/2018AJ....156..123A} {156, 123}

\bibitem[\protect\citeauthoryear{{Beattie}, {Krumholz}, {Federrath}, {Sampson}
  \& {Crocker}}{{Beattie} et~al.}{2022}]{Beattie22a}
{Beattie} J.~R.,  {Krumholz} M.~R.,  {Federrath} C.,  {Sampson} M.,   {Crocker}
  R.~M.,  2022, arXiv e-prints, \href
  {https://ui.adsabs.harvard.edu/abs/2022arXiv220313952B} {p. arXiv:2203.13952}

\bibitem[\protect\citeauthoryear{{Bisbas}, {Papadopoulos}  \& {Viti}}{{Bisbas}
  et~al.}{2015}]{Bisbas15a}
{Bisbas} T.~G.,  {Papadopoulos} P.~P.,   {Viti} S.,  2015, \mn@doi [\apj]
  {10.1088/0004-637X/803/1/37}, \href
  {http://adsabs.harvard.edu/abs/2015ApJ...803...37B} {803, 37}

\bibitem[\protect\citeauthoryear{{Blumenthal} \& {Gould}}{{Blumenthal} \&
  {Gould}}{1970}]{Blumenthal70a}
{Blumenthal} G.~R.,  {Gould} R.~J.,  1970, \mn@doi [Reviews of Modern Physics]
  {10.1103/RevModPhys.42.237}, \href
  {https://ui.adsabs.harvard.edu/abs/1970RvMP...42..237B} {42, 237}

\bibitem[\protect\citeauthoryear{{Chan}, {Kere{\v{s}}}, {Hopkins}, {Quataert},
  {Su}, {Hayward}  \& {Faucher-Gigu{\`e}re}}{{Chan} et~al.}{2019}]{Chan19a}
{Chan} T.~K.,  {Kere{\v{s}}} D.,  {Hopkins} P.~F.,  {Quataert} E.,  {Su} K.~Y.,
   {Hayward} C.~C.,   {Faucher-Gigu{\`e}re} C.~A.,  2019, \mn@doi [\mnras]
  {10.1093/mnras/stz1895}, \href
  {https://ui.adsabs.harvard.edu/abs/2019MNRAS.488.3716C} {488, 3716}

\bibitem[\protect\citeauthoryear{{Cherenkov Telescope Array Consortium}
  et~al.,}{{Cherenkov Telescope Array Consortium} et~al.}{2019}]{CTA19a}
{Cherenkov Telescope Array Consortium} et~al., 2019, {Science with the
  Cherenkov Telescope Array}.
World Scientific Publishing: Signapore, \mn@doi{10.1142/10986}

\bibitem[\protect\citeauthoryear{{Crocker}, {Krumholz}  \&
  {Thompson}}{{Crocker} et~al.}{2021a}]{Crocker21a}
{Crocker} R.~M.,  {Krumholz} M.~R.,   {Thompson} T.~A.,  2021a, \mn@doi
  [\mnras] {10.1093/mnras/stab148}, \href
  {https://ui.adsabs.harvard.edu/abs/2021MNRAS.502.1312C} {502, 1312}

\bibitem[\protect\citeauthoryear{{Crocker}, {Krumholz}  \&
  {Thompson}}{{Crocker} et~al.}{2021b}]{Crocker21b}
{Crocker} R.~M.,  {Krumholz} M.~R.,   {Thompson} T.~A.,  2021b, \mn@doi
  [\mnras] {10.1093/mnras/stab502}, \href
  {https://ui.adsabs.harvard.edu/abs/2021MNRAS.503.2651C} {503, 2651}

\bibitem[\protect\citeauthoryear{{Dirac}}{{Dirac}}{1930}]{Dirac30a}
{Dirac} P. A.~M.,  1930, \mn@doi [Mathematical Proceedings of the Cambridge
  Philosophical Society] {10.1017/S0305004100016091}, 26, 361

\bibitem[\protect\citeauthoryear{{Draine}}{{Draine}}{2011}]{Draine11a}
{Draine} B.~T.,  2011, {Physics of the Interstellar and Intergalactic Medium}.
Princeton University Press: Princeton, NJ

\bibitem[\protect\citeauthoryear{{Dullemond}}{{Dullemond}}{2012}]{Dullemond12a}
{Dullemond} C.~P.,  2012, Astrophysics Source Code Library, \href
  {http://adsabs.harvard.edu/abs/2012ascl.soft02015D} {p.~2015}

\bibitem[\protect\citeauthoryear{{En{\ss}lin}, {Pfrommer}, {Springel}  \&
  {Jubelgas}}{{En{\ss}lin} et~al.}{2007}]{Enslin07a}
{En{\ss}lin} T.~A.,  {Pfrommer} C.,  {Springel} V.,   {Jubelgas} M.,  2007,
  \mn@doi [\aap] {10.1051/0004-6361:20065294}, \href
  {http://adsabs.harvard.edu/abs/2007A\%26A...473...41E} {473, 41}

\bibitem[\protect\citeauthoryear{{Evoli}, {Gaggero}, {Vittino}, {Di Bernardo},
  {Di Mauro}, {Ligorini}, {Ullio}  \& {Grasso}}{{Evoli}
  et~al.}{2017}]{Evoli17a}
{Evoli} C.,  {Gaggero} D.,  {Vittino} A.,  {Di Bernardo} G.,  {Di Mauro} M.,
  {Ligorini} A.,  {Ullio} P.,   {Grasso} D.,  2017, \mn@doi [\jcap]
  {10.1088/1475-7516/2017/02/015}, \href
  {https://ui.adsabs.harvard.edu/abs/2017JCAP...02..015E} {2017, 015}

\bibitem[\protect\citeauthoryear{Fehlberg}{Fehlberg}{1970}]{Fehlberg70a}
Fehlberg E.,  1970, \mn@doi [Computing] {10.1007/BF02241732}, 6, 61

\bibitem[\protect\citeauthoryear{{Galassi}, {Davies}, {Theiler}, {Gough},
  {Jungman}, {Alken}, {Booth}  \& {Rossi}}{{Galassi} et~al.}{2009}]{Galassi09a}
{Galassi} M.,  {Davies} J.,  {Theiler} J.,  {Gough} B.,  {Jungman} G.,  {Alken}
  P.,  {Booth} M.,   {Rossi} F.,  2009, GNU Scientific Library Reference
  Manual.
3rd ed. edn

\bibitem[\protect\citeauthoryear{Garc\'ia-Portugu\'es}{Garc\'ia-Portugu\'es}{2022}]{Garcia-Portugues22a}
Garc\'ia-Portugu\'es E.,  2022, Notes for Nonparametric Statistics.
\url {https://bookdown.org/egarpor/NP-UC3M/}

\bibitem[\protect\citeauthoryear{{Gardiner}}{{Gardiner}}{2009}]{Gardiner09a}
{Gardiner} C.,  2009, {Stochastic Methods: A Handbook for the Natural and
  Physical Sciences}, 4 edn.
Springer, Berlin

\bibitem[\protect\citeauthoryear{{Ginsburg} et~al.,}{{Ginsburg}
  et~al.}{2016}]{Ginsburg16a}
{Ginsburg} A.,  et~al., 2016, \mn@doi [\aap] {10.1051/0004-6361/201526100},
  \href {http://adsabs.harvard.edu/abs/2016A%26A...586A..50G} {586, A50}

\bibitem[\protect\citeauthoryear{{Girichidis} et~al.,}{{Girichidis}
  et~al.}{2016}]{Girichidis16a}
{Girichidis} P.,  et~al., 2016, \mn@doi [\apjl] {10.3847/2041-8205/816/2/L19},
  \href {https://ui.adsabs.harvard.edu/abs/2016ApJ...816L..19G} {816, L19}

\bibitem[\protect\citeauthoryear{{Girichidis}, {Pfrommer}, {Pakmor}  \&
  {Springel}}{{Girichidis} et~al.}{2022}]{Girichidis22a}
{Girichidis} P.,  {Pfrommer} C.,  {Pakmor} R.,   {Springel} V.,  2022, \mn@doi
  [\mnras] {10.1093/mnras/stab3462}, \href
  {https://ui.adsabs.harvard.edu/abs/2022MNRAS.510.3917G} {510, 3917}

\bibitem[\protect\citeauthoryear{{Gould}}{{Gould}}{1972}]{Gould72a}
{Gould} R.~J.,  1972, \mn@doi [Physica] {10.1016/0031-8914(72)90227-3}, \href
  {https://ui.adsabs.harvard.edu/abs/1972Phy....60..145G} {60, 145}

\bibitem[\protect\citeauthoryear{Harris et~al.,}{Harris
  et~al.}{2020}]{Numpy20a}
Harris C.~R.,  et~al., 2020, \mn@doi [Nature] {10.1038/s41586-020-2649-2}, 585,
  357

\bibitem[\protect\citeauthoryear{{Hopkins} et~al.,}{{Hopkins}
  et~al.}{2020}]{Hopkins20a}
{Hopkins} P.~F.,  et~al., 2020, \mn@doi [\mnras] {10.1093/mnras/stz3321}, \href
  {https://ui.adsabs.harvard.edu/abs/2020MNRAS.492.3465H} {492, 3465}

\bibitem[\protect\citeauthoryear{{Hopkins}, {Squire}, {Butsky}  \&
  {Ji}}{{Hopkins} et~al.}{2021a}]{Hopkins21c}
{Hopkins} P.~F.,  {Squire} J.,  {Butsky} I.~S.,   {Ji} S.,  2021a, arXiv
  e-prints, \href {https://ui.adsabs.harvard.edu/abs/2021arXiv211202153H} {p.
  arXiv:2112.02153}

\bibitem[\protect\citeauthoryear{{Hopkins}, {Chan}, {Squire}, {Quataert}, {Ji},
  {Kere{\v{s}}}  \& {Faucher-Gigu{\`e}re}}{{Hopkins}
  et~al.}{2021b}]{Hopkins21a}
{Hopkins} P.~F.,  {Chan} T.~K.,  {Squire} J.,  {Quataert} E.,  {Ji} S.,
  {Kere{\v{s}}} D.,   {Faucher-Gigu{\`e}re} C.-A.,  2021b, \mn@doi [\mnras]
  {10.1093/mnras/staa3692}, \href
  {https://ui.adsabs.harvard.edu/abs/2021MNRAS.501.3663H} {501, 3663}

\bibitem[\protect\citeauthoryear{{Hopkins}, {Butsky}, {Panopoulou}, {Ji},
  {Quataert}, {Faucher-Giguere}  \& {Keres}}{{Hopkins}
  et~al.}{2022}]{Hopkins22a}
{Hopkins} P.~F.,  {Butsky} I.~S.,  {Panopoulou} G.~V.,  {Ji} S.,  {Quataert}
  E.,  {Faucher-Giguere} C.-A.,   {Keres} D.,  2022, \mnras~in review, \href
  {https://ui.adsabs.harvard.edu/abs/2021arXiv210909762H} {p. arXiv:2109.09762}

\bibitem[\protect\citeauthoryear{{Hunter}}{{Hunter}}{2007}]{Hunter07a}
{Hunter} J.~D.,  2007, \mn@doi [Computing in Science \& Engineering]
  {10.1109/MCSE.2007.55}, 9, 90

\bibitem[\protect\citeauthoryear{{Ivlev}, {Silsbee}, {Padovani}  \&
  {Galli}}{{Ivlev} et~al.}{2021}]{Ivlev21a}
{Ivlev} A.~V.,  {Silsbee} K.,  {Padovani} M.,   {Galli} D.,  2021, \mn@doi
  [\apj] {10.3847/1538-4357/abdc27}, \href
  {https://ui.adsabs.harvard.edu/abs/2021ApJ...909..107I} {909, 107}

\bibitem[\protect\citeauthoryear{{Jones}}{{Jones}}{1968}]{Jones68a}
{Jones} F.~C.,  1968, \mn@doi [Physical Review] {10.1103/PhysRev.167.1159},
  \href {https://ui.adsabs.harvard.edu/abs/1968PhRv..167.1159J} {167, 1159}

\bibitem[\protect\citeauthoryear{{Kafexhiu}}{{Kafexhiu}}{2016}]{Kafexhiu16a}
{Kafexhiu} E.,  2016, \mn@doi [\prc] {10.1103/PhysRevC.94.064603}, \href
  {https://ui.adsabs.harvard.edu/abs/2016PhRvC..94f4603K} {94, 064603}

\bibitem[\protect\citeauthoryear{{Kafexhiu}, {Aharonian}, {Taylor}  \&
  {Vila}}{{Kafexhiu} et~al.}{2014}]{Kafexhiu14a}
{Kafexhiu} E.,  {Aharonian} F.,  {Taylor} A.~M.,   {Vila} G.~S.,  2014, \mn@doi
  [\prd] {10.1103/PhysRevD.90.123014}, \href
  {https://ui.adsabs.harvard.edu/abs/2014PhRvD..90l3014K} {90, 123014}

\bibitem[\protect\citeauthoryear{{Kelner}, {Aharonian}  \& {Bugayov}}{{Kelner}
  et~al.}{2006}]{Kelner06a}
{Kelner} S.~R.,  {Aharonian} F.~A.,   {Bugayov} V.~V.,  2006, \mn@doi [\prd]
  {10.1103/PhysRevD.74.034018}, \href
  {https://ui.adsabs.harvard.edu/abs/2006PhRvD..74c4018K} {74, 034018}

\bibitem[\protect\citeauthoryear{{Kim} \& {Rudd}}{{Kim} \&
  {Rudd}}{1994}]{Kim94a}
{Kim} Y.-K.,  {Rudd} M.~E.,  1994, \mn@doi [\pra] {10.1103/PhysRevA.50.3954},
  \href {https://ui.adsabs.harvard.edu/abs/1994PhRvA..50.3954K} {50, 3954}

\bibitem[\protect\citeauthoryear{{Kim}, {Santos}  \& {Parente}}{{Kim}
  et~al.}{2000}]{Kim00b}
{Kim} Y.-K.,  {Santos} J.~P.,   {Parente} F.,  2000, \mn@doi [\pra]
  {10.1103/PhysRevA.62.052710}, \href
  {https://ui.adsabs.harvard.edu/abs/2000PhRvA..62e2710K} {62, 052710}

\bibitem[\protect\citeauthoryear{{Kissmann}}{{Kissmann}}{2014}]{Kissmann14a}
{Kissmann} R.,  2014, \mn@doi [Astroparticle Physics]
  {10.1016/j.astropartphys.2014.02.002}, \href
  {https://ui.adsabs.harvard.edu/abs/2014APh....55...37K} {55, 37}

\bibitem[\protect\citeauthoryear{{Knudsen}, {Brun-Nielsen}, {Charlton}  \&
  {Poulsen}}{{Knudsen} et~al.}{1990}]{Knudsen90a}
{Knudsen} H.,  {Brun-Nielsen} L.,  {Charlton} M.,   {Poulsen} M.~R.,  1990,
  \mn@doi [Journal of Physics B Atomic Molecular Physics]
  {10.1088/0953-4075/23/21/026}, \href
  {https://ui.adsabs.harvard.edu/abs/1990JPhB...23.3955K} {23, 3955}

\bibitem[\protect\citeauthoryear{{Kopp}, {B{\"u}sching}, {Strauss}  \&
  {Potgieter}}{{Kopp} et~al.}{2012}]{Kopp12a}
{Kopp} A.,  {B{\"u}sching} I.,  {Strauss} R.~D.,   {Potgieter} M.~S.,  2012,
  \mn@doi [Computer Physics Communications]
  {https://doi.org/10.1016/j.cpc.2011.11.014}, 183, 530

\bibitem[\protect\citeauthoryear{{Krumholz}, {Crocker}, {Xu}, {Lazarian},
  {Rosevear}  \& {Bedwell-Wilson}}{{Krumholz} et~al.}{2020}]{Krumholz20a}
{Krumholz} M.~R.,  {Crocker} R.~M.,  {Xu} S.,  {Lazarian} A.,  {Rosevear}
  M.~T.,   {Bedwell-Wilson} J.,  2020, \mn@doi [\mnras]
  {10.1093/mnras/staa493}, \href
  {https://ui.adsabs.harvard.edu/abs/2020MNRAS.493.2817K} {493, 2817}

\bibitem[\protect\citeauthoryear{{Lacki} \& {Thompson}}{{Lacki} \&
  {Thompson}}{2010}]{Lacki10b}
{Lacki} B.~C.,  {Thompson} T.~A.,  2010, \mn@doi [\apj]
  {10.1088/0004-637X/717/1/196}, \href
  {https://ui.adsabs.harvard.edu/abs/2010ApJ...717..196L} {717, 196}

\bibitem[\protect\citeauthoryear{{Lacki}, {Thompson}  \& {Quataert}}{{Lacki}
  et~al.}{2010}]{Lacki10a}
{Lacki} B.~C.,  {Thompson} T.~A.,   {Quataert} E.,  2010, \mn@doi [\apj]
  {10.1088/0004-637X/717/1/1}, \href
  {http://adsabs.harvard.edu/abs/2010ApJ...717....1L} {717, 1}

\bibitem[\protect\citeauthoryear{{Mathis}, {Mezger}  \& {Panagia}}{{Mathis}
  et~al.}{1983}]{Mathis83a}
{Mathis} J.~S.,  {Mezger} P.~G.,   {Panagia} N.,  1983, \aap, \href
  {http://adsabs.harvard.edu/abs/1983A%26A...128..212M} {128, 212}

\bibitem[\protect\citeauthoryear{{Merten}, {Becker Tjus}, {Fichtner},
  {Eichmann}  \& {Sigl}}{{Merten} et~al.}{2017}]{Merten17a}
{Merten} L.,  {Becker Tjus} J.,  {Fichtner} H.,  {Eichmann} B.,   {Sigl} G.,
  2017, \mn@doi [\jcap] {10.1088/1475-7516/2017/06/046}, \href
  {https://ui.adsabs.harvard.edu/abs/2017JCAP...06..046M} {2017, 046}

\bibitem[\protect\citeauthoryear{{M{\"u}cke}, {Rachen}, {Engel}, {Protheroe}
  \& {Stanev}}{{M{\"u}cke} et~al.}{1999}]{Mucke99a}
{M{\"u}cke} A.,  {Rachen} J.~P.,  {Engel} R.,  {Protheroe} R.~J.,   {Stanev}
  T.,  1999, \mn@doi [\pasa] {10.1071/AS99160}, \href
  {https://ui.adsabs.harvard.edu/abs/1999PASA...16..160M} {16, 160}

\bibitem[\protect\citeauthoryear{{Narayanan} \& {Krumholz}}{{Narayanan} \&
  {Krumholz}}{2017}]{Narayanan17a}
{Narayanan} D.,  {Krumholz} M.~R.,  2017, \mn@doi [\mnras]
  {10.1093/mnras/stw3218}, \href
  {https://ui.adsabs.harvard.edu/abs/2017MNRAS.467...50N} {467, 50}

\bibitem[\protect\citeauthoryear{{Narayanan} et~al.,}{{Narayanan}
  et~al.}{2021}]{Narayanan21a}
{Narayanan} D.,  et~al., 2021, \mn@doi [\apjs] {10.3847/1538-4365/abc487},
  \href {https://ui.adsabs.harvard.edu/abs/2021ApJS..252...12N} {252, 12}

\bibitem[\protect\citeauthoryear{{O'Neill}}{{O'Neill}}{2014}]{ONeill14a}
{O'Neill} M.~E.,  2014, Technical Report HMC-CS-2014-0905, PCG: A Family of
  Simple Fast Space-Efficient Statistically Good Algorithms for Random Number
  Generation.
Harvey Mudd College, Claremont, CA

\bibitem[\protect\citeauthoryear{{Oka}, {Geballe}, {Goto}, {Usuda}, {Benjamin},
  {McCall}  \& {Indriolo}}{{Oka} et~al.}{2019}]{Oka19a}
{Oka} T.,  {Geballe} T.~R.,  {Goto} M.,  {Usuda} T.,  {Benjamin} {McCall} J.,
  {Indriolo} N.,  2019, \mn@doi [\apj] {10.3847/1538-4357/ab3647}, \href
  {https://ui.adsabs.harvard.edu/abs/2019ApJ...883...54O} {883, 54}

\bibitem[\protect\citeauthoryear{{Papadopoulos}}{{Papadopoulos}}{2010}]{Papadopoulos10a}
{Papadopoulos} P.~P.,  2010, \mn@doi [\apj] {10.1088/0004-637X/720/1/226},
  \href {http://adsabs.harvard.edu/abs/2010ApJ...720..226P} {720, 226}

\bibitem[\protect\citeauthoryear{{Pattle}}{{Pattle}}{1959}]{Pattle59a}
{Pattle} R.~E.,  1959, \mn@doi [Q.~J.~Mech.~App.~Math.]
  {10.1093/qjmam/12.4.407}, 12, 407

\bibitem[\protect\citeauthoryear{{Patwary} et~al.,}{{Patwary}
  et~al.}{2016}]{Patwary16a}
{Patwary} M. M.~A.,  et~al., 2016, arXiv e-prints, \href
  {https://ui.adsabs.harvard.edu/abs/2016arXiv160708220P} {p. arXiv:1607.08220}

\bibitem[\protect\citeauthoryear{{Peretti}, {Blasi}, {Aharonian}  \&
  {Morlino}}{{Peretti} et~al.}{2019}]{Peretti19a}
{Peretti} E.,  {Blasi} P.,  {Aharonian} F.,   {Morlino} G.,  2019, \mn@doi
  [\mnras] {10.1093/mnras/stz1161}, \href
  {https://ui.adsabs.harvard.edu/abs/2019MNRAS.487..168P} {487, 168}

\bibitem[\protect\citeauthoryear{{Peskin} \& {Schroeder}}{{Peskin} \&
  {Schroeder}}{1995}]{Peskin95a}
{Peskin} M.~E.,  {Schroeder} D.~V.,  1995, {An Introduction to Quantum Field
  Theory}.
Addison-Wesley

\bibitem[\protect\citeauthoryear{{Prantzos} et~al.,}{{Prantzos}
  et~al.}{2011}]{Prantzos11a}
{Prantzos} N.,  et~al., 2011, \mn@doi [Reviews of Modern Physics]
  {10.1103/RevModPhys.83.1001}, \href
  {https://ui.adsabs.harvard.edu/abs/2011RvMP...83.1001P} {83, 1001}

\bibitem[\protect\citeauthoryear{{Roth}, {Krumholz}, {Crocker}  \&
  {Celli}}{{Roth} et~al.}{2021}]{Roth21a}
{Roth} M.~A.,  {Krumholz} M.~R.,  {Crocker} R.~M.,   {Celli} S.,  2021, \mn@doi
  [\nat] {10.1038/s41586-021-03802-x}, \href
  {https://ui.adsabs.harvard.edu/abs/2021Natur.597..341R} {597, 341}

\bibitem[\protect\citeauthoryear{{Rudd}, {Kim}, {Madison}  \& {Gay}}{{Rudd}
  et~al.}{1992}]{Rudd92a}
{Rudd} M.~E.,  {Kim} Y.~K.,  {Madison} D.~H.,   {Gay} T.~J.,  1992, \mn@doi
  [Reviews of Modern Physics] {10.1103/RevModPhys.64.441}, \href
  {https://ui.adsabs.harvard.edu/abs/1992RvMP...64..441R} {64, 441}

\bibitem[\protect\citeauthoryear{{Salem} \& {Bryan}}{{Salem} \&
  {Bryan}}{2014}]{Salem14a}
{Salem} M.,  {Bryan} G.~L.,  2014, \mn@doi [\mnras] {10.1093/mnras/stt2121},
  \href {https://ui.adsabs.harvard.edu/abs/2014MNRAS.437.3312S} {437, 3312}

\bibitem[\protect\citeauthoryear{{Sampson}, {Beattie}, {Krumholz}, {Crocker},
  {Federrath}  \& {Seta}}{{Sampson} et~al.}{2022}]{Sampson22a}
{Sampson} M.~L.,  {Beattie} J.~R.,  {Krumholz} M.~R.,  {Crocker} R.~M.,
  {Federrath} C.,   {Seta} A.,  2022, arXiv e-prints, \href
  {https://ui.adsabs.harvard.edu/abs/2022arXiv220508174S} {p. arXiv:2205.08174}

\bibitem[\protect\citeauthoryear{{Schultz}, {Krstic}, {Lee}  \&
  {Raymond}}{{Schultz} et~al.}{2008}]{Schultz08a}
{Schultz} D.~R.,  {Krstic} P.~S.,  {Lee} T.~G.,   {Raymond} J.~C.,  2008,
  \mn@doi [\apj] {10.1086/533579}, \href
  {https://ui.adsabs.harvard.edu/abs/2008ApJ...678..950S} {678, 950}

\bibitem[\protect\citeauthoryear{{Skilling}}{{Skilling}}{1975}]{Skilling75a}
{Skilling} J.,  1975, \mn@doi [\mnras] {10.1093/mnras/172.3.557}, \href
  {https://ui.adsabs.harvard.edu/abs/1975MNRAS.172..557S} {172, 557}

\bibitem[\protect\citeauthoryear{{Socrates}, {Davis}  \&
  {Ramirez-Ruiz}}{{Socrates} et~al.}{2008}]{Socrates08a}
{Socrates} A.,  {Davis} S.~W.,   {Ramirez-Ruiz} E.,  2008, \mn@doi [\apj]
  {10.1086/590046}, \href {http://adsabs.harvard.edu/abs/2008ApJ...687..202S}
  {687, 202}

\bibitem[\protect\citeauthoryear{{Strong} \& {Moskalenko}}{{Strong} \&
  {Moskalenko}}{1998}]{Strong98a}
{Strong} A.~W.,  {Moskalenko} I.~V.,  1998, \mn@doi [\apj] {10.1086/306470},
  \href {https://ui.adsabs.harvard.edu/abs/1998ApJ...509..212S} {509, 212}

\bibitem[\protect\citeauthoryear{{Strong}, {Moskalenko}  \& {Ptuskin}}{{Strong}
  et~al.}{2007}]{Strong07a}
{Strong} A.~W.,  {Moskalenko} I.~V.,   {Ptuskin} V.~S.,  2007, \mn@doi [Annual
  Review of Nuclear and Particle Science]
  {10.1146/annurev.nucl.57.090506.123011}, \href
  {https://ui.adsabs.harvard.edu/abs/2007ARNPS..57..285S} {57, 285}

\bibitem[\protect\citeauthoryear{{Tanaka}, {Nagai}  \& {Kamegai}}{{Tanaka}
  et~al.}{2021}]{Tanaka21a}
{Tanaka} K.,  {Nagai} M.,   {Kamegai} K.,  2021, \mn@doi [\apj]
  {10.3847/1538-4357/ac004c}, \href
  {https://ui.adsabs.harvard.edu/abs/2021ApJ...915...79T} {915, 79}

\bibitem[\protect\citeauthoryear{{Thompson}, {Quataert}, {Waxman}, {Murray}  \&
  {Martin}}{{Thompson} et~al.}{2006}]{Thompson06a}
{Thompson} T.~A.,  {Quataert} E.,  {Waxman} E.,  {Murray} N.,   {Martin} C.~L.,
   2006, \mn@doi [\apj] {10.1086/504035}, \href
  {http://adsabs.harvard.edu/abs/2006ApJ...645..186T} {645, 186}

\bibitem[\protect\citeauthoryear{{Uhlig}, {Pfrommer}, {Sharma}, {Nath},
  {En{\ss}lin}  \& {Springel}}{{Uhlig} et~al.}{2012}]{Uhlig12a}
{Uhlig} M.,  {Pfrommer} C.,  {Sharma} M.,  {Nath} B.~B.,  {En{\ss}lin} T.~A.,
  {Springel} V.,  2012, \mn@doi [\mnras] {10.1111/j.1365-2966.2012.21045.x},
  \href {http://adsabs.harvard.edu/abs/2012MNRAS.423.2374U} {423, 2374}

\bibitem[\protect\citeauthoryear{{Werhahn}, {Pfrommer}, {Girichidis}  \&
  {Winner}}{{Werhahn} et~al.}{2021a}]{Werhahn21b}
{Werhahn} M.,  {Pfrommer} C.,  {Girichidis} P.,   {Winner} G.,  2021a, \mn@doi
  [\mnras] {10.1093/mnras/stab1325}, \href
  {https://ui.adsabs.harvard.edu/abs/2021MNRAS.505.3295W} {505, 3295}

\bibitem[\protect\citeauthoryear{{Werhahn}, {Pfrommer}  \&
  {Girichidis}}{{Werhahn} et~al.}{2021b}]{Werhahn21c}
{Werhahn} M.,  {Pfrommer} C.,   {Girichidis} P.,  2021b, \mn@doi [\mnras]
  {10.1093/mnras/stab2535}, \href
  {https://ui.adsabs.harvard.edu/abs/2021MNRAS.508.4072W} {508, 4072}

\bibitem[\protect\citeauthoryear{{Wibking} \& {Krumholz}}{{Wibking} \&
  {Krumholz}}{2022}]{Wibking22b}
{Wibking} B.~D.,  {Krumholz} M.~R.,  2022, \mnras~in review, \href
  {https://ui.adsabs.harvard.edu/abs/2021arXiv210504136W} {p. arXiv:2105.04136}

\bibitem[\protect\citeauthoryear{{Wiener}, {Pfrommer}  \& {Oh}}{{Wiener}
  et~al.}{2017}]{Wiener17a}
{Wiener} J.,  {Pfrommer} C.,   {Oh} S.~P.,  2017, \mn@doi [\mnras]
  {10.1093/mnras/stx127}, \href {http://ads.nao.ac.jp/abs/2017MNRAS.467..906W}
  {467, 906}

\bibitem[\protect\citeauthoryear{{Yang}, {Kafexhiu}  \& {Aharonian}}{{Yang}
  et~al.}{2018}]{Yang18a}
{Yang} R.-z.,  {Kafexhiu} E.,   {Aharonian} F.,  2018, \mn@doi [\aap]
  {10.1051/0004-6361/201730908}, \href
  {https://ui.adsabs.harvard.edu/abs/2018A&A...615A.108Y} {615, A108}

\bibitem[\protect\citeauthoryear{{Yoast-Hull}, {Gallagher}, {Zweibel}  \&
  {Everett}}{{Yoast-Hull} et~al.}{2014}]{Yoast-Hull14a}
{Yoast-Hull} T.~M.,  {Gallagher} J.~S. I.,  {Zweibel} E.~G.,   {Everett} J.~E.,
   2014, \mn@doi [\apj] {10.1088/0004-637X/780/2/137}, \href
  {https://ui.adsabs.harvard.edu/abs/2014ApJ...780..137Y} {780, 137}

\bibitem[\protect\citeauthoryear{{Yoast-Hull}, {Gallagher}  \&
  {Zweibel}}{{Yoast-Hull} et~al.}{2016}]{Yoast-Hull16a}
{Yoast-Hull} T.~M.,  {Gallagher} J.~S.,   {Zweibel} E.~G.,  2016, \mn@doi
  [\mnras] {10.1093/mnrasl/slv195}, \href
  {http://adsabs.harvard.edu/abs/2016MNRAS.457L..29Y} {457, L29}

\bibitem[\protect\citeauthoryear{{Zweibel}}{{Zweibel}}{2017}]{Zweibel17a}
{Zweibel} E.~G.,  2017, \mn@doi [Physics of Plasmas] {10.1063/1.4984017}, \href
  {https://ui.adsabs.harvard.edu/abs/2017PhPl...24e5402Z} {24, 055402}

\makeatother
\end{thebibliography}




\appendix

\section{Details of the tree algorithm}
\label{app:kdtree}

Here we describe the algorithm we use to evaluate \autoref{eq:kde} to reconstruct integrals over the CR distribution function and their gradients. For convenience, we rewrite this equation here as
\begin{equation}
q = \sum_{s,i} g(\mathbf{x}_{si}) \Theta_{si} \Upsilon_{si} w_{si}
\label{eq:fieldqty}
\end{equation}
where $q$ is one of the quantities -- $n$, $P$, or $U$ -- on the left hand side of \autoref{eq:kde}, $w$ is a corresponding quantity computed from the properties of each CR packet -- $1$, $v_{si} p_{si}$, and $T_{si}$, and $g(\mathbf{x}_{si}) = K_{\tensor{H}}\left(\mathbf{x}-\mathbf{x}_{si}\right)$. We shall refer to the quantities $q$ appearing on the left hand as field quantities, and the quantities $w$ appearing on the right hand side as field weights. Note that the contribution from each packet is therefore the product of a purely geometric term $g(x_{\rm si})$ that depends only on packet position, a step indicator $\Theta_{si}$ that depends only on the relative rigidities of the packet for which the field quantity is being computed and the packets contributing to it, and a term $\Upsilon_{si} w_{si}$ that depends only on other properties of the contributing packet. Our algorithm is based on this decomposition. Further note that \autoref{eq:kdegrad}, describing the gradients of field quantities, can be written in a completely analogous fashion, simply by changing the geometric term to $g(\mathbf{x}_{si}) = -\tensor{H}_{\nabla}^{-1} \left(\mathbf{x}-\mathbf{x}_{si}\right) K_{\tensor{H}_\Delta}\left(\mathbf{x}-\mathbf{x}_{si}\right)$. We can therefore apply the same algorithm to gradients of field quantities, with only very minor modifications that we discuss below.

Given this discussion, we first describe how we construct the kd-tree in \aref{app:kdtree_build}, the algorithm we use to evaluate \autoref{eq:kde} in non-distributed memory calculations in \aref{app:kdtree_serial}, and then the extension to the distributed memory case in \aref{app:kdtree_parallel}.

\subsection{Building the kd-tree}
\label{app:kdtree_build}

We construct a balanced kd-tree by standard methods, and assign each leaf a bandwidth tensor $\tensor{H}$ as described in \autoref{ssec:reconstruction}. We then carry out an additional step: for each leaf $L$ in the tree, we compute $q_{R_n} = \sum_{s,i \in L}\Theta(R_{s,i}-R_n) \Upsilon_{si} w_{si}$ for a series of rigidities $R_n$ uniformly spaced in logarithm from the largest to the smallest packet rigidity present in the volume. In words, this sum represents the total maximum possible contribution that the packets in the leaf could make to a field quantity $q$ for a CR with rigidity $R_n$.

Once we have evaluated these sums for every leaf, we recursively compute the corresponding sums for every other node in the tree, by simply summing the results from that node's two children. In this way, for every node in the tree we record the maximum possible contribution $q_{R_n}$ that packets contained in that node could make to the field quantities of packets with rigidities $>R_n$.

\subsection{Shared memory}
\label{app:kdtree_serial}

We evaluate \autoref{eq:fieldqty} for all packets in the tree by processing one leaf at a time, using a separate OpenMP thread for each leaf. For each CR packet in the leaf, we compute its rigidity $R$ and find the corresponding rigidities $R_n$ that bracket it, i.e., we find $n$ such that $R_n < R < R_{n+1}$. The algorithm we apply relies on the \texttt{nodeList}: a list of all the non-leaf nodes of the tree we have examined so far, along with, for each node a central estimate $q_{\rm node}$ and an error $e_{\rm node}$ for the contribution that packets contained in that node make to the field quantities. The central estimates and errors have the property that the true contribution of the packets in each tree node, defined by \autoref{eq:fieldqty} evaluated for the packets within the node, lies strictly in the range $(q_{\rm node}-e_{\rm node}, q_{\rm node} + e_{\rm node})$. 

\textbf{(1)} Start the algorithm by adding the root node of the tree to \texttt{nodeList}. For each CR packet in the leaf whose field quantities are being computed, evaluate the minimum and maximum possible contributions $q_{\rm min}$ and $q_{\rm max}$ to the field quantities made by packets contained within the root node; the minimum is simply given by $q_{\rm min} = \min(g(\mathbf{x})) q_{R_{n+1}}$, and the maximum by $q_{\rm max} = \max(g(\mathbf{x})) q_{R_{n}}$, where here the minimum and maximum of $g$ are evaluated over the bounding box of the node. In words, we find the minimum possible contribution $q_{\rm min}$ by assuming that all the packets in the node are at the location $\mathbf{x}$ that makes the geometric factor $g(\mathbf{x})$ as small as possible, and that the target packet is at the largest possible rigidity, $R_{n+1}$; similarly, the maximum possible contribution arises if all the packets in the node are at the location where $g(\mathbf{x})$ has its maximum, and the target packet is at its smallest possible rigidity, $R_n$. We can then take $q_{\rm node} = (q_{\rm min} + q_{\rm max})/2$ and $e_{\rm node} = (q_{\rm max} - q_{\rm min})/2$. We add these quantities to \texttt{nodeList}.

\textbf{(2)} Evaluate the sum of $q_{\rm node}$ and $e_{\rm node}$ over all the nodes in \texttt{nodeList} and all packets in the leaf we are processing. Define the maximum possible relative error for each packet in the leaf by $\mathrm{RE} = \sum e_{\rm node} / (\sum q_{\rm node} - \sum e_{\rm node})$, where the maximum is over all packets; the true value of $q$ is guaranteed to differ from $\sum q_{\rm node}$ by at most a factor of RE for the packet with the largest error. If RE is below a user-specified tolerance for all packets in the leaf, terminate iteration and set $q = \sum q_{\rm node}$.

\textbf{(3)} If the RE exceeds the tolerance for any packet in the leaf, search through \texttt{nodeList} and find the node that makes the largest contribution to RE for the packet with the largest total RE. We then ``open'' this node by removing it from \texttt{nodeList}, and replacing it with its two children. If those children are not leaves, we compute $q_{\rm node}$ and $e_{\rm node}$ for them exactly as for the root node in step 1. If they are leaves, we compute $q_{\rm node}$ for them by evaluating the sum in \autoref{eq:fieldqty} directly for the packets those leaves contain, and setting the corresponding error $e_{\rm node}$ to zero.

\textbf{(4)} Go back to step 2, and repeat until RE is below the tolerance for all packets.

When applying this algorithm to the gradients of field quantities, in \texttt{nodeList} we have not only estimates of field quantities $q_{\rm node}$, but estimates of their gradients $\nabla q_{\rm node}$, and the corresponding uncertainties $e_{\rm node}$ and $\nabla e_{\rm node}$. Since $\nabla q$ is a vector quantity, we define the relative error as $\mathrm{RE} = |\sum \nabla e_{\rm node}| h_{\rm max} / (\sum q_{\rm node} - \sum e_{\rm node})$, where $h_{\rm max}$ is the largest eigenvalue of the bandwidth tensor $\tensor{H}$.

\subsection{Distributed memory}
\label{app:kdtree_parallel}

As described in \autoref{ssec:parallel}, in a distributed memory parallel calculation, on each rank we have a tree, some of whose nodes may have children that reside on a different MPI rank. It is therefore possible that, finding the node that makes the largest contribution to the relative errors (step 3 in \autoref{app:kdtree_serial}), we will find that node cannot be opened because its children are only available on another MPI rank. If this occurs, we remove the node from the node list and do not add any children, but we record the MPI rank on which those children live, and we keep track of the total central estimate $q_{\rm ext}$ and uncertainty $e_{\rm ext}$ contributed by such ``external'' nodes, and we define the relative error to include their contributions: $\mathrm{RE} = (\sum e_{\rm node} + e_{\rm ext}) / (\sum q_{\rm node} - \sum e_{\rm node} - e_{\rm ext})$, and similarly for gradients. We distinguish this from the ``local'' relative error $\mathrm{RE}_{\rm loc} = \sum e_{\rm node} / (\sum q_{\rm node} - \sum e_{\rm node})$.

If at any point while iterating, we reach a state where $\mathrm{RE}_{\rm loc}$ is below our target tolerance for all packets, but $\mathrm{RE}$ exceeds our tolerance for at least some of them, we conclude that we cannot complete those packets without access to the information contained on other MPI ranks. We therefore send an MPI request to all the ranks that contribute to $e_{\rm ext}$, and request that they evaluate their contribution to the target packets; they do so using the same algorithm as given in \autoref{app:kdtree_serial}, but rather than starting from the root node of the tree, they start from the roots of their own local trees, i.e., evaluating only their own contributions. Once they have computed these contributions with sufficient accuracy that the total estimate will be below the tolerance, they send these estimates back to the rank that made the request. This rank then sums the contributions from all external ranks to arrive at a final estimate for the field quantities.


\bsp	
\label{lastpage}
\end{document}